\documentclass[eqsecnum,preprint]{revtex4}

\makeatletter       

\renewcommand{\p@subsection}{}

\renewcommand{\p@subsubsection}{}

\makeatother

\newcommand{\lbfig}[1]{\refstepcounter{fig} \label{#1} }
\newcounter{fig}

\usepackage{amssymb}
\usepackage{amsmath}
\usepackage{bbm}
\usepackage{epsf}
\usepackage{epsfig}
\usepackage{psfrag}
\usepackage{citesort}

\newcommand{\beq}{\begin{equation}}
\newcommand{\eeq}{\end{equation}}
\newcommand{\beqa}{\begin{eqnarray}}
\newcommand{\eeqa}{\end{eqnarray}}
\newcommand{\Tr}{\text{Tr}}

\newcommand{\del}{\partial}
\newcommand{\delk}{\partial_{k_z}}
\newcommand{\tk}{\tilde{k}_0}
\newcommand{\tw}{\tilde{\omega}}

\newcommand{\deldag}{\mathbin{\partial\mkern-10.5mu\big/}}

\newcommand{\kdag}{\mathbin{k\mkern-9.8mu\big/}}

\newcommand{\bX}{\mathbf{X}}

\def\Slash#1{#1\kern-0.55em\raise.05ex\hbox{/}}
\def\slash#1{#1\kern-0.5em\raise.05ex\hbox{{$\scriptstyle /$}}}

\begin{document}

\pagenumbering{Roman}


\begin{titlepage}

\vskip 0.3in

\preprint{BNL-72342-2004-JA, HD-THEP-04-22}

\vskip 0.6in

\centerline{\LARGE \bf Transport equations for chiral fermions to order $\hbar$}
\vskip 0.2in
\centerline{\LARGE \bf and electroweak baryogenesis: Part~II}

\vskip 0.4in

\begin{center}

  {\Large \bf
   Tomislav Prokopec$^{\diamond}$,
   Michael G. Schmidt$^\diamond$ \\
   \vskip 0.2in
   and Steffen Weinstock$^\bullet$}

\vskip 0.3in
  {\it $^\diamond$ Institut f\"ur Theoretische Physik, Universit\"at Heidelberg\\
           Philosophenweg 16, D-69120 Heidelberg, Germany} \\

  \vskip 0.1in
  {\it $^\bullet$ Nuclear Theory Group, Brookhaven National Laboratory, Upton, NY 11973-5000, USA}

\end{center}

\vskip 0.4in

\centerline{\bf Abstract}
\vskip 0.2in

This is the second in a series of two papers.
While in Paper~I~\cite{PSW_1} we derive
semiclassical Boltzmann transport equations and study
their flow terms, here we make use of the results from that paper
and address the collision terms.
We use a model Lagrangean, in which fermions couple to scalars through
Yukawa interactions and approximate the self-energies 
by the one-loop expressions.
This approximation already contains important aspects of thermalization
and scatterings required for quantitative studies of transport in plasmas.  
We compute the CP-violating contributions to both the scalar and
the fermionic collision term.
CP-violating sources appear at order $\hbar$
in gradient and a weak coupling expansion, as a consequence
of a CP-violating split in the fermionic dispersion relation.
These represent the first controlled calculations of CP-violating collisional 
sources, thus establishing the existence of `spontaneous' baryogenesis sources
in the collision term.
The calculation is performed by
the use of the spin quasiparticle states established in Paper~I.
Next, we present the relevant leading order calculation of the relaxation
rates for the spin states and relate them to the more standard relaxation
rates for the helicity states. We also analyze the associated fluid equations,
and make a comparison between the source arising from the semiclassical
force in the flow term and the collisional source,
and observe that the semiclassical source
tends to be more important in the limit of inefficient diffusion/transport.

\vskip 0.4in

\footnoterule
\vskip 3truemm
{\small\tt
\noindent
$^\diamond$T.Prokopec@, M.G.Schmidt@thphys.uni-heidelberg.de \\
$^\bullet$weinstock@bnl.gov
}

\end{titlepage}

\pagenumbering{arabic}


\tableofcontents


\cleardoublepage
\section{Introduction}

This is the second part in a series of two closely related papers, where we study
the emergence of Boltzmann transport equations from quantum field theory
for particles with a space-time dependent mass, which often arises 
{\it via} a coupling to a varying background field. 
This setting is typically realised at phase transitions, an 
important example being electroweak scale baryogenesis, in which case 
particles acquire mass at the electroweak phase transition
through the Higgs mechanism. 
This paper is an immediate continuation of Paper~I~\cite{PSW_1}, 
to which we refer for a general introduction and motivation. 
We frequently refer to sections and equations of Paper~I, and 
indicate this by adding a prefix ``I.'' to the respective reference.

\vskip 0.05in

We work in a model~(I.2.13-I.2.14) 
with chiral fermions $\psi_{L,R}$ that couple to scalars
$\phi$ {\it via} a Yukawa interaction ${\cal L}_{\rm int}$ 
(see Eq.~(\ref{Yukawa_part2}) below):
\begin{eqnarray}
  {\cal L} &=&
              i \bar{\psi} \deldag \psi
            - \bar{\psi}_L m \psi_R - \bar{\psi}_R m^* \psi_L
            + \left(\del_\mu\phi \right)^\dagger\left( \del^\mu\phi\right)
            - \phi^\dagger M^2\phi
            + {\cal L}_{\rm int}
\,.
\label{lagrangean_part2}
\end{eqnarray}
Due to the interaction with some background field the masses 
$m$ and $M^2$ are space-time dependent, and they may contain complex elements
to allow for CP-violation.
In Paper~I, based on a 2PI effective action we derive
the Schwinger-Dyson equations for this system, from which we 
arrive at the Kadanoff-Baym equations~(I.2.54-I.2.55)
for the scalar and fermionic Wightman functions $\Delta^{<,>}$ and $S^{<,>}$, 
which in the Wigner representation read
\begin{eqnarray}
  \Big(  k^2 - \frac{1}{4} \del^2 + ik\cdot\del 
       - M^2{\rm e}^{-\frac{i}{2}\stackrel{\leftarrow}{\del}\cdot\del_k}
  \Big)\Delta^{<,>}
       - {\rm e}^{-i\diamond}\{\Pi_h\}\{ \Delta^{<,>}\}
       - {\rm e}^{-i\diamond}\{\Pi^{<,>}\}\{\Delta_h\}
   =     {\cal C}_\phi
\label{Wigner-space:scalar_eom_part2}
\\
  \Big(     \kdag
          + \frac i2 \deldag 
          - m_h
            {\rm e}^{-\frac{i}{2}\stackrel{\leftarrow}{\del}\cdot\,\del_k}
          - i\gamma^5m_a
            {\rm e}^{-\frac{i}{2}\stackrel{\leftarrow}{\del}\cdot\,\del_k}
  \Big)S^{<,>}
       -   {\rm e}^{-i\diamond}\{\Sigma_h\}\{ S^{<,>}\}
       -   {\rm e}^{-i\diamond}\{\Sigma^{<,>}\}\{ S_h\}
   =       {\cal C}_\psi
\,,
\label{Wigner-space:fermionic_eom_part2}
\end{eqnarray}
where $\Pi^{<,>}$, $\Pi_h$ ($\Sigma^{<,>}$, $\Sigma_h$) denote 
the scalar (fermionic) self-energies. 
The scalar and fermionic collision terms are of the form
\begin{eqnarray}
  {\cal C}_\phi &=&  \frac 12 {\rm e}^{-i\diamond}
                     \big( \{\Pi^>\} \{\Delta^<\} 
                         - \{\Pi^<\} \{\Delta^>\} 
                     \big)
\label{Cphi_part2}
\\
  {\cal C}_\psi &=&  \frac 12 {\rm e}^{-i\diamond}
                     \big( \{\Sigma^>\}\{S^<\} 
                         - \{\Sigma^<\}\{S^>\}
                     \big)
\,,
\label{Cpsi_part2}
\end{eqnarray}
where $\diamond\{a\}\{b\} \equiv (1/2)(\partial_x a\cdot\partial_k b 
                            - \partial_k a\cdot\partial_x b)$.
In section I.3 we argue that the last two terms on the left hand side of the
equations, which represent collisional broadening due to multiparticle effects, 
can be neglected. 

 We then show that to first order in gradients (equivalently, to linear order in 
$\hbar$) the scalar Wigner function 
is solved by the classical on-shell distribution function~(I.4.12)
\beq
   i\Delta^<(k,x)
 = \frac{\pi}{\omega_\phi}
                 \left[ 
                    \delta(k_0-\omega_\phi)
                   -\delta(k_0+\omega_\phi)
                 \right]n^\phi(k,x)
\,,
\label{ce-diag-solution_part2}
\eeq
where the dispersion relation is given by
$\omega_\phi = (\vec k^{\,2}+M^2)^{1/2}$. 
This solution applies to the one field case, but also to 
the diagonal entries of the Wightman function in the mass eigenbasis 
in the case with mixing scalars, provided no eigenvalues $M_i^2$
of $M^2$ are nearly degenerate, that is $\hbar k\cdot \partial\ll M_{i}^2-M_{j}^2
\;(\,\forall i,j)$.
From~(\ref{ce-diag-solution_part2}) it then follows that
particle and antiparticle densities are the on-shell 
projections~(I.4.31, I.4.34)

\beqa
  f^\phi_+(\vec k,x) 
  &=&  n^\phi \big(\omega_\phi(\vec{k},x),\vec{k},x\big)
\label{scalar_distribution_function_part2}
\\
  f^\phi_-(\vec k,x)
  &=&  n^{\phi cp} \big(\omega_\phi, \vec{k}, x \big)
   =  - \big[ 1 +  n^\phi \big(-\omega_\phi, -\vec{k}, x \big)
             \big] 
\,.
\label{particle-antiparticle_part2}
\eeqa
When working to order $\hbar$ accuracy, one gets the following 
Boltzmann equation for the CP-violating density,
$\delta f^\phi = f^\phi_+ - f^\phi_-$~(I.4.36)
\beqa
   \left(  \del_t
         + \frac{\vec{k}\cdot{\nabla}}{\omega_\phi}
         - \frac{(\partial_z{M^2}(z))}{2\omega_\phi}
           \partial_{k_z}\!
   \right)
   \delta f^\phi(\vec{k},x)
\!=\! \int_0^\infty\!\! \frac{dk_0}{2\pi} 
                        \big[
                  {\cal C}_\phi(k, x)
               +  {\cal C}^\dagger_\phi(k, x)
               +\,{\cal C}_\phi(- k, x)
               +  {\cal C}^\dagger_\phi(- k, x)
             \big]
.\quad\;
\label{scalar_pos_freq_deltaCP_part2}
\eeqa
%

\vskip 0.05in

The discussion of the fermionic kinetic equation in Paper~I begins with the
observation (see Refs.~\cite{KainulainenProkopecSchmidtWeinstock:2001} 
and~\cite{KainulainenProkopecSchmidtWeinstock:2002}) that, if masses only depend
on the z-coordinate ({\it e.g.} caused by planar phase interfaces),
the following spin operator~(I.5.18)
\beq
   S_z(k)
 = \frac{1}{\tilde k_0}
    \left(k_0 \gamma^0 - \vec k_\| \cdot \vec\gamma\right)\gamma^3\gamma^5
\,
\label{Sz_part2}
\eeq
is a conserved quantity. Here 
$\tilde{k}_0 = {\rm sign}(k_0)(k_0^2 - k_x^2 - k_y^2)^{1/2}$ 
is obtained by boosting $k_0$ to a frame, in which $k_x$ and $k_y$ vanish.
With the spin projector
$P_s(k) = (\mathbbm{1}+sS_z(k))/2\;$ $(s=\pm 1)$ the fermionic Wigner
function can accordingly be cast to a spin diagonal form~(I.5.25)
\beq
   S_s
 = iP_s \left[  s\gamma^3\gamma^5 g^{s}_0 
              - s\gamma^3         g^{s}_3 
              + {\mathbbm 1}      g^{s}_1 
              - i\gamma^5         g^{s}_2    \right]
\,.
\label{S<_decomposition2_part2}
\eeq
In stationary situations the densities
$g^s_a\; (a=0,1,2,3)$ are (real) functions of $z$ and momentum. 
Only one of these functions is independent. Indeed, by making use of the 
constraint equations, one can express $g^s_{1,2,3}$ in terms of $g^s_0$. 

 In Paper~I we show that, when the collisional broadening (width) is 
neglected, quasiparticle picture remains valid to order $\hbar$
(see also~\cite{KainulainenProkopecSchmidtWeinstock:2001} 
and~\cite{KainulainenProkopecSchmidtWeinstock:2002}), such that
the constraint equation for the vector density $g^s_0$ 
is solved by the following on-shell form~(I.5.71)
\beqa
     g_0^{<s}(k,x)
 &=& 2\pi \sum_{\pm} \frac{\delta(k_0\mp\omega_{\pm s})}
                          {2\omega_{\pm s}Z_{\pm s}} \,
     |\tilde{k}_0|\,n_s
\,,
\label{spectral-solution:fermions_part2}
\eeqa
where $n_s = n_s(k,x)$ is a distribution function,
to be determined by solving the kinetic equation, and 
$Z_{s}$ is the rate of change of the energy-squared around the poles
(divided by $2\omega_s$). 
The dispersion relation $\omega_s=\omega_s(k,z)$
acquires a CP violating, spin dependent,
shift due to the interaction with the background field~(I.5.72):
\beq
  \omega_s =  \omega_0
             - \frac{s}{2\tilde \omega_0\omega_0}
               |m|^2\partial_z {\theta}
\,.
\label{dispersion-relation:fermions_part2}
\eeq
Here $\theta$ is the phase of the complex mass and
%
$
         \omega_0 = (\vec{k}^2 + |m|^2)^\frac 12,
$
$
   \tilde\omega_0 = (k_z^2    + |m|^2)^\frac 12
.
$
%
The densities of particles and anti-particles with spin $s$ are the
on-shell projections~(I.5.120-121)
\beqa
     f_{s+}(\vec{k}, x) 
 &=& n_s(\omega_s, \vec{k}, x)
\label{onshell-density_part2}
\\
     f_{s-}(\vec{k}, x)
 &=& n^{cp}_s(\omega_s, \vec{k}, x)
  =  1 - n_{-s}(-\omega_s, -\vec{k}, x)
\,.
\label{onshell-density-cp_part2}
\eeqa
Besides an equilibrium part and a CP-even deviation from the
equilibrium density, these particle densities
contain spin dependent contributions $\delta f_{s\pm}$.
From these we form two combinations, which are related to vector and
axial density~(I.5.131-132):
\beqa
   \delta f^v_s &\equiv& \delta f_{s+} - \delta f_{s-}
\label{distribution-fv_part2}
\\
   \delta f^a_s &\equiv& \delta f_{s+} - \delta f_{-s-}
\,.
\label{distribution-fa_part2}
\eeqa
While the Boltzmann equation for the vector part does not have a
semiclassical source~(I.5.133)
\beq
   \big(  \del_t
        + \vec v_0 \cdot \nabla_{\vec x} 
        + F_0 \del_{k_z}
   \big) \delta f^v_s 
 = \int _0^\infty \frac{dk_0}{\pi}
    \big( {\cal K}^s_0(k) - {\cal K}^{cps}_0(k) \big)
\,,
\label{boltzmann:deltaf-v_part2}
\eeq
there is a CP-violating semiclassical source $s{\cal S}^{\rm flow}$~(I.5.135) 
in the equation for the distribution function related to the 
axial density~(I.5.134):
\beq
  \big(  \del_t
         + \vec v_0 \cdot \nabla_{\vec x} 
         + F_0 \partial_{k_z}
  \big)
       \delta f^a_s 
  + s{\cal S}^{\rm flow}
=  \int _0^\infty \frac{dk_0}{\pi}
      \left( {\cal K}^s_0(k) - {\cal K}^{cp-s}_0(k) \right)
\,,
\label{boltzmann:deltaf-a_part2}
\eeq
where $\vec v_0 = \vec k/\omega_0$ and $F_0 = - (\partial_z|m|^2)/(2\omega_0)$.
The collisional contributions on the right hand side are essentially
the traces of the spin projected fermionic collision term.


\vskip 0.05in

In Paper~I we formally kept the contributions of the collision term in all
expressions, as can be seen in Eqs.~(\ref{scalar_pos_freq_deltaCP_part2})
and~(\ref{boltzmann:deltaf-v_part2},\ref{boltzmann:deltaf-a_part2}), but
we never evaluated them. This will be done here. In the actual calculations
we truncate the self-energies in the collision term at the one-loop order.

%

\vskip 0.05in

Section~\ref{Collision term} is devoted to a calculation of CP-violating 
sources both in the scalar and in the fermionic collision term.
We emphasize that this represents the first ever calculation of CP-violating
sources arising from nonlocal (loop) effects. It would be incorrect to 
associate or identify our calculation with, for example,
Ref.~\cite{Riotto:1998}, where the author evaluates the self-energy for stops
and charginos by taking account of the contributions from two mass insertions,
which represent an approximation of the stop and charginos interactions
with the Higgs background.
Formally, all mass insertions are to be considered as singular
contributions to the self-energy, and hence ought to be viewed as part
of the (renormalized) mass term. We emphasize 
that in the formalism presented here all contributions to the 
self-energy arising from the mass insertions (resummed) are fully included
in the fermion masses in the flow term. Treating mass insertions 
in any other way is {\it not} in the spirit of Kadanoff-Baym equations,
and hence it is not guaranteed to yield correct results.

\vskip 0.05in

Making use of the Wigner functions obtained in Paper~I,
in section~\ref{Scalar collision term} we compute a CP-violating
collision source in the scalar kinetic equation. The source originates
from the fermionic Wigner functions, which we model by the CP-violating
equilibrium solutions accurate to the first order in $\hbar$. The source
appears in the fluid continuity equation. We also briefly discuss
the case of mixing fermions. This represents a first calculation of a 
CP-violating source in the scalar kinetic equation, and can be used, 
for example, for baryogenesis scenarios mediated by scalar particles,
{\it e.g.} Higgs particles or squarks. 

\vskip 0.05in

CP-violating collisional sources in the fermionic transport equations
are the subject of section~\ref{Fermionic collision term}.
We calculate the leading order CP-violating source, which appears first
when the self-energies are approximated at the one-loop order. 
The source is of first order in gradients, 
it is suppressed by the wall velocity, and 
its parametric dependence is qualitatively different from the semiclassical
force source. For example, the collisional source is suppressed by 
the Yukawa coupling squared, and it exhibits a kinematic mass threshold -- 
it vanishes when $M \leq 2|m|$, where $M$ denotes the scalar mass.
Moreover, the source appears in the fluid Euler equation, to be contrasted
with the semiclassical force, which sources the continuity equation. 
The source we compute arises from 
a nonlocal (one-loop) contribution to the collision term, and it 
differs qualitatively from the spontaneous baryogenesis sources obtained
so far in the literature~\cite{HuetNelson:1995+1996,Riotto:1995,Riotto:1998,
CarenaQuirosRiottoViljaWagner:1997,CarenaMorenoQuirosSecoWagner:2000,
CarenaQuirosSecoWagner:2002}, where the source is computed from the (local)
tree level mass insertions. 

\vskip 0.05in
 
 In order to study equilibration, 
in section~\ref{Relaxation to equilibrium} we use a linear response 
approximation to compute relaxation rates for spin quasiparticles
from the one-loop self-energies. 
When viewed in the Boltzmann collision term, these rates
capture an on-shell emission or absorption of a scalar particle by a fermion,
such that, as a consequence of the vertex energy-momentum conservation, 
the rates vanish in the limit of vanishing masses. Moreover,
for finite fermion and scalar masses, the rates are kinematically suppressed,
illustrating the shortcomings of the one loop approximation to the 
self-energies. Since the two-loop rates, which include the radiative vertex
corrections to the above processes, but also the (off-shell) scalar exchange, 
do not vanish in the massless limit, they typically dominate relaxation 
to equilibrium, even when Yukawa couplings are quite small. 
To calculate these rates in the collision term of the Kadanoff-Baym
equations would require an extensive calculational effort,
and it is beyond the scope of this work. 
We begin section~\ref{Relaxation to equilibrium}
by showing that thermalization rates do not depend on whether one uses
equilibrium Wigner functions (approximated at leading order in gradients), 
or the standard spinors (with a definite spin). 
Next, we point out that in the ultrarelativistic limit the spin rates reduce
to the (more standard) helicity rates, provided helicity is identified 
as the spin in the direction of propagation, which accords with our findings
in section~\ref{Kinetics of fermions: tree-level analysis}.
This  establishes a connection between the relaxation rates for spin and 
helicity states, which are almost exclusively quoted in literature. 

\vskip 0.05in

By taking moments of the transport equations obtained 
in section~\ref{Kinetics of fermions: tree-level analysis},
in section~\ref{Fluid equations}
we derive and analyze the fermionic fluid equations
for CP-violating densities and velocities.  
We observe that the continuity equation contains the CP-violating
source arising from the semiclassical force, while the Euler
equation -- which expresses the fluid momentum conservation -- contains
the collisional source. 
The earlier works on baryogenesis, which were based on
a heuristic WKB approach, have used helicity states, 
for which the semiclassical source appears in the Euler equation. 
Of course, helicity states can be obtained by a coherent superposition
of spin states, such that, in principle, it should not matter
which states one uses. In a quasiparticle picture, however,
 the information on quantum coherence is lost, 
and the dynamics of helicity states differ in general from that of 
spin states. Namely, at the order $\hbar$, the interactions with a
CP-violating Higgs background coherently mix the states of opposite helicity, 
which cannot be described in a quasiparticle picture.
Formally, one can combine the Boltzmann equations for different spin and 
sign of momentum (in the direction of the wall propagation),  
such that the source in the fluid equations appears in the Euler equation, 
just as if helicity were conserved. These new equations
correspond to taking different moments of the Boltzmann equations. 
Based on the above comments, it is clear that it would be incorrect
to identify these as the fluid equations for quasiparticles of
conserved helicity. In fact, this identification becomes legitimate
only in the extreme relativistic limit. 
Taking different moments of the Boltzmann equations,
such that the semiclassical source appears in different fluid equations,
results formally in {\it a priori} equally accurate fluid equations.
This observation lead us to conclude that these two sets of equations
can be used as a testing ground for the accuracy of fluid equations, 
which could otherwise be tested only by a comparison with the exact solution
of the Boltzmann equation. We perform this test
in section~\ref{Fluid equations}, and find that the two sets of 
diffusion equations give results that differ at about 10 percent level.
Finally, we show how, starting with fluid equations,
one arrives at a diffusion equation which contains both sources, 
and thus allows for a comparison of the strength of the semiclassical and
collisional sources at the one-loop level. 
Our conclusion is that unless diffusion is very effective the semiclassical
force dominates. A more detailed look at the sources and rates from
the one-loop collision term analysis indicates a mild suppression of 
the collisional source with respect to the flow term source. We expect 
that this suppression persists at higher loops as well. However,
to get a more quantitative estimate for the strength -- and thus relevance --
of the collisional sources, 
one would have to include the contributions from the two-loop self-energies,
which is beyond the scope of this work.


\cleardoublepage
\section{Collision term}
\label{Collision term}

In Paper~I we have considered tree-level interactions
of particles in a plasma with spatially dependent scalar and pseudoscalar 
mass terms. In this section we extend our analysis to include 
CP-violating contributions from the collision terms of both scalars and
fermions. Our study is mainly motivated by an ongoing dispute in literature
concerning the question whether the dominant CP-violating sources for
electroweak baryogenesis arise from the flow term or from the collision term. 
While here we present just a first attempt to address this tantalizing
question in a controlled manner, we believe that our analysis, which is based
on the collision term with the self-energies truncated at one loop,
serves as a good indicator for the typical strength of the collisional sources,
and may serve as a good guideline for a comparison of the 
baryogenesis sources from the flow term (semiclassical force) 
and collisional (spontaneous) sources. Some, but not all,
of the results discussed in this section have already been presented at 
conferences~\cite{KainulainenProkopecSchmidtWeinstock:2002b,
Prokopec:2002,ProkopecKainulainenSchmidtWeinstock:2003}.
 
\vskip 0.1in

The simple picture of collisional baryogenesis is based on the idea of
spontaneous baryogenesis~\cite{CohenKaplanNelson:1991}. According to this
idea, since particles and the corresponding CP-conjugate antiparticles
with opposite spin
see different energy-momentum dispersion relations in the presence of a
CP-violating bubble wall, rescatterings have a tendency to preferably
populate one type of particle states compared to the other,
 and thus source a CP-violating current. This current
then biases sphalerons, leading to baryon production. This idea has been
further developed and adapted to electroweak baryogenesis mediated by
supersymmetric particles, mostly charginos and neutralinos of the
MSSM~\cite{HuetNelson:1995+1996,Riotto:1995,Riotto:1998,
CarenaQuirosRiottoViljaWagner:1997,CarenaMorenoQuirosSecoWagner:2000,
CarenaQuirosSecoWagner:2002} and NMSSM~\cite{DaviesFroggattMoorhouse:1996}.
This approach is to be contrasted with the semiclassical force
mechanism~\cite{JoyceProkopecTurok:1995,JoyceProkopecTurok:1996,
ClineKainulainen:2000,ClineJoyceKainulainen:2000+2001,HuberSchmidt:2000,HuberSchmidt:2001,HuberJohnSchmidt:2001} 
discussed in detail in Paper~I.

\vskip 0.1in

In the classical Boltzmann equation the flow term, which is first order in
gradients, is balanced by the collision term of zeroth order in gradients,
and hence both can formally be considered to be of the order $\hbar^0$
in gradient expansion. We shall assume that our truncation of the flow
term at second order in gradients (formally at first order in $\hbar$) 
can be considered equivalent to a truncation of the collision term
at first order in gradients, so we will consistently neglect all higher
order terms here.

\vskip 0.1in

All Wigner functions in our work are on-shell and can eventually be
reduced to particle distribution functions~(\ref{ce-diag-solution_part2},
\ref{spectral-solution:fermions_part2}). These distribution functions
can be split up into a thermal distribution, the Bose-Einstein distribution
$n^\phi_{\rm eq}$ in the scalar case and the Fermi-Dirac distribution
$n_{\rm eq}$ for the fermions, and a correction function $\delta n^\phi$
and $\delta n$, respectively. Let us for the study of the collision term
assume that these deviations from the thermal distributions are suppressed
by at least one order in gradients in comparison to them. 
After expanding the sums $n^\phi_{\rm eq}+\delta n^\phi$ and
$n_{\rm eq}+\delta n$ in the collision term, we can organize the
contributions into two groups: those which contain correction
functions, and those which only contain the equilibrium distributions.
The terms in the first group are the relaxation terms one usually finds
on the right hand side of a Boltzmann equation. They try to drive the
deviations to zero and vanish for $\delta n^\phi = \delta n = 0$. We will
study them in more detail in section~\ref{Relaxation to equilibrium}.
The terms of the second group owe their presence to the fermionic Wigner
function, which contains first order effects besides the non-thermal
distributions. These terms are present even for
$\delta n^\phi = \delta n = 0$
and therefore precisely constitute the above mentioned collisional
sources for the correction functions.
There is a caveat, however: by assuming that the deviations are
suppressed by at least one gradient compared to the thermal distributions
we ignore possible contributions from the dynamical CP-even deviations
from equilibrium, which are formally of the order $O(v_w\hbar^0)$, and
hence may also contribute to the source in the collision term
(recall that in section~\ref{Boltzmann transport equation for
CP-violating fermionic densities}
we discussed how a CP-even deviation from equilibrium can source
a CP-violating deviation in the flow term~(\ref{boltzmann:deltaf-a_even})).

\vskip 0.1in

\begin{figure}[tbp]
\centerline{\hspace{.in} 
\epsfig{file=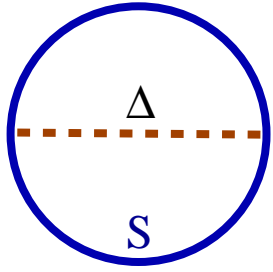, width=1.4in,height=1.4in}
}
\vskip -0.1in
\caption{\small
The two-loop diagram corresponding to the Yukawa interaction~(\ref{Yukawa_part2})
contributing to the 2PI effective
action~(\ref{Gamma2_part2}).
}
\lbfig{figure:2pi-2loop}
\end{figure}

We shall work in the model of chiral fermions
$\psi_{L,R}$~(\ref{lagrangean_part2}),
which couple {\it via} Yukawa interactions to scalars $\phi$,
\beqa
     {\cal L}_{\rm int}
 &=& -  \bar{\psi}_L  y\phi          \psi_R
     -  \bar{\psi}_R (y\phi)^\dagger \psi_L    
\nonumber\\
 &=& - \bar{\psi}\left( P_R \otimes y\phi + P_L \otimes (y\phi)^\dagger
                 \right) \psi 
\,,
\label{Yukawa_part2}
\eeqa
where $P_{L,R} = (1\mp \gamma^5)/2$ denote the chiral projectors, 
$y$ is the Yukawa coupling, which is a matrix (vector) in the scalar
(fermionic) flavor space, and $\otimes$ denotes a direct product of 
the spinor and flavor spaces. 
Provided the momenta are not too small, the dynamics of fermions and scalars
at the electroweak scale is expected to be perturbative,
and the simplest nontrivial contribution to the effective action 
$\Gamma_2$ in~(\ref{2PI_effective_action})
from the two-particle irreducible vacuum graphs
arises from the two-loop diagram shown in figure~\ref{figure:2pi-2loop}.
A perturbative treatment is justified as long as we are interested
in modelling the dynamics of nearly thermal quasiparticles, 
which are predominantly responsible for the sources and thermalisation
processes discussed in this work, since a majority of quasiparticles
carry nearly thermal momenta. After an easy calculation we find 
\begin{equation}
 \Gamma_{\rm 2}  
      =  - \int_{{\cal C}} d^4 u \, d^4 v \;
           {\rm Tr}\big[P_R \otimes  y^l             S(u,v)
                        P_L \otimes {y^{l'}}^\dagger S(v,u)
                   \big] 
                          \, \Delta_{ll'}(u,v)
\label{Gamma2_part2}
\,,
\end{equation}
from which we get the following one-loop expressions for the 
self-energies:
\begin{eqnarray}
        \Pi^{ac}_{ll'}(u,v)
  &=& \;\;\, iac \frac{\delta\Gamma_2[\Delta,S]}{\delta \Delta^{ca}_{l'l}(v,u)} 
\nonumber\\
  &=& - i {\rm Tr}\,\big[P_R \otimes y^{l'}        S^{ca}(v,u)
                         P_L \otimes {y^l}^\dagger S^{ac}(u,v) \big]
\label{Pi:ab2}
\\
\Sigma^{ac}(u,v)
     &=& - iac \frac{\delta\Gamma_2[\Delta,S]}{\delta S^{ca}(v,u)} 
\nonumber\\
  &=& i \big[ P_L \otimes {y^l}^\dagger S^{ac}(u,v)
              P_R \otimes y^{l'} \Delta_{l'l}^{ca}(v,u) 
            + P_R \otimes y^l S^{ac}(u,v)
              P_L \otimes {y^{l'}}^\dagger \Delta_{ll'}^{ac}(u,v)
\big]
\,.
\label{Sigma:ab2}
\end{eqnarray}
The indices $l$ and $l^\prime$ denote scalar flavor.
For the treatment of the collision term we need the self-energies
\beqa
            \Pi^{<,>}(u,v)
  &\equiv&  \Pi^{+-,-+}(u,v)
  =         -i {\rm Tr}\,\big[ P_R \otimes y S^{>,<}(v,u)
                               P_L \otimes y^\dagger S^{<,>}(u,v)\big]
\label{Pi-<}
\\
            \Sigma^{<,>}(u,v)
  &\equiv & \Sigma^{+-,-+}(u,v)
  =         i \big[ P_L \otimes y^\dagger S^{<,>}(u,v) 
                    P_R \otimes y \Delta^{>,<}(v,u) 
\label{Sigma-<}
\\
&&\hphantom{XXXXXXXXX}
                  + P_R \otimes y S^{<,>}(u,v)
                    P_L \otimes y^\dagger \Delta^{<,>}(u,v)
              \big]
\,.
\nonumber
\eeqa

Equations~(\ref{Wigner-space:scalar_eom_part2}-\ref{Wigner-space:fermionic_eom_part2})
together with~(\ref{Pi:ab2}-\ref{Sigma:ab2}) represent a closed set
of equations for the quantum dynamics of scalar and fermionic fields 
described by the Lagrangean~(\ref{lagrangean_part2}). 
This description is accurate in the weak coupling limit,
in which the Yukawa couplings satisfy $\| y \| \ll 1$. 
One can in fact study the quantum dynamics of scalar fields also
in the strongly coupled regime by making use of the 1/N expansion,
where N denotes the number of scalar field 
components~\cite{Berges:2002,BergesSerreau:2002,
                 AartsAhrensmeierBaierBergesSerreau:2002}.
Rather than studying this rather general problem, we proceed to 
develop approximations which are appropriate in the presence of slowly
varying scalar field condensates close to thermal equilibrium. In this case 
a gradient expansion can be applied to simplify the problem.

\vskip 0.1in

By making use of~(\ref{Gtt_by_G<>}) it is easy to show 
that the self-energies $\Pi^{t}$ and $\Pi^{\bar t}$
satisfy~(\ref{Pit-Pi<>}) with $\Pi^{\tt sg} = 0$, representing
a nontrivial consistency check of our loop approximation scheme
for the self-energies. Of course this check works as well for the fermionic
self-energies.

\subsection{Scalar collision term}
\label{Scalar collision term}

 We begin with the analysis of the collision term~(\ref{Cphi_part2})
of the scalar kinetic equation in the Wigner representation.
We expand the diamond operator to first order in gradients and
obtain two contributions, in the following referred to as zeroth
and first order collision term, respectively (although the ``zeroth
order'' contribution in fact is of first order in gradients, as we
will see):
\beqa
      {\cal C}_\phi &=& {\cal C}_{\phi 0} + {\cal C}_{\phi 1}
\nonumber\\
         &=& - \frac 12 \Big( i\Pi^> i\Delta^< - i\Pi^< i\Delta^> \Big)
          + \frac i2 \diamond 
         \Big(  \left\{ i\Pi^> \right\}\left\{ i\Delta^< \right\}
               -\left\{ i\Pi^< \right\}\left\{ i\Delta^> \right\} \Big)
\,.
\label{ScalarColl}
\end{eqnarray}
We shall now consider the scalar collision term for a single scalar
and fermionic field, and then generalise our analysis to the case of
scalar and fermionic mixing in the next section.

\subsubsection{The scalar collision term with no mixing}
\label{The scalar collision term with no mixing}

The one-loop expression for the single-field scalar self-energy
(vacuum polarization)~(\ref{Pi-<})
in the Wigner representation reads
\begin{equation}
   i\Pi^{<,>}(k,x)  
 = - |y|^2\, \int \frac{d^4k'd^4k''}{(2\pi)^8} (2\pi)^4\delta^4(k+k'-k'')
   {\rm Tr}\,\big[P_R iS^{>,<}(k',x) P_L iS^{<,>}(k'',x)\big]
\,,
\label{Pi-<>}
\end{equation}
where $P_{L,R} = (\mathbbm{1}\mp \gamma^5)/2$ are the projectors on
left- and right-handed chirality states.
A graphical representation of this one-loop self-energy is shown
in figure~\ref{fig:scalar-1loop0}. 
The hermiticity property~(\ref{S_herm_config}) of the fermionic
Wigner function immediately implies the hermiticity relation
\begin{equation}
\big( i\Pi^{<,>}(k,x) \big)^\dagger = i\Pi^{<,>}(k,x)
\,,
\label{Pi-<>_hermiticity}
\end{equation}
which holds in general for reasonable truncations of the scalar self-energy. 
\begin{figure}[tbp]
\centerline{
\epsfig{figure=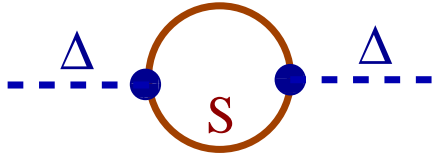, height=0.8in,width=2.3in}
}
\vskip -0.1in
\caption{\small
The one-loop scalar vacuum polarization diagram~(\ref{Pi-<>})
from the Yukawa interaction~(\ref{Yukawa_part2}),
which contributes to the collision term
of the scalar kinetic equation~(\ref{Wigner-space:scalar_eom_part2}). 
}
\lbfig{fig:scalar-1loop0}
\end{figure}
With the above expression for the self-energy we can write the
zeroth order collision term in~(\ref{ScalarColl}) as
\beqa
{\cal C}_{\phi 0} 
     &=& \frac {|y|^2}{2} \, \int \frac{d^4k'd^4k''}{(2\pi)^8}
         (2\pi)^4\delta^4(k+k'-k'')
\nonumber
\\
&\times&\Big\{
    {\rm Tr}\,\big[P_R iS^<(k') P_L iS^{>}(k'')\big]i\Delta^<(k)
- {\rm Tr}\,\big[P_R iS^{>}(k') P_L iS^{<}(k'')\big]i\Delta^>(k)
\Big\},
\label{C0-1}
\eeqa
where we suppressed the (local) $x$-coordinate dependence for
notational simplicity.
We insert the spin-diagonal decomposition~(\ref{S<_decomposition2_part2})
of the fermionic Wigner function and take the trace. Because of the
chiral projectors within the trace, the component functions $g_1^s$
and $g_2^s$ do not contribute. With the help of the relation
\begin{equation}
{\rm Tr}\, \Big[ P_R P_{s'}(k') P_{s''}(k'')\Big] 
     = \frac  12 \Bigg(1 + s's''
\frac{k_0'k_0'' - \vec k_\|'\cdot \vec k_\|''}{\tilde {k_0'}\tilde {k_0''}}
                \Bigg)
.
\label{TrSSP}
\end{equation}
we then find
\begin{eqnarray}
      {\cal C}_{\phi 0} 
  &=& \frac {|y|^2}{4} \,\sum_{s's''}
      \int \frac{d^4k'd^4k''}{(2\pi)^8} (2\pi)^4\delta^4(k+k'-k'')
      s's'' \Bigg(  1
                  + s's'' \frac{k_0' k_0'' -  \vec k_\|'\cdot \vec k_\|''}
                               {\tilde {k_0'}\tilde {k_0''}}
            \Bigg)
\nonumber\\
  &&
     \times \,
      \Big\{
            \Big(g_{0}^{s'<}(k')   - g_{3}^{s'<}(k')   \Big)
            \Big(g_{0}^{s''>}(k'') + g_{3}^{s''>}(k'') \Big) i\Delta^<(k)
\nonumber\\
  &&\hphantom{X}
     -      \Big(g_{0}^{s'>}(k')   - g_{3}^{s'>}(k') \Big)
            \Big(g_{0}^{s''<}(k'') + g_{3}^{s''<}(k'') \Big) i\Delta^>(k)
      \Big\}
\,.
\label{C0-2}
\end{eqnarray}

\medskip

In our tree-level analysis of fermions in
section~\ref{Kinetics of fermions: tree-level analysis} we found
relations that enabled us to express all fermionic component functions
in terms of $g^s_0$. Here we need only equation~(\ref{ced:diag3b})
for $g^s_3$, which for one fermionic particle reduces to
\beq
    g_3^{s<,>}
  = \frac{1}{\tilde k_0}
    \Big[
           sk_z \,         g_0^{s<,>}
         + \frac{1}{2\tilde k_0}|m|^2 \theta'
            \partial_{k_z} g_0^{s<,>}
         + \frac 12 {\cal C}_{\psi s3}
    \Big]
\,.
\label{coll_3by0}
\eeq
In the following we neglect the collisional contribution in this relation,
since it would result in terms which are of higher order in the coupling
constant. 
We explained in the beginning of this section that the CP-violating
sources, which contribute at first order in gradients to the collision
term, are obtained by replacing the distribution functions in the
on-shell {\it Ansatz} of the Wigner functions by the respective thermal
distributions.
For the remaining component $g^s_0$ of the fermionic Wigner function
we therefore use
\begin{eqnarray}
     g_{0\,\rm eq}^{s<}
 &=& \hphantom{-}
     2\pi \delta\Big(k^2 - |m|^2 + \frac{1}{\tilde k_0}s|m|^2\theta'\Big) 
     |\tilde {k}_0| n_{\rm eq}(\hat k_0)
\nonumber\\
     g_{0\,\rm eq}^{s>}
 &=& - 2\pi \delta\Big(k^2 - |m|^2 + \frac{1}{\tilde k_0}s|m|^2\theta'\Big)
     |\tilde {k}_0| (1 - n_{\rm eq}(\hat k_0))
\,,
\label{g0<>1}
\end{eqnarray}
with the Fermi-Dirac distribution
\beq
   n_{\rm eq}(\hat k_0)
 = \frac{1}{e^{\beta \hat k_0}+1}
\label{FermiDirac2}
\eeq 
instead of the full distribution function.
Note that we are working in the wall frame, which moves with the velocity
$v_w$ within the rest frame of the plasma. Boosting the fermionic
equilibrium Wigner function from the plasma frame to the wall frame 
results in the appearance of
\beq
   \hat{k}_0 \equiv \gamma_w (k_0 + v_w k_z)
\label{hat k0}
\eeq
in the argument of the equilibrium distribution.
%
%
The CP-violating first order effects in the fermionic Wigner functions that
have the potential to create collisional sources are the derivative
term in the relation~(\ref{coll_3by0}) between $g^s_3$ and $g^s_0$ and
the energy shift in~(\ref{g0<>1}).
Since in the scalar Wigner function there is no first order correction
other than the non-thermal distribution, we can here simply use the
equilibrium form
\beqa
     i\Delta_{\rm eq}^<(k)
 &=& 2\pi \delta(k^2-M^2) {\rm sign}(k_0) n^\phi_{\rm eq}(\hat{k}_0)
\label{Green_scalar_eq_<2}
\\
     i\Delta_{\rm eq}^>(k)
 &=& 2\pi \delta(k^2-M^2) {\rm sign}(k_0) (1+ n^\phi_{\rm eq}(\hat{k}_0))
\,,
\label{Green_scalar_eq_>2}
\eeqa
again with the Bose-Einstein distribution
\beq
  n^\phi_{\rm eq}(\hat{k}_0) = \frac{1}{{\rm e}^{\beta \hat{k}_0} - 1}
\label{BoseEinstein2}
\eeq
evaluated at $\hat{k}_0$.
All terms in~(\ref{C0-2}) without derivatives acting on $g^s_{0\,\rm eq}$
then drop out because of the KMS relations
({\it cf.} section~\ref{Equilibrium Green functions})
\beqa
     \Delta^>_{\rm eq}
 &=& {\rm e}^{\beta \hat{k}_0} \Delta^<_{\rm eq} 
\nonumber\\
     g_{0\,\rm eq}^{s>}
 &=& - e^{\beta \hat k_0} g_{0\,\rm eq}^{s<}
\label{g0-kms}
\eeqa
and the energy conserving delta function:
\begin{eqnarray}
     {\cal C}_{\phi 0} 
 &=& \frac {|y|^2}{8} |m|^2\theta' \,\sum_{s's''}
     \int \frac{d^4k'd^4k''}{(2\pi)^8} (2\pi)^4\delta^4(k+k'-k'')
     \Bigg(s's'' + \frac{k_0' k_0'' -  \vec k_\|'\cdot \vec k_\|''}
                        {\tilde {k_0'}\tilde {k_0''}}
     \Bigg)
\label{C0-3}
\\
&\times&
  \Bigg\{
   \bigg[ \frac{1}{\tilde {k_0'}^2} \partial_{k_z'}   g_{0\,\rm eq}^{s'<}(k')
          \Big(1-\frac{s''k_z''}{\tilde {k_0''}}\Big) g_{0\,\rm eq}^{s''>}(k'')
         -\Big(1+\frac{s' k_z' }{\tilde {k_0' }}\Big) g_{0\,\rm eq}^{s'<}(k')
          \frac{1}{\tilde {k_0''}^2}\partial_{k_z''}  g_{0\,\rm eq}^{s''>}(k'')
   \bigg] i\Delta^<_{\rm eq}(k)
\nonumber\\
&&\hphantom{i} -
   \bigg[ \frac{1}{\tilde {k_0'}^2} \partial_{k_z'}   g_{0\,\rm eq}^{s'>}(k') 
          \Big(1-\frac{s''k_z''}{\tilde {k_0''}}\Big) g_{0\,\rm eq}^{s''<}(k'')
         -\Big(1+\frac{s' k_z' }{\tilde {k_0' }}\Big) g_{0\,\rm eq}^{s'>}(k') 
          \frac{1}{\tilde {k_0''}^2}\partial_{k_z''}  g_{0\,\rm eq}^{s''<}(k'')
   \bigg] i\Delta^>_{\rm eq}(k)
  \Bigg\}
\,.
\qquad
\nonumber
\end{eqnarray}
This expression for the scalar collision term is already explicitly
of first order in gradients, so we can neglect the spin-dependent
shift in the spectral delta functions in~(\ref{g0<>1}). The resulting
spin-independent function, living on the unperturbed mass shell
$\delta(k^2-|m|^2)$, is denoted by $g_{0\,\rm eq}$.  We can now sum over
spins and, taking into account correction terms caused by the momentum
derivatives
\beq
   \del_{k_z} g_{0\,\rm eq}^{s>}(k)
 = - e^{\beta \hat k_0}
     \left(   \del_{k_z} g_{0\,\rm eq}^{s<} 
            + \beta \gamma_w v_w g_{0\,\rm eq}^{s<} 
     \right)
\,,
\label{KMS_deriv}
\eeq
use the KMS relation for the remaining terms to obtain
\begin{eqnarray}
     {\cal C}_{\phi 0} 
 &=& - \frac{|y|^2}{2}  \beta\gamma_w v_w |m|^2\theta' 
       \int \frac{d^4k'd^4k''}{(2\pi)^8} (2\pi)^4\delta^4(k+k'-k'')
\nonumber\\
&\times&
     \frac{k_0'k_0'' - \vec k_\|'\cdot\vec k_\|''}
          {\tilde {k_0'}\tilde {k_0''}}
     g^<_{0\,\rm eq}(k') g^>_{0\,\rm eq}(k'') i\Delta^<_{\rm eq}(k)
     \bigg(   \frac{1}{({\tilde k_0' })^2}
            + \frac{1}{({\tilde k_0''})^2}
     \bigg)
\,.
\label{C0-4}
\end{eqnarray}
This collisional source is proportional to the wall velocity $v_w$. This
displays the fact that for a wall at rest the leading order Wigner
functions indeed form an equilibrium solution.

At linear order in $v_w$ the equilibrium relation
\beq
   g_{0\,\rm eq}^{<,>}(-k_0')
 = - g_{0\,\rm eq}^{>,<}(k_0')
\label{g0_eq:momentum-flip}
\eeq
allows one to make the replacement $k'\leftrightarrow -k''$ in the
last term within brackets, so that Eq.~(\ref{C0-4}) can be rewritten as
\begin{eqnarray}
     {\cal C}_{\phi 0} (k)
 &=& - |y|^2 \frac{v_w |m|^2\theta'}{T} \,  
      i\Delta^<_{\rm eq}(k) {\cal I}^{>}_{\phi 0}(k) 
\label{C-phi0-e}
\\
     {\cal I}^{<,>}_{\phi 0}(k)
 &=& \int \frac{d^4k'd^4k''}{(2\pi)^4} \delta^4(k+k'-k'')
     \frac{1}{(\tilde {k_0'})^2}
     \frac{k_0' k_0'' -  \vec k_\|'\cdot \vec k_\|''}
          {\tilde {k_0'}\tilde {k_0''}} 
     g_{0\,\rm eq}^{>,<}(k') g_{0\,\rm eq}^{<,>}(k'')  
\,.
\label{calI}
\end{eqnarray}
Recall also that as a consequence of the KMS condition 
${\cal I}^{>}_{\phi 0}(k) i\Delta^<(k)
       = {\cal I}^{<}_{\phi 0}(k) i\Delta^>(k)$. 
In Appendix~\ref{A:CP-violating sources in the scalar collision term} 
we show how, by performing all but one of the integrations
in~(\ref{calI}), ${\cal I}_{\phi 0}\equiv {\cal I}_{\phi 0}^{>}$ can be
reduced to a single integral~(\ref{calI-f}), such that the scalar
collisional source simplifies to 
\begin{eqnarray}
{\cal C}_{\phi 0}  &=&
     - |y|^2 \frac{v_w |m|^2\theta'}{8T} \,  
       \frac{\theta(M-2|m|)}{\omega_\phi}
       \int_0^\infty \frac{dk'}{\omega_0'}\theta(k^2{k'}^2-\beta_0^2)
\Big[
  \theta(k_0)\delta(k_0-\omega_\phi)
  +\theta(-k_0)\delta(k_0+\omega_\phi)
\Big]
\nonumber\\
&\times&
\Big[
     \frac{M^2}{2|m|}\Re \frac{1}{\sqrt{(k_z - c_-)(c_+ - k_z)}}
  +  \Im \frac{ k_z}{\sqrt{(k_z - c_-)(c_+ - k_z)}} 
  -  \frac{k'}{k}
\Big]
\nonumber\\
&\times& f^\phi_0(1-\bar f_0')
        \Big[ \theta(\omega_0^{\prime\prime})f_0'' 
                   |_{\omega_0''=\omega_0^\prime-\omega_\phi}
            - \theta(\omega_0^{\prime\prime})(1-f_0'') 
                   |_{\omega_0''=\omega_\phi-\omega_0^\prime}
\Big]
\,.
\label{C-phi0-f}
\end{eqnarray}
The $c_\pm$ and $\beta_0$ are defined in Eqs.~(\ref{c+c-}) and~(\ref{a-pm})
in Appendix~\ref{A:CP-violating sources in the scalar collision term}, and 
\beq
          f^\phi_0
   \equiv n_{\rm eq}^\phi(\omega_\phi)
   =      \frac{1}{{\rm e}^{\beta \omega_\phi} - 1}
\,,\quad
          f_0 
   =      \bar{f}_0
   \equiv n_{\rm eq}(\omega_0)
   =      \frac{1}{{\rm e}^{\beta \omega_0} + 1}
\,.
\label{f_Phi:barf_Phi}
\eeq
Note that the source contains a mass threshold: it is nonzero only if
\begin{equation}
  M > 2|m|
\,.
\end{equation}
The physical reason for this threshold is that in the one loop self-energy
only on-shell absorption and emission processes are allowed. Together with
the energy-momentum conservation, this leads to the above condition.

\bigskip

We now make a slight digression and consider the first order collision
term in~(\ref{ScalarColl}). Because of the derivative in the diamond operator
this term is already of first order in gradients and we can use the
leading order equilibrium Wigner functions for its computation. Then the
KMS relations~(\ref{g0-kms}) induce a similar relation for the scalar
self-energy,
\beq
   \Pi_{\rm eq}^>(k)
 = {\rm e}^{\beta \hat k_0}\Pi_{\rm eq}^<(k)
\,,
\label{KMSPi}
\eeq
and we find
\beqa
     {\cal C}_{\phi 1}
 &=& -\frac 14       \Big(\del_{k_z} \beta\hat{k}_0
                     \Big)
              \del_z \Big( \Pi^>_{\rm eq}(k) i\Delta^<_{\rm eq}(k)
                     \Big)
\nonumber\\
 &=& \frac{|y|^2}{4}\beta\gamma_wv_w \partial_z
     \int \frac{d^4k'd^4k''}{(2\pi)^8}(2\pi)^4 \delta(k+k'-k'') 
     {\rm Tr}[P_R S_{\rm eq}^<(k') P_L S_{\rm eq}^>(k'')]
     i\Delta_{\rm eq}^<(k)
\,.
\label{C-phi1-a}
\eeqa
Following a similar procedure as in the leading order term, 
this can be reduced to
\begin{equation}
   {\cal C}_{\phi 1} 
 = -i\frac {|y|^2}{2}\beta\gamma_wv_w \partial_z 
    \int \frac{d^4k'd^4k''}{(2\pi)^8} (2\pi)^4\delta(k+k'-k'')
    \frac{k_0' k_0'' -  \vec k'\cdot \vec k''}{\tilde {k_0'}\tilde {k_0''}}
    g_{0\,\rm eq}^<(k') g_{0\,\rm eq}^>(k'') i\Delta^<_{\rm eq}(k)
\,.
\label{C-phi1-c}
\end{equation}
It is now clear
from Eqs.~(\ref{scalar_pos_freq_deltaCP_part2}) 
and~(\ref{scalar_pos_freq_deltaCP})
that this first order term
{\it contributes only to the constraint equation}, since 
$({\cal C}_{\phi 1})^\dagger = - {\cal C}_{\phi 1}$ is antihermitean.
Furthermore, one can show that this contribution to the constraint equation
results in a subdominant contribution in the kinetic equation, and hence
can be consistently neglected.

In anticipation of the discussion of fluid equations in
section~\ref{Fluid equations} we remark that the CP-violating sources
in the fluid equations are obtained by taking moments of the scalar
kinetic equation~(\ref{scalar_pos_freq_deltaCP_part2})
for the CP-violating distribution
function.
We note that the zeroth order collisional source
${\cal C}_{\phi 0}$~(\ref{C0-4})
is real and symmetric in the momentum argument:
\beq
{\cal C}_{\phi 0}^\dagger =  {\cal C}_{\phi 0}^* = {\cal C}_{\phi 0}
\,,\qquad
{\cal C}_{\phi 0}(-k,x) = {\cal C}_{\phi 0}(k,x)
\,.
\label{C0=C0}
\eeq
The symmetry is a simple consequence of the relation
\begin{equation}
\delta(k_0+k_0'-k_0'') g_{0\,\rm eq}^>(k_0') g_{0\,\rm eq}^<(k_0'') 
 \;\rightarrow\; \delta(-k_0+k_0'-k_0'') 
                       g_{0\,\rm eq}^<(k_0'') g_{0\,\rm eq}^>(k_0'), 
\label{KMS2}
\end{equation}
where we made use of Eq.~(\ref{g0_eq:momentum-flip}).
We furthermore recall that the first order source~(\ref{C-phi1-c})
is antihermitean, such that the following relation holds:
\begin{equation}
\frac 12 \big[
                  {\cal C}_{\phi}(k, x)
               +  {\cal C}^\dagger_{\phi}(k, x)
               +  {\cal C}^*_{\phi}(- k, x)
               +  {\cal C}^T_{\phi}(- k, x)
         \big]
               =  2{\cal C}_{\phi 0}(k, x)
\,.           
\label{Cphi_sum_CPviol}
\end{equation}
From Eq.~(\ref{calI-c}) and~(\ref{delta-omega2}) it follows that 
the integrand function in the scalar collisional source
${\cal C}_{\phi} = {\cal C}_{\phi 0}$ is of the form
$\int dk_z' F(k_zk_z',k_z^2,{k_z'}^2)$, so that simple changes of
variables imply
\beqa
\int_{-\infty}^0 d k_z \int dk_z' F(k_zk_z',k_z^2,{k_z'}^2) 
 &=& \;\;\;\int_0^{\infty} d k_z \int dk_z' F(k_zk_z',k_z^2,{k_z'}^2) 
\nonumber\\
\int_{-\infty}^0 d k_z \,k_z\, \int dk_z' F(k_zk_z',k_z^2,{k_z'}^2) 
 &=& - \int_0^{\infty} d k_z \,k_z \, \int dk_z' F(k_zk_z',k_z^2,{k_z'}^2) 
\,,
\label{scalar:collision:source:Form}
\eeqa
and we immediately conclude that the source in the scalar 
Euler fluid equation, which is obtained by taking the first moment of
Eq.~(\ref{scalar_pos_freq_deltaCP_part2}), vanishes:
\begin{equation} 
\int\frac{d^4k}{4\pi^4}\theta(k_0)\frac{k_z}{\omega_\phi}
                           {\cal C}_{\phi} = 0
\,.
\label{scalar:collision:sources}
\end{equation} 
The same conclusion is reached upon observing that ${\cal C}_\phi$ 
is symmetric under $k_z \rightarrow -k_z$. 
\begin{figure}[tbp]
\centerline{\hspace{.in} 
\epsfig{file=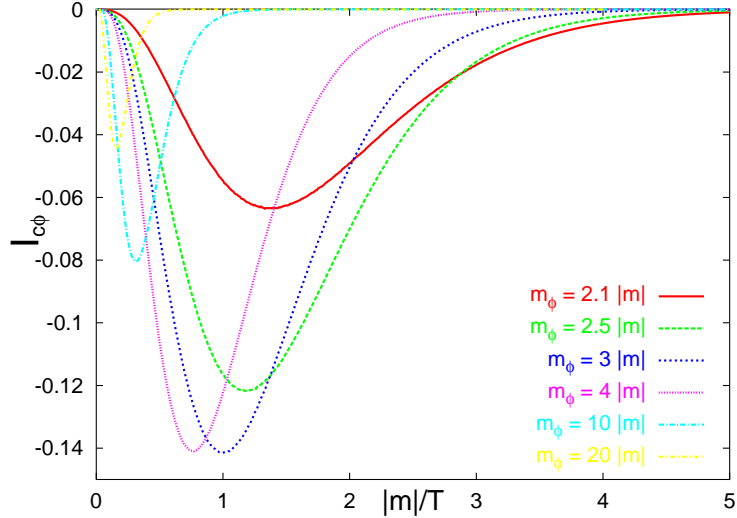, width=4.in}
}
\vskip -0.15in
\caption{\small
The integral in the CP-violating collisional source~(\ref{C-phi0-g})
of the scalar continuity fluid equation as a function 
of the mass parameters, $M/T$ and $|m|/T$.
}
\lbfig{fig:scalar_coll_source_1field}
\end{figure}
The source in the continuity equation, which is the 0th moment 
of~(\ref{scalar_pos_freq_deltaCP_part2}), does not vanish, however.
Indeed, upon performing the $k_0$, $k_z$ and angular
(azimuthal) integrations in~(\ref{C-phi0-f}), we obtain
\begin{eqnarray}
{\cal S}_\phi^{\rm coll}  &\equiv& 2\int\frac{d^4k}{(2\pi)^4} {\cal C}_{\phi 0} 
    = - \frac{|y|^2}{8\pi^3} v_w (|m|^2\theta')\,T \, \theta(M-2|m|)
         \; {\cal I}_{c\phi}
\nonumber\\
{\cal I}_{c\phi} &=&
     \beta^2 \int_0^\infty\frac{kdk}{\omega_\phi}
      \int_0^\infty \frac{dk'}{\omega_0'}\theta(k^2{k'}^2-\beta_0^2)
\nonumber\\
&\times&
\bigg[
  \frac{M^2}{2|m|}\arctan\Big(\frac{k'}{|m|}\Big)
- \frac{\beta_0}{k'}\bigg(1-\frac{|m|}{k'}\arctan\Big(\frac{k'}{|m|}\Big)\bigg)
- k'
\bigg]
\nonumber\\
&\times& f^\phi_0(1-\bar f_0')
        \Big[ \theta(\omega_0^{\prime\prime})(1-f_0'') 
                   |_{\omega_0''=\omega_0^\prime-\omega_\phi}
            - \theta(\omega_0^{\prime\prime})(1-f_0'') 
                   |_{\omega_0''=\omega_\phi-\omega_0^\prime}
\Big]
\,.
\label{C-phi0-g}
\end{eqnarray}
We now evaluate this collisional source numerically as a
function of the mass parameters, $M/T$ and $|m|/T$ ($M \geq 2|m|$).
The result is shown in figure~\ref{fig:scalar_coll_source_1field}.
The source peaks at $|m|\approx 1T$ and $M \approx (3-4) T$,
and it is (exponentially) Boltzmann suppressed when $|m|,M \gg T$. 
Note that the nature of the collisional source~(\ref{C-phi0-g}) is quite
different from the source~(\ref{boltzmann:deltaf-a_source})
arising from the semiclassical force in the flow term.
In particular, the source~(\ref{C-phi0-g}) is a consequence of 
CP-violation in the fermionic Wigner function, which circles in the one-loop 
self-energy shown in figure~\ref{fig:scalar-1loop0}, 
and it is hence suppressed by the Yukawa coupling squared, $|y|^2$. 
Moreover, the source exhibits a kinematic mass threshold, $M \geq 2|m|$,
which is a consequence of the on-shell approximation in 
the fermionic Wigner functions. The threshold is of course absent
in the semiclassical force source.
To get a rough estimate of how strong the source can be, 
we take $v_w \sim 0.03$, $|m|\sim T$, $M \sim 3T$, $|y|\sim 1$,  
$\theta'\sim 0.1T$ (maximum CP-violation), 
such that ${\cal S}_\phi^{\rm coll}\sim 10^{-6} T^4$.

\subsubsection{The scalar collision term with mixing}
\label{The scalar collision term with mixing}

We shall now study the scalar collision term in the case of scalar
and fermionic mixing, for which the Lagrangean~(\ref{Yukawa_part2}) reads
\begin{equation}
   {\cal L}'_{\rm Yu}
 = - \phi^l \bar{\psi}_L y^l \psi_R + {\rm h.c.} \,
 = - \bar{\psi} \left( P_R \otimes \phi^ly^l 
                      +P_L \otimes {\phi^l}^*{y^l}^\dagger \right) \psi
\,,
\label{Yukawa_mixing}
\end{equation}
where $y^l = (y^l_{ij})$ is a matrix in the fermionic flavor space and $l$ 
denotes the different scalar particles.
The one-loop scalar self-energy is a straightforward generalization
of~(\ref{Pi-<>}), 
\begin{eqnarray}
  i\Pi^{<,>}_{ll'}(k,x)
 &=& - \int \frac{d^4k'd^4k''}{(2\pi)^8} (2\pi)^4 \delta^4(k+k'-k'')
{\rm Tr}\Big(P_R \otimes y^l iS^{>,<}(k',x)
             P_L \otimes y^{l'\dagger} iS^{<,>}(k'',x) \Big)
,\qquad
\label{Pi_mixing}
\end{eqnarray}
where the trace is now to be taken both in spinor and in fermionic flavor
space. The mixing in the bosonic sector was handled in the flow term by
a unitary rotation into the basis in which the mass is diagonal
({\it cf.} section~\ref{Kinetics of scalars: tree-level analysis}),
resulting in Eq.~(\ref{scalar_pos_freq_deltaCP_part2}), so that an identical
rotation has to be performed also in the collision term~(\ref{ScalarColl}).

Let us consider first the influence of the scalar mixing.
The scalar collision term is now to be understood 
as a product of two matrices in the scalar flavor space.
After the rotation into the diagonal basis the form is maintained.
The self-energy gets rotated as 
$i\Pi^{<,>} \rightarrow i\Pi^{<,>}_d = Ui\Pi^{<,>}U^\dagger$,
which corresponds to the following redefinition of the couplings,
$y \rightarrow y_d = Uy$ and 
$y^\dagger \rightarrow {y}^{\,\dagger}_d = y^\dagger U^\dagger$
in~(\ref{Pi_mixing}). The diagonal contributions to the zeroth order
collision term contain the terms
$y_d^l {y_d^{l'}}^{\,\dagger}\, i\Delta^{<,>}_{d,l'l}$,
such that the off-diagonals $(1-\delta_{ll'})i\Delta^{<,>}_{d,l'l}$
may give rise to a CP-violating contribution when either the Yukawa
couplings $y_d^{l}$ or $i\Delta^{<,>}_{d,l'l}$ are CP-violating.
Treating the latter contributions
properly would require the dynamical treatment of the off-diagonals, which
is beyond the scope of this work. Note that there is no contribution
at order $\hbar$ coming from the diagonals of $i\Delta^{<,>}_{d}$.
For a diagrammatic illustration of the problem see 
figure~\ref{fig:scalar-1loop} below.
The diagonal elements of the first order collision term are imaginary,
and therefore do not contribute to the kinetic equation.

Next, we consider the question of fermionic mixing
({\it cf.} section~\ref{Flavor diagonalization}).
In this case the mass
$m$ in the free part of the fermionic Lagrangean~(\ref{lagrangean_part2}) 
is a non-hermitean matrix in flavor space. The diagonal basis is
defined by
%
\begin{equation}
     m_{d} \equiv UmV^\dagger
           = {\rm diag}\Big[|m_j|{\rm e}^{i\theta_j}\Big], 
\label{m-df}
\end{equation}
where $U$ and $V$ are the unitary matrices that diagonalize $m$.
Since the self-energy~(\ref{Pi_mixing}) is a trace in the fermionic flavor
space, its form stays invariant and we just have to replace the fermionic
Wigner function and the coupling matrix by the rotated versions. 
For notational simplicity we shall work with only one scalar
particle, and drop the superfluous indices $d$. 
The fermionic Wigner function may contain both diagonal and off-diagonal
elements in the flavor diagonal basis.

Let us first consider the contribution from the diagonal elements.
To this purpose we write the fermionic Wigner functions as 
$S^{<,>}_{ij} \rightarrow \delta_{ij}S^{<,>}_{i} + {\cal O}(\hbar)$, 
so that we have:
\begin{equation}
 i\Pi^{<,>}(k,x) = \sum_{ij}y_{ij}y^*_{ij}
     \int \frac{d^4k'd^4k''}{(2\pi)^8} (2\pi)^4 \delta^4(k+k'-k'') 
\nonumber\\
        {\rm Tr}\Big( P_R S^{>,<}_j(k',x) P_L S_i^{<,>}(k'',x) \Big).
\label{Pi_mixing-2}
\end{equation}
The trace is now to be taken in the spinor space only. We can go through
the same steps as in the non-mixing case of
section~\ref{The scalar collision term with no mixing}, always keeping
track of the indices attached to the fermionic functions.
One difference in comparison to the non-mixing case is that in
relation~(\ref{ced:diag3b}) between $g^s_3$ and $g^s_0$ there is an
additional term from the fermionic flavor rotation.
The expression analogous to~(\ref{C0-4}) is then
\begin{eqnarray}
  {\cal C}_{\phi 0}(k,x) &=&  -
\frac 12 \beta v_w \int\frac{d^4k'd^4k''}{(2\pi)^8} (2\pi)^4 \delta^4(k+k'-k'')
  \frac{k'_0k''_0 - \vec{k}'_\|\cdot\vec{k}''_\|}{\tilde{k}_0'\tilde{k}_0''}
\label{C0-4_mixing}
\\
&\times&
 \sum_{ij} |y_{ij}|^2 
  g^>_{0i\,\rm eq}(k') g^<_{0j\,\rm eq}(k'') i\Delta^>_{0\,\rm eq}(k)
        \left(  \frac{\big(|m|^2(\theta'+2\Delta_z)\big)_i}{\tilde{k}_0'^2}
  + \frac{\big(|m|^2(\theta'+2\Delta_z)\big)_j}
                    {\tilde{k}_0^{\prime\prime 2}} \right),
\nonumber
\end{eqnarray}
where $\Delta_z = \frac i2 (V\partial_zV^\dagger - U\partial_zU^\dagger)$.
Upon performing a similar change of variables as in the non-mixing case, 
$k_0'\leftrightarrow -k_0''$ and $ i\leftrightarrow j$, we have 
\beq
 \delta^4(k_0+k_0'-k_0'')g_{0i\,\rm eq}^{s'>}(k_0')g_{0j\,\rm eq}^{s''<}(k_0'') 
 \;\rightarrow\; 
\delta^4(-k_0+k_0'-k_0'')g_{0j\,\rm eq}^{s'<}(k_0'')g_{0i\,\rm eq}^{s''>}(k_0'), 
\label{KMS2i}
\eeq
such that Eq.~(\ref{C0-4_mixing}) can be recast as
\begin{eqnarray}
{\cal C}_{\phi 0}  &=&
     \sum_{ij}\frac {|y_{ij}|^2 + |y_{ji}|^2}{2} 
     \frac{v_w (|m|^2\theta')_i}{T} \,  
      {\cal I}^{>}_{\phi 0ij}(k) i\Delta^<_{\rm eq}(k)
\label{C-phi0-e:mixing}
\nonumber\\
 {\cal I}^{<,>}_{\phi 0ij}(k) &=&
      \int \frac{d^4k'd^4k''}{(2\pi)^8} (2\pi)^4\delta(k+k'-k'')
\frac{k_0' k_0'' -  \vec k_\|'\cdot \vec k_\|''}
   {\tilde {k_0'}\tilde {k_0''}} 
      g_{0i\,\rm eq}^{>,<}(k') g_{0j\,\rm eq}^{<,>}(k'')  
                 \frac{1}{(\tilde {k_0'})^2}
\,.
\label{calI:mixing}
\end{eqnarray}
This integral can be evaluated in part; upon performing analogous steps as 
in the one field case, one arrives at the result
\begin{eqnarray}
{\cal I}_{\phi 0ij}(k) &=& 
  -  \frac{\theta(M-|m_i|-|m_j|)}{8\pi^2}
       \int k'dk' \theta(k^2{k'}^2-\beta_0^2)
\int_{k'_{z_1}}^{k'_{z_2}} \frac{dk'_z}{(\tilde \omega_{0i}')^2}
    \frac{{k_z^\prime}^2 + k_z k_z^\prime 
        - \frac{M^2 - |m_i|^2 - |m_j|^2 }{2}}
         {\Big[(k'_{z_2} - k'_z)(k'_z - k'_{z_1})\Big]^\frac 12}
\nonumber\\
&\times&\Bigg[ 
         \theta(\omega_{0i}'-\omega_\phi)
         \left[\theta(-k_0) f_{0i}' (1-f_{0j}'') 
           +   \theta(k_0) (1-\bar f_{0i}') \bar f_{0j}'' \right]
\nonumber\\
&&
  -  \theta(\omega_\phi-\omega_{0i}')
      \left[\theta(-k_0)f_{0i}'\bar f_{0j}'' 
          + \theta(k_0)(1-\bar f_{0i}')(1-f_{0j}'')\right]
\Bigg]
\,,
\label{calI-mixing:2}
\end{eqnarray}
which, in the one field limit, reduces to Eq.~(\ref{calI-e}).
By making use of the integrals~(\ref{integrals-f-kz'}), the 
$k_z^\prime$-integral can be evaluated. The final $k'$ and $k$ integrals 
have to be performed numerically, the results are very similar
to the ones presented in figure~\ref{fig:scalar_coll_source_1field}.

\bigskip

Like in the non-mixing case the first order collision term is imaginary
and therefore does not contribute to the collisional sources at order
$\hbar$. The self-energy
stays imaginary, and the derivative from the diamond operator acts on the
self-energy as a whole and does not notice the rotation.
Finally, we remark on the contributions from the off-diagonal elements of the
Wigner functions, $(1-\delta_{ij}) iS^{<,>}_{ij}$, which are suppressed by
at least one power of $\hbar$. In the case when the fermionic interaction and
flavor bases do not coincide, it is clear that the off-diagonal elements of 
$iS^{<,>}_{ij}$ could in principle contribute to a CP-violating source at
order $\hbar$ in the bosonic collision term~(\ref{ScalarColl}). 
A proper treatment of this problem would require a dynamical treatment
of the off-diagonal elements of $iS^{<,>}_{ij}$
(in the flavor diagonal basis) and it is beyond the scope of this work.

\subsubsection{Discussion of scalar collisional sources}
\label{Discussion of scalar collisional sources}

\begin{figure}[tbp]
\centerline{
\epsfig{figure=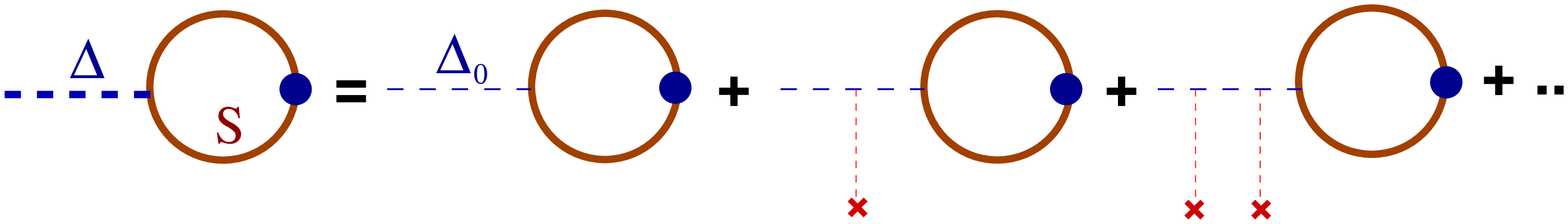, height=0.9in,width=5.8in}
}
\vskip -0.1in
\caption{\small
The one-loop collision term in the scalar kinetic equation 
in a theory with the Yukawa interaction~(\ref{Yukawa_part2}). 
The interactions of the scalar propagator with the scalar condensate
are of the same nature as the tree-level interactions discussed 
in section~\ref{Kinetics of scalars: tree-level analysis}. 
Just like in the tree-level case, at order $\hbar$,
scalars do {\it not} feel the condensate.}
\lbfig{fig:scalar-1loop}
\end{figure}
Contributions to the scalar CP-violating collisional source could, in 
principle, arise from the scalar or from the fermionic sector of 
the theory~(\ref{lagrangean_part2},\ref{Yukawa_part2}). 
In this section we have shown that, to order $\hbar$
in an expansion in gradients, there are no CP-violating contributions
to the collision term of the scalar kinetic equation arising from 
the scalar sector. Our proof includes the more general case of mixing scalars.
A simple qualitative argument in support of this is shown
in figure~\ref{fig:scalar-1loop}.

On the other hand, as we have argued in 
sections~\ref{The scalar collision term with no mixing}
and~\ref{The scalar collision term with mixing}, 
the fermionic contributions to the scalar one-loop self-energy
do result in CP-violating effects which contribute
to the collision term of the scalar kinetic equation. 
As illustrated in figure~\ref{fig:scalar-1loop-f}, these effects
arise from CP-violating contributions to the fermionic propagators
which run in the self-energy loop, and can be thought of as radiative
contributions to the CP-violating scatterings off the Higgs condensate. 
These contributions can source CP-violating scalar currents
in the Higgs or sfermionic sectors, for example, and thus may be 
of relevance for electroweak baryogenesis. From our one-loop study
of the CP-violating source shown in figure~\ref{fig:scalar_coll_source_1field}
it follows that: (a) the source has a very different parametric dependence
than the semiclassical source shown in figure~\ref{figure:flow-source},
and (b) the source 
tends to be large in the case of fermions whose mass is of the order
of the temperature,
and where the scalar mass is a few times bigger. With $T\sim 100~{\rm GeV}$,
popular extensions of the Standard Model, which include supersymmetric 
theories, typically contain such particles. 
\begin{figure}[tbp]
\centerline{
\epsfig{figure=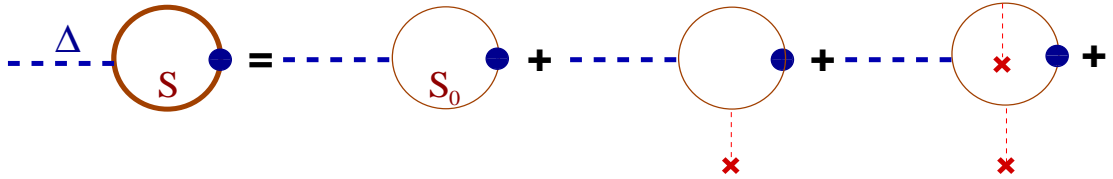, height=0.9in,width=5.8in}
}
\vskip -0.1in
\caption{\small
The collision term in the scalar kinetic equation as in 
figure~\ref{fig:scalar-1loop}. 
The interactions of the fermionic (loop) propagators 
with the scalar field condensate that can contribute 
in CP-odd manner at the order $\hbar$ to the scalar kinetic equation. 
}
\lbfig{fig:scalar-1loop-f}
\end{figure}
%
%
%


\subsection{Fermionic collision term}
\label{Fermionic collision term}

Now we turn our attention to the collision term for fermions.
Again, we expand in gradients, and distinguish between the zeroth and first
order collision terms,
\begin{equation}
  {\cal C}_\psi = {\cal C}_{\psi 0} + {\cal C}_{\psi 1} + ..
\,,
\label{C:psi}
\end{equation}
where 
\begin{eqnarray}
      {\cal C}_{\psi 0}
                &=& \frac 12\Big( \Sigma^>S^< - \Sigma^<S^> \Big)
\label{coll_ferm:psi0}
\\
{\cal C}_{\psi 1} &=& 
- \frac i2 \diamond 
                         \Big( \left\{\Sigma^>\right\}\left\{S^<\right\}
                          - \left\{\Sigma^<\right\}\left\{S^>\right\} \Big)
 \,.
\label{coll_ferm:psi1}
\end{eqnarray}
CP-violating contributions to the collision term appear, like 
in the scalar case, at first order in gradients. Since the calculational 
procedure of the sources is very similar to that in the scalar case, 
we shall outline only its main steps. 

\subsubsection{Fermionic collision term with no mixing}

The single-field fermionic self-energy, when approximated by 
the one-loop expression~(\ref{Sigma:ab2}) and written
in Wigner space, reads
\beqa
  i\Sigma^{<,>} &=& |y|^2
    \int\frac{d^4k'\,d^4k''}{(2\pi)^4}\, \delta^4(k-k'+k'')
\nonumber\\
&&\hskip 2.7cm \times\;   \big[ i\Delta^{>,<}(k'')  P_L iS^{<,>}(k') P_R
            + i\Delta^{<,>}(-k'') P_R iS^{<,>}(k') P_L   \big]  
\,.
\label{selfenergy_ferm_Wigner}
\eeqa
A graphical representation of this self-energy
is shown in figure~\ref{fig:fermionic-1loop}. 
\begin{figure}[tbp]
\centerline{
\epsfig{figure=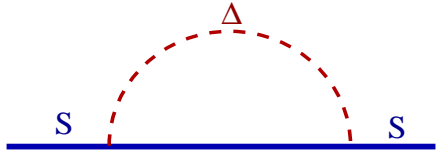, height=0.8in,width=2.3in}
}
\vskip -0.1in
\caption{\small
The one-loop fermionic vacuum polarization 
diagram ({\it cf.} Eq.~(\ref{selfenergy_ferm_Wigner}))
from the Yukawa interaction~(\ref{Yukawa_part2}),
which contributes to the collision term~(\ref{C:psi})
of the fermionic kinetic equation~(\ref{Wigner-space:fermionic_eom_part2}). 
}
\lbfig{fig:fermionic-1loop}
\end{figure}
We again use the spin diagonal 
{\it Ansatz}~(\ref{S<_decomposition2_part2}) for the
fermionic Wigner function, which in the self-energy shows up between two
chiral projectors. This can be simplified by
\begin{equation}
   P_L \sum_{s^\prime} S^{<,>}_{s^\prime}(k^\prime)P_R 
 = \sum_{s^\prime} iP_{s^\prime}(k^\prime)s^\prime P_L 
                \bigl[\gamma^3\gamma^5 g^{s^\prime<,>}_0(k^\prime)
                    - \gamma^3         g^{s^\prime<,>}_3(k^\prime)
                \bigr]  
\,,
\label{sandwich_b}
\end{equation}
and similarly for $P_L \leftrightarrow P_R$, such that the self-energy 
becomes
\begin{equation}
  i\Sigma^{<,>}(k) = -|y|^2 \sum_{s'} s'
    \int\frac{d^4k'\,d^4k''}{(2\pi)^4}\, \delta^4(k-k'+k'')
      P_{s'}(k') i \Delta^{>,<}(k'') 
        \left(\gamma^3\gamma^5 g^{<,>s'}_0(k')
           -  \gamma^3         g^{<,>s'}_3(k')           
        \right) 
\,.\quad\,
\label{selfenergy_ferm_first_explicit}
\end{equation}
Again we use thermal distributions for the Wigner functions, and with
\begin{equation}
 i\Delta^<_{\rm eq}(-k'') = i\Delta^>_{\rm eq}(k'')
\label{Delta<=Delta>}
\end{equation}
the chiral projectors drop out of the zeroth order collision term:
\begin{eqnarray}
 {\cal C}_{\psi 0} &=&
 \frac {|y|^2}{2} \sum_{s'}
\int \frac{d^4k'd^4k''}{(2\pi)^4}\,\delta^4(k-k'+k'') 
   \sum_{s,s^\prime}P_{s'}(k') P_s(k) s'
\label{calCpsi0}
\\
&&\times\;\Big[
     i\Delta^<_{\rm eq}(k'')
                     \Big( \gamma^3\gamma^5 g_{0\,\rm eq}^{s'>}(k')
                          -\gamma^3         g_{3\,\rm eq}^{s'>}(k')
                     \Big) 
                     \Big(s\gamma^3\gamma^5 g_{0\,\rm eq}^{s<}(k) 
                          - s\gamma^3       g_{3\,\rm eq}^{s<}(k) 
                          +                 g_{1\,\rm eq}^{s<}(k)
                          - i\gamma^5       g_{2\,\rm eq}^{s<}(k)
                     \Big) 
\nonumber\\
&& -\, i\Delta^>_{\rm eq}(k'')
                     \Big( \gamma^3\gamma^5 g_{0\,\rm eq}^{s'<}(k') 
                          -\gamma^3         g_{3\,\rm eq}^{s'<}(k')
                     \Big) 
                     \Big(s\gamma^3\gamma^5 g_{0\,\rm eq}^{s>}(k) 
                          - s\gamma^3       g_{3\,\rm eq}^{s>}(k) 
                          +                 g_{1\,\rm eq}^{s>}(k)
                          - i\gamma^5       g_{2\,\rm eq}^{s>}(k)
                     \Big) 
\Big]
\,.
\nonumber
\end{eqnarray}
It is not hard to see that, at leading order in gradients, the collision 
term vanishes.

The further procedure is now analogous to the scalar case. We use
equations~(\ref{ced:diag1b}-\ref{ced:diag3b}), simplified to the case
of only one fermionic particle, to express the functions $g^s_{i\,\rm eq}$
($i=1,2,3$) in terms of $g^s_{0\,\rm eq}$. 
The terms without derivatives acting on
$g^s_{0\,\rm eq}$ vanish because of the KMS relations~(\ref{g0-kms}) and
the energy-momentum conserving $\delta$-function, while Eq.~(\ref{KMS_deriv})
shows that all surviving terms are wall velocity suppressed
- as they should be:
\begin{eqnarray}
{\cal C}_{\psi 0} &=&
 \frac {|y|^2}{4}\beta\gamma_w v_w  
   \int \frac{d^4k'd^4k''}{(2\pi)^4}\,\delta(k-k'+k'') 
     \sum_{ss'} P_{s'}(k') P_s(k)
         g_{0\,\rm eq}^<(k)
         g_{0\,\rm eq}^>(k')
         i\Delta^<_{\rm eq}(k'')
\nonumber\\
&&\hskip 3.5cm\times\;\bigg\{\mathbbm{1}
                        \bigg[|m|^2\theta'
                         \bigg(
                           - s'\frac{k_z}{\tilde k_0}\frac{1}{(\tilde k_0')^2} 
                           + s\frac{1}{\tilde k_0^2}\frac{k_z'}{\tilde k_0'}
                         \bigg)
                        \bigg]
\nonumber\\
&&\hskip 4.cm +\;s\gamma^3\gamma^5
                  \bigg[
                      - iss'|m|^2\theta' \frac{m_a}{\tilde k_0(\tilde k_0')^2} 
                      -  s' m_a'\frac{k_z}{\tilde k_0^2}
                      +  im_h'\frac{k_z}{\tilde k_0^2}\frac{k_z'}{\tilde k_0'}
                        \bigg]
\nonumber\\
&&\hskip 4.cm -\;is\gamma^3\bigg[iss'|m|^2\theta'
                                \frac{m_h}{\tilde k_0}\frac{1}{(\tilde k_0')^2}
                             +  s'm_h'\frac{k_z}{\tilde k_0^2}
                             +  im_a'\frac{k_z}{\tilde k_0^2}
                                     \frac{k_z'}{\tilde k_0'}
                          \bigg]
\nonumber\\
&&\hskip 4.cm -\; \gamma^5\bigg[-ss'|m|^2\theta'
                             \Big(\frac{1}{\tilde k_0^2} 
                               +  \frac{1}{(\tilde k_0')^2}
                             \Big)
                            \bigg]
\bigg\}
\,.
\label{calCpsi0sum}
\end{eqnarray}
Since this expression is explicitly of first order, we switched to the
spin-independent functions $g_{0\,\rm eq}$.
Let us now have a look at the first order collision term~(\ref{coll_ferm:psi1}).
The fermionic equilibrium self-energy satisfies the KMS relation
\beq
\Sigma_{\rm eq}^>(k) = -{\rm e}^{\beta \hat k_0}\Sigma_{\rm eq}^<(k)
\,,\qquad \hat k_0 = \gamma_w (k_0 + v_w k_z)
\,,
\label{KMS:Sigma}
\eeq
so that - just as in the scalar case - we can write:
\begin{equation}
  {\cal C}_{\psi 1}
    = - \frac i4 \beta\gamma_w v_w
                \del_z \, \big(  \Sigma^< S^> \big)
\,.
\label{coll_ferm_first}
\end{equation}

Eventually we are interested in the contribution from the collision term
to the fermionic kinetic equation~(\ref{boltzmann:deltaf-a_part2}).
To obtain this contribution we have to multiply the collision term with
the spin projector $P_s(k)$ and take the trace. We find
\beq
    \Tr \,\big[P_{s'}(k') P_s(k)\big]
  = 1 + ss' \frac{k_0 k_0' - \vec k_\|\cdot\vec k_\|'}{\tilde k_0\tilde k_0'}
\,,
\eeq
while the traces containing additional factors of $\gamma^3$ and $\gamma^5$
vanish.

The trace of the first order collision term is imaginary and so does not
contribute to the kinetic equation. We find
\begin{eqnarray}
     {\cal K}^s_0(k)
 &=& - \Re \Tr \big[\mathbbm{1} P_s(k){\cal C}_{\psi}\big]
\nonumber\\
 &=& \frac {|y|^2}{2}\beta\gamma_w v_w  s|m|^2\theta'
      \int \frac{d^4k'd^4k''}{(2\pi)^4}\,\delta^4(k-k'+k'') 
         g_{0\,\rm eq}^<(k)
         g_{0\,\rm eq}^>(k')
         i\Delta_{\rm eq}^<(k'')
\nonumber\\
&&\hskip 5.5cm\times\; 
   \bigg(
       \frac{1}{\tilde k_0^2}\frac{k_z'}{\tilde k_0'} 
    -   \frac{k_z}{\tilde k_0}\frac{1}{(\tilde k_0')^2} 
        \frac{k_0 k_0' - \vec k_\|\cdot\vec k_\|'}{\tilde k_0\tilde k_0'}
   \bigg)
\,.
\label{C:psi:K}
\end{eqnarray}
This collisional source is symmetric under $k\rightarrow - k$, therefore
it contributes to equation~(\ref{boltzmann:deltaf-a_part2}) for
$\delta f^a_s$, where the collisional contribution is
\beq
   s{\cal S}^{\rm coll}
 = \int_0^\infty \frac{dk_0}{\pi}
    \left[ {\cal K}^s_0(k) - {\cal K}^{cp-s}_0(k) \right]_{src}
 = \int_0^\infty \frac{dk_0}{\pi}
    \left[ {\cal K}^s_0(k) + {\cal K}^{s}_0(-k) \right]_{src}
 = \int_0^\infty \frac{dk_0}{\pi}
    2 \left[ {\cal K}^s_0(k) \right]_{src}
\,,
\label{ferm-coll-src}
\eeq
but not to equation~(\ref{boltzmann:deltaf-v_part2}) for $\delta f^v_s$.
To further simplify this expression we keep only the terms that
contribute at linear order in the wall velocity $v_w$.
The $\delta$-functions in $g_{0\,\rm eq}$ and $i\Delta_{\rm eq}$ project
onto the classical shell, $k_0=\pm\omega_0$, $k'_0=\pm\omega'_0$ and
$k''_0=\pm\omega''_\phi$. Together with the energy-momentum conservation
(expressed through the delta function in Eq.~(\ref{C:psi:K})), this
leads to the constraint
\begin{equation}
  \omega_0 \omega'_0 - |m|^2 \ge \vec{k}\cdot\vec{k}' 
\,.
\end{equation}
After multiplying by $-2$ and adding $\vec{k}^2+\vec{k}'^2$ this can be
rewritten as
\begin{equation}
      \left( \omega_0 - \omega'_0 \right)^2 
  \le \bigl(\vec{k}-\vec{k}'\bigr)^2
  <   \bigl(\vec{k}-\vec{k}'\bigr)^2 + M^2
\,,
\end{equation}
where we assumed $M^2>0$. With the spatial part of the momentum conserving
$\delta$-function we recognize the last term as the square of $\omega''_\phi$
and finally conclude
\beq
  \pm( \omega_0 - \omega'_0 ) < \omega''_\phi  \,.
\eeq
This means, however, that after performing the $k'_0$, $k''_0$ and $\vec{k}''$
integrals the contributions with $\delta(\omega_0-\omega'_0\pm\omega''_\phi)$
vanish, as well as the ones which contain
$\delta(\pm(\omega_0+\omega'_0+\omega''_\phi))$, such that we find
\begin{eqnarray}
  {\cal K}^s_{0}
    &=& - \frac{|y|^2}{32\pi} s|m|^2\theta' \, \beta v_w
       \int d^3k'
      \frac{\tilde{\omega}_0\tilde{\omega}'_0}{\omega_0\omega'_0\omega''_\phi}
     \delta(\omega_0+\omega'_0-\omega''_\phi)       \nonumber\\
&&\hphantom{XXXXXXXXXXXXXXX}
  \times\;
    \Big[ \delta(k_0+\omega_0)    f_0     f'_0 (1+f^\phi_0)
         -\delta(k_0-\omega_0) (1-f_0) (1-f'_0)   f^\phi_0  \Big]  \nonumber\\
&&\hphantom{XXXXXXXXXXXXXXX}
  \times\;
         \left(\frac{k'_z}{\tw_0^2\tw_0'}
               +\frac{k_z}{\tw_0^2\tw_0^{'3}}(\omega_0\omega'_0 + \vec{k}_\|\cdot\vec{k}'_\|)
         \right) \,.
\label{coll_ferm_src_0_a}
\end{eqnarray}
With $\omega''_\phi=\omega_0+\omega'_0$, the following relation holds
\begin{equation}
   (1-f_0) (1-f'_0) f^\phi_0  =  f_0 f'_0 (1+f^\phi_0) \,,
\end{equation}
and after inserting this into~(\ref{coll_ferm_src_0_a}), we immediately
see that the collisional source is
\begin{eqnarray}
  {\cal K}^s_0  
    &=& -\frac{|y|^2}{32\pi}  s|m|^2\theta' \, \beta v_w
    \left[ \delta(k_0+\omega_0) - \delta(k_0-\omega_0) \right]  \nonumber\\
&\times& \int d^3k'
      \frac{1}{\omega_0\omega'_0\omega''_\phi\tw_0}
     \delta(\omega_0+\omega'_0-\omega''_\phi)
       f_0 f'_0 (1+f_0^\phi)  
         \left( k'_z
               +k_z \frac{\omega_0\omega'_0 + \vec{k}_\|\cdot\vec{k}'_\|}{\tw_0^{'2}}
         \right)
\,.
\label{coll_ferm_src_0_b}
\end{eqnarray}

Just like in the scalar case, parts of the $\vec{k}'$-integration
can be done analytically. The calculation is performed in 
Appendix~\ref{Collisional source in the fermionic kinetic equation}.

In section~\ref{Fluid equations} we derive fluid
equations for the CP-violating part of the fermionic distribution function.
For this we need the zeroth and first $k_z$-moment of the collisional source.
In appendix~\ref{Collisional source in the fermionic kinetic equation}
we show that the collisional source is odd under $k_z\leftrightarrow k'_z$,
so the zeroth moment vanishes. The first moment
\beq
    \int \frac{d^3k}{(2\pi)^3} \frac{k_z}{\omega_0}
    s{\cal S}^{\rm coll} (\vec{k})
  = 4 \int_{k_0>0} \frac{d^4k}{(2\pi)^4} \frac{k_z}{\omega_0} {\cal K}^s_0
  = v_w s|m|^2\theta' y^2 \frac{T^3{\cal I}}{16\pi^3|m|^2}
\,.
\label{1st_moment_of_coll_source}
\eeq
is nonzero, however. In figure~\ref{fig_coll} we plot the dimensionless
integral ${\cal I}$ as a function of the fermion mass $|m|$;
the scalar mass is chosen to be proportional to $|m|$.
The source vanishes for small values of the masses, suggesting that 
the expansion in gradients used here yields the dominant sources. 
Note that the source contributes only in the kinematically allowed region,
$M\geq 2 |m|$. When the masses are large, $|m|, M \gg T$, 
the source is, as expected, Boltzmann-suppressed. 
\begin{figure}[bh]
  \unitlength=1in
  \begin{center}
    \psfrag{xlab}{$\scriptstyle |m|/T \rightarrow$}    
    \psfrag{ylab}{$\scriptstyle {\cal I}$}    
    \psfrag{m21}[r]{$\scriptstyle M = 2.1|m|$}
    \psfrag{m25}[r]{$\scriptstyle M = 2.5|m|$}
    \psfrag{m3}[r]{$\scriptstyle M = 3|m|$}
    \psfrag{m4}[r]{$\scriptstyle M = 4|m|$}
    \psfrag{m10}[r]{$\scriptstyle M = 10|m|$}
    \psfrag{m20}[r]{$\scriptstyle M = 20|m|$}
    \includegraphics[width=3.5in,angle=-90]{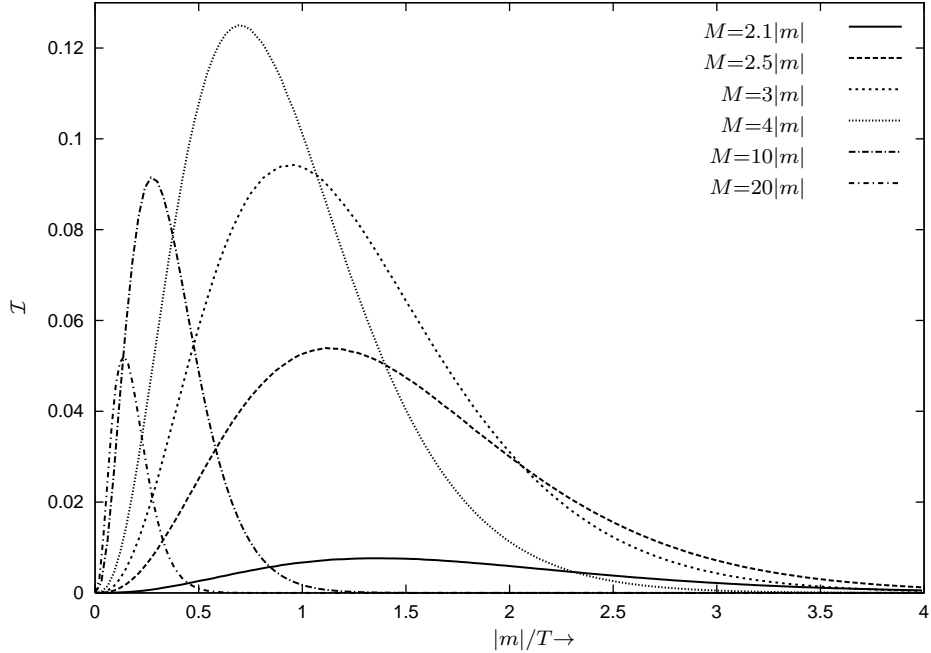}
  \end{center}
\lbfig{fig_coll}
\caption{\small
First moment of the collisional source (\ref{coll_ferm_src_0_b}) as a
function of the rescaled mass $|m|/T$. The scalar mass is set to be a multiple
of $|m|$.
}
\end{figure}

\subsubsection{Fermionic collision term with mixing}
\label{Fermionic collision term with mixing}

For several mixing scalar and fermionic particles the procedure
is very similar to what we have done in
section~\ref{The scalar collision term with mixing}
for the scalar collision term:
we apply flavor rotations to diagonalize the mass matrix in the
flow term of our equations and study the effects of these rotations
on the collision term.
The mixing Lagrangean~(\ref{Yukawa_mixing}) yields a fermionic
self-energy of the form
\beqa
     \Sigma^{<,>}(k,x)
 &=& \int\frac{d^4k'\,d^4k''}{(2\pi)^8} (2\pi)^4\delta^4(k-k'+k'')
\nonumber\\
&&\hphantom{XX}
   \Big( i\Delta^{>,<}_{ll'}(k'',x)
         P_L\otimes y^{l'\dagger} S^{<,>}(k',x) P_R\otimes y^l
\nonumber\\
&&\hphantom{XX}
       + i\Delta^{<,>}_{l'l}(k'',x)
         P_R\otimes y^{l'} S^{<,>}(k',x) P_L\otimes y^{l\dagger}
   \Big)
\,.
\label{self_energy_fermionic_mixing}
\eeqa 

The quantities with scalar flavor indices form a trace and so
their form does not change under the scalar flavor rotation, we only have
to replace $y\rightarrow y_d$ and $i\Delta \rightarrow i\Delta_d$.
We again neglect the off-diagonal components of the scalar Wigner
function, $i\Delta_{d,l'l} \rightarrow \delta_{ll'}i\Delta_{d,l}$,
and so the self-energy simply becomes a sum of self-energies without
mixing, with one contribution for each scalar mass-eigenstate. 
Since scalar mixing only appears within the self-energy, the derivatives
appearing in the first order collision term do not change this result.

It is therefore sufficient to study fermionic mixing with only one scalar
particle. After the fermionic flavor rotation, the diagonal components 
of the zeroth order collision term read
\beqa
 \big( \bX {\cal C}_{\psi0}  \bX^\dagger  \big)_{\tt ii}
      &=&   \int\frac{d^4k'\,d^4k''}{(2\pi)^8} (2\pi)^4\delta^4(k-k'+k'')
\label{coll_ferm_zeroth_fermix}\\
&&\hphantom{XX}
  \frac 12 \sum_j \Big(
       \big[ y^*_{ji}y_{ji}i\Delta^<(k'',x) \, P_L \,
                           S_j^>(k',x) \, P_R
\nonumber\\
&&\hphantom{XXXXXX}
            +y^*_{ij}y_{ij}i\Delta^>(k'',x) \, P_R \, 
                           S_j^>(k',x) \, P_L \big] S^<_i(k,x)   \nonumber\\
       &&\hphantom{XXXXXXXXXXXXXXXXXXXXXXX}
                 - ( < \leftrightarrow > )  \quad  \Big)
\,.
\nonumber
\eeqa
Here $y$ and $S$ denote already the rotated quantities, and we neglected
the flavor off-diagonal elements of the fermionic Wigner function,
$S_{ij} \rightarrow \delta_{ij}S_i$. With the same steps as in the
non-mixing case we arrive at the expression analogous
to~(\ref{C:psi:K}):
\beqa
    {\cal K}^s_{0\tt ii}
&=& \frac 14 \beta \gamma_w v_w
    \int\frac{d^4k'\,d^4k''}{(2\pi)^8} (2\pi)^4\delta^4(k-k'+k'')
    i\Delta^>_{\rm eq}(k'')
\label{coll_ferm_zeroth_src_trace_mixing}\\
&&
    \sum_j \frac{|y_{ij}|^2+|y_{ji}|^2}{2}
    g^<_{0j\rm eq}(k') g^>_{0i\rm eq}(k)
\nonumber\\
&&
    \sum_{s'} \Re \Tr P_{s'}(k')P_s(k)
  \bigg[ - s \frac{k_z'}{\tilde{k}_0'}
             \frac{1}{\tilde{k}_0^2}
             \left[|m|^2 (\theta' +2 \Delta_z) \right]_i
         + s'\frac{k_z}{\tilde{k}_0}
             \frac{1}{{\tilde{k_0'}}^2}
             \left[|m|^2 (\theta' +2 \Delta_z) \right]_j
  \bigg]
\,.
\nonumber
\eeqa
Like in the non-mixing case the contributions from the leading order functions
$g^<_{eq}$ and $i\Delta^>_{eq}$ vanish, if they lead to
$\delta(\omega_{0i}-\omega'_{0j}\pm\omega''_\phi)$ or
$\delta(\pm(\omega_{0i}+\omega'_{0j}+\omega''_\phi))$.
We find
\beqa
  {\cal K}^s_{0\tt ii}  
    &=& \frac 14 \beta v_w
    \left[ \delta(k_0+\omega_{0i}) - \delta(k_0-\omega_{0i}) \right]
\label{coll_ferm_src_0_b_mixing}\\
&&
      \int\frac{d^3k'}{(2\pi)^3} \frac{\pi^2}{2}
       \sum_j   \frac{|y_{ij}|^2+|y_{ji}|^2}{2}
      \frac{1}{\omega''_\phi}
     \delta(\omega_{0i}+\omega'_{0j}-\omega''_\phi)
       f_{0i} f'_{0j} (1+f^\phi_0)   \nonumber\\
&&
         \left( s\big(|m|^2(\theta'+2\Delta_z)\big)_i
                  \frac{k'_z}{\omega_{0i}\tw_{0i}\omega'_{0j}}
               +s\big(|m|^2(\theta'+2\Delta_z)\big)_j
                  \frac{k_z}{ \omega_{0i}\tw_{0i}\omega'_{0j}}
                           \frac{\omega_{0i}\omega'_{0j}
                                 + \vec{k}_\|\cdot\vec{k}'_\|}{\tw_{0j}^{'2}}
         \right)   \,.
\nonumber
\eeqa
From here on one can proceed as we did in the non-mixing case and perform
some of the integrals. This is shown in
appendix~\ref{Collisional source in the fermionic kinetic equation}.

In the first order collision term for the fermionic equation the
diagonalization leads to extra terms with spatial derivatives acting on the
rotation matrices. But none of them contributes to the relevant trace for
the kinetic equation for $g^s_0$.


\cleardoublepage
\section{Relaxation to equilibrium}
\label{Relaxation to equilibrium}

In the beginning of the last chapter we explained that the collision term
consists of two parts: the sources, which are caused by the structure
of the fermionic Wigner function,
and relaxation terms due to the deviation of the distribution functions
from thermal distributions. This chapter deals with these relaxation terms.
In a first section, considering as example the scalar collision term,
we show that the relaxation part of the collision term is equivalent
to the right hand side of a usual Boltzmann equation in our one-loop
approximation.
In a second section we give explicit expressions for the relaxation terms
relevant for the kinetic
equations~(\ref{scalar_pos_freq_deltaCP_part2}) 
and~(\ref{boltzmann:deltaf-v_part2}--\ref{boltzmann:deltaf-a_part2}), and 
comment on their consistency with the flow terms.
In this chapter we work with only one scalar and one fermionic particle.
The extension to several mixing particle species is quite
straightforward.

\subsection{Boltzmann collision term}

Since we only keep terms up to first order in gradients in the collision
term, we can for the sake of calculating the relaxation term use the leading
order expressions~(\ref{Green_scalar_eq_<}--\ref{Green_scalar_eq_>})
for the scalar Wigner functions, however with the thermal distributions
$n^\phi_{\rm eq}= 1/({\rm e}^{\beta \hat k_0}-1)$ replaced by the function 
$n^\phi = n^\phi_{\rm eq} + \delta n^\phi$, which is a sum of the thermal 
distribution and an unknown first order correction,
$\delta n^\phi = \delta n^\phi(k,x)$.
For the fermionic Wigner function we use the spin-diagonal expressions
({\it cf.} Eqs~(\ref{Green_fermionic_eq_<}--\ref{Green_fermionic_eq_>}))
\beqa
 S^<(k,x)  &=&   \hphantom{-}
              2\pi i \sum_s P_s(k)
              (\kdag + m_h - i\gamma^5 m_a)
              \delta(k^2 - |m|^2) {\rm sign}(k_0) 
              n_s(k,x)
\label{Green_fermionic_eq_s<}
\\
 S^>(k,x)  &=&  -
              2\pi i \sum_s P_s(k)
              (\kdag + m_h - i\gamma^5 m_a)
              \delta(k^2 - |m|^2) {\rm sign}(k_0) 
              (1-n_s(k,x))
\,,
\label{Green_fermionic_eq_s>} 
\eeqa
where 
\begin{equation}
  n_s = \frac{1}{{\rm e}^{\beta\hat k_0}+1} + \delta n_s
\,,\qquad  \hat k_0 = \gamma_w (k_0 + v_w k_z)
\label{n_s}
\end{equation}
denotes the Fermi-Dirac distribution plus an unknown, spin
dependent, correction, $\delta n_s=\delta n_s(k,x)$. 
Note that the spin projection operator {\it per} construction
commutes with $\kdag$.

Since the scalar first order collision term contains an explicit
derivative, any first order correction from the distribution functions
would lead to second order terms and therefore can be neglected.
With the above Wigner functions the scalar zeroth order collision
term~(\ref{C0-1}) becomes
\beqa
     {\cal C}_{\phi 0}
 &=& -\frac {|y|^2}{2} \, \int \frac{d^4k'd^4k''}{(2\pi)^8}
         (2\pi)^4\delta^4(k+k'-k'')
          \sum_{s's''}
          \Tr \left[ P_R P_{s'}(k') \kdag' P_{s''}(k'') \kdag'' \right]
\nonumber\\
 &&      2\pi \delta({k'}^2-|m|^2)  {\rm sign}(k'_0) \,\,
         2\pi \delta({k''}^2-|m|^2) {\rm sign}(k''_0)\,\,
         2\pi \delta(k^2-M^2)       {\rm sign}(k_0)
\nonumber\\
 &&      \Big[  n_{s'}(k')
                \left(1-n_{s''}(k'')\right)
                n^\phi(k)
               -\left(1-n_{s'}(k')\right)
                n_{s''}(k'')
                \left(1+n^\phi(k)\right)
         \Big]
\,.
\label{ScalCollRel}
\eeqa
We introduce the notation
\beq
   F_{s's''}(k',k'')
 = \Tr \left[ P_R P_{s'}(k') \kdag' P_{s''}(k'') \kdag'' \right]
\label{trace-function}
\eeq
for the trace appearing in this expression and note that this object is
antisymmetric in both of its momentum arguments.
%
Upon having a closer look at the Boltzmann
equation~(\ref{scalar_pos_freq_deltaCP_part2}) for scalars that we derived in 
section~\ref{Boltzmann transport equation for CP-violating scalar densities},
we see that the collision term appearing on the right hand side effectively is
\beq
   \frac{1}{\pi}\int_0^\infty dk_0 \, {\cal C}_{\phi}(k,x)
\,,
\label{Cphi:2}
\eeq
where ${\cal C}_{\phi}$ is defined in~(\ref{ScalarColl}).
After performing the $k'_0$-and $k''_0$-integrations and
making some rearrangements, the relaxation contribution
to~(\ref{ScalCollRel}--\ref{Cphi:2}) can be written as
\beqa
&&
 \frac{1}{\pi}\int_0^\infty dk_0 \, {\cal C}_{\phi 0}(k,x)
 = -\frac {|y|^2}{2} \,
       \int \frac{d^3k'd^3k''}{(2\pi)^6 2\omega_0'2\omega_0''}
        \sum_{s's''} F_{s's''}(\omega_0',\vec{k}',\omega_0'',\vec{k}'')
\label{ScalCollRel2}\\
&&\hphantom{x}
  \times
      \Big\{
         (2\pi)^4\delta^4(k+k'-k'') 
         \left[    f_{s'+}(\vec{k}') (1-f_{s''+}(\vec{k}'')   f^\phi(\vec{k})
               -(1-f_{s'+}(\vec{k}'))   f_{s''+}(\vec{k}'') (1+f^\phi(\vec{k}))
         \right]
\nonumber\\
&&\hphantom{X2}
 +       (2\pi)^4\delta^4(k-k'-k'') 
         \left[ (1-f_{s'-}(\vec{k}'))(1-f_{s''+}(\vec{k}''))   f^\phi(\vec{k})
               -   f_{s'-}(\vec{k}')    f_{s''+}(\vec{k}'') (1+f^\phi(\vec{k}))
         \right]
\nonumber\\
&&\hphantom{X2}\,
 +       (2\pi)^4\delta^4(k-k'+k'') 
         \left[ (1-f_{s'-}(\vec{k}'))   f_{s''-}(\vec{k}'')    f^\phi(\vec{k})
               -   f_{s'-}(\vec{k}') (1-f_{s''-}(\vec{k}''))(1+f^\phi(\vec{k}))
         \right]
      \Big\}
\,,
\nonumber
\eeqa
where $f_{s\pm}$ and $f^\phi$ are the on-shell densities for fermionic
particles and antiparticles, and scalar particles, respectively.

\begin{figure}[tbp]
\centerline{
  \psfrag{a}[r]{$\scriptstyle (a)$}    
  \psfrag{b}[r]{$\scriptstyle (b)$}    
  \psfrag{c}[r]{$\scriptstyle (c)$}    
  \psfrag{d}[r]{$\scriptstyle (d)$}    
  \psfrag{1}[l]{$\scriptstyle k$}    
  \psfrag{2}[l]{$\scriptstyle k',s'$}    
  \psfrag{3}[l]{$\scriptstyle k'',s''$}    
\epsfig{file=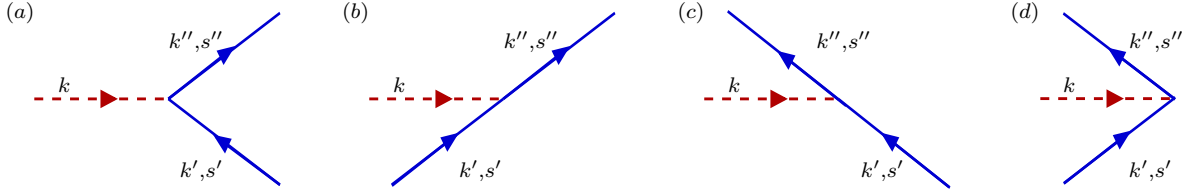, height=1.0in,width=6.in}
 }
\vskip 0.0in
\lbfig{figure:scalar-reactions}
\caption{\small
The scattering diagrams for an incoming scalar particle.
Particles and momenta run from the left to the right.
}
\end{figure}

\vskip 0.05in

Let us now, for comparison, consider the collision term one would
usually expect for a scalar particle.
In figure~\ref{figure:scalar-reactions} we show all tree level processes for
an incoming scalar particle given by the interaction Lagrangean
\beq
   {\cal L}_{\rm int}
 = - \bar{\psi}\left( P_R y\phi + P_L y^*\phi^* \right) \psi 
\,.
\label{Yukawa_Int}
\eeq
The matrix element for the first process, the transition of the
scalar particle into a fermion-antifermion pair, is
\beq
  {\cal M}^{(a)} = -iy \bar{u}_{s''}(k'') P_R v_{s'}(k')
\,,
\eeq
with the notations and conventions for the spinors as used
in Ref.~\cite{ItzyksonZuber:1980}. Calculating the absolute square of the
matrix element is done as usual, but with the complication that
we do not sum over spins. We find
\beq
    |{\cal M}^{(a)}_{s',s''}(k',k'')|^2
  = |y|^2\, \Tr \left[ P_R P_{s'}(k') \kdag' P_{s''}(k'') \kdag'' \right]
  = |y|^2 F_{s's''}(k',k'')
\,.
\label{matrix_element_2}
\eeq
This does not look immediately like an absolute square, since by sending
$k'$ to $-k'$ the sign of the trace changes, but we have to keep in mind
that the particles have to be on-shell and the above expression has to
be multiplied by the energy-momentum conserving $\delta$-function,
$\delta^4(k-k'-k'')$, so that eventually~(\ref{matrix_element_2}) is positive
definite (one can also check this by actually performing the trace).
In a Boltzmann collision term this matrix element appears with a minus
sign and is multiplied by the distribution functions $(1-f_+)(1-f_-)$
for the outgoing fermion and antifermion, respectively, and $f^\phi$ for
the incoming scalar particle.
We can identify exactly this contribution in the second line within the
curly brackets in Eq.~(\ref{ScalCollRel2}). Subtracted from it is the inverted
process where a fermion and an antifermion annihilate to form a scalar
({\it cf.} Ref.~\cite{Weldon:1983} for similar analyses).
In the same way we can find the contribution for the process shown
in~\ref{figure:scalar-reactions}(b), absorption of a scalar by a fermion,
and~\ref{figure:scalar-reactions}(c), absorption by an antifermion, in the
first and third line of~(\ref{ScalCollRel2}), respectively.
Subtracted are the inverted processes, where the scalar comes out.
The process shown in~\ref{figure:scalar-reactions}(d) is kinematically
forbidden and therefore does not appear in~(\ref{ScalCollRel2}).
So the relaxation part of the scalar collision term is exactly what
one would expect to find on the right hand side of a classical
Boltzmann equation for a scalar particle. It can be shown that
the same is true for the fermionic collision term and the Boltzmann
equation for a particle with definite spin.

\subsection{Relaxation rates}

In section~\ref{Boltzmann transport equation for CP-violating scalar densities}
we derived the kinetic equation~(\ref{scalar_pos_freq_deltaCP_part2}) for the
scalar CP-violating particle density $\delta f_\phi$.
Obviously,  expression~(\ref{ScalCollRel})
for the relaxation term is real, so that the right hand side
of the kinetic equation becomes
\beq
 \int_0^\infty \frac{dk_0}{2\pi}
  2\Big[ {\cal C}_\phi(k) + {\cal C}_\phi(-k) \Big]
\,.
\eeq
We insert~(\ref{ScalCollRel}) and send $k'\rightarrow -k'$ and
$k''\rightarrow -k''$ in the second term. The trace stays invariant
under this change, so that we find
\beqa
&&
 \int_0^\infty \frac{dk_0}{2\pi}
  2\Big[ {\cal C}_\phi(k) + {\cal C}_\phi(-k) \Big]
\nonumber\\
&=&
  -\frac {|y|^2}{2} \, \int \frac{d^4k'd^4k''}{(2\pi)^6}
         (2\pi)^4\delta^4(k+k'-k'')
          \sum_{s's''} F_{s's''}(k',k'')
          2 \int_0^\infty dk_0 \,
\nonumber\\
 &&      \delta({k'}^2-|m|^2)  {\rm sign}(k'_0) \,\,
         \delta({k''}^2-|m|^2) {\rm sign}(k''_0)\,\,
         \delta(k^2-M^2)       {\rm sign}(k_0)
\nonumber\\
 &&      \Big\{ n_{s'}(k')
                 \left(1-n_{s''}(k'')\right)
                 n^\phi(k)
                -\left(1-n_{s'}(k')\right)
                 n_{s''}(k'')
                 \left(1+n^\phi(k)\right)
\nonumber\\
 &&             -n_{s'}(-k')
                 \left(1-n_{s''}(-k'')\right)
                 n^\phi(-k)
                +\left(1-n_{s'}(-k')\right)
                 n_{s''}(-k'')
                 \left(1+n^\phi(-k)\right)
         \Big\}
\,.
\eeqa
The terms in the last line can be rewritten as antiparticle densities
based on the relations~(\ref{particle-antiparticle_part2})
and~(\ref{onshell-density-cp_part2}).
Then we split up the distribution functions into a thermal distribution plus a
first order correction term,
\beqa
  n^\phi(k)      = n^\phi_{\rm eq}(\hat{k}_0) + \delta n^\phi(k)
  \,,\quad&&\quad
  n^{\phi cp}(k) = n^\phi_{\rm eq}(\hat{k}_0) + \delta n^{\phi cp}(k)
\label{distributions:a}
\\
  n_s(k)      = n_{\rm eq}(\hat{k}_0) + \delta n_s(k)
  \,,\quad&&\quad
  n_s^{cp}(k) = n_{\rm eq}(\hat{k}_0) + \delta n_s^{cp}(k)
\,,
\label{distributions:b}
\eeqa
and linearize with respect to the correction terms. This constitutes
a rather standard {\it linear response} approach to equilibration,
which is believed to lead to quantitatively correct results for
systems close to thermal equilibrium, provided they are not 
prone to (hydrodynamic turbulent) 
instabilities~\cite{HuetKajantieLeighLiuMcLerran:1992}.

Note that the Bose-Einstein and Fermi-Dirac 
distributions in~(\ref{distributions:a}--\ref{distributions:b}) are evaluated at
$\hat{k}_0 = \gamma_w(k_0+v_wk_z)$, since the wall frame moves with respect
to the plasma.
Under extensive use of the explicit form of the  equilibrium
distributions~(\ref{FermiDirac2}), (\ref{BoseEinstein2}) and
energy-momentum conservation we can rewrite the first term in curly
brackets as
\beqa
&&  \left( 1-n_{\rm eq}(\hat{k}_0'') \right) \,
    n^\phi_{\rm eq}(\hat{k}_0) \,
    \left( 1-n_{\rm eq}(\hat{k}_0') \right)^{-1} \,
    \delta n_{s'}(k')
\nonumber\\
&&\hphantom{XXXXXX} -\; n_{\rm eq}(\hat{k}_0') \,
    n^\phi_{\rm eq}(\hat{k}_0) \,
    {n_{\rm eq}(\hat{k}_0'')}^{-1} \,
    \delta n_{s''}(k'')
\nonumber\\
&&\hphantom{XXXXXX} +\; n_{\rm eq}(\hat{k}_0') \,
    \left( 1-n_{\rm eq}(\hat{k}_0'') \right) \,
    \left( 1+n^\phi_{\rm eq}(\hat{k}_0) \right)^{-1} \,
    \delta n_\phi(k)
\,.
\eeqa
Exactly the same manipulations are then applied to the $CP$-conjugate
densities, and since the equilibrium distributions are spin independent,
we can combine the two expressions to
\beqa
&&\!\!\!\!\!\!\!
 \int_0^\infty \frac{dk_0}{2\pi}
  2\Big[ {\cal C}_\phi(k) + {\cal C}_\phi(-k) \Big]
\label{relaxation-delta-scalar}
\\
&&\hphantom{\,}
 = -|y|^2 \, \int_0^\infty dk_0 \, \int \frac{d^4k'd^4k''}{(2\pi)^6}
         (2\pi)^4\delta^4(k+k'-k'')
         \sum_{s's''} F_{s's''}(k',k'')
\nonumber\\
&&\hphantom{\,}
  \times\;       \delta({k'}^2-|m|^2)\,  {\rm sign}(k'_0) \,\,
         \delta({k''}^2-|m|^2)\, {\rm sign}(k''_0)\,\,
         \delta(k^2-M^2)      \, {\rm sign}(k_0)
         n_{\rm eq}(\hat{k}_0') 
         \big(1-n_{\rm eq}(\hat{k}_0'')\big)
         n^\phi_{\rm eq}(\hat{k}_0)
\nonumber\\
&&\hphantom{XX}\times\;
         \Bigg\{ 
  \frac{\delta n^{s'}(k') - \delta n^{-s'}_{cp}(k')}
       {n_{\rm eq}(\hat{k}_0') \big(1- n_{\rm eq}(\hat{k}_0')\big)}
- \frac{\delta n^{s''}(k'') - \delta n^{-s''}_{cp}(k'')}
       {n_{\rm eq}(\hat{k}_0'') \big(1- n_{\rm eq}(\hat{k}_0'')\big)}
+ \frac{\delta n^\phi(k) - \delta n^\phi_{cp}(k)}
       {n^\phi_{\rm eq}(\hat{k}_0) \big(1+n^\phi_{\rm eq}(\hat{k}_0)\big)}
         \Bigg\}
\,.
\nonumber
\eeqa

For the fermionic collision term the same procedure can be applied.
For the combination of collision terms we eventually need in the
kinetic equation~(\ref{boltzmann:deltaf-a_part2}) for the CP-violating
distribution function $\delta f^a_s$, we find
\beqa
 \int_0^\infty \frac{dk_0}{\pi}
  \Big[ {\cal K}^s_0(k) + {\cal K}^s_0(-k) \Big]
&=&
  -|y|^2 \, \int \frac{d^4k'd^4k''}{(2\pi)^6}
          \sum_{s'} \int_0^\infty dk_0 \,
\label{relaxation-delta-ferm}
\\
 &&\hskip -0.8in \times\;
         \delta({k'}^2-|m|^2) \, {\rm sign}(k'_0) \,\,
         \delta({k''}^2-|m|^2)\, {\rm sign}(k''_0)\,\,
         \delta(k^2-M^2)      \, {\rm sign}(k_0)
\nonumber\\
 &&\hskip -0.8in \times\;
   \bigg\{
              (2\pi)^4\delta^4(k-k'+k'')
              F_{ss'}(k',k'')
              n_{\rm eq}(\hat{k}_0) \left(1-n_{\rm eq}(\hat{k}_0')\right)
               n^\phi_{\rm eq}(\hat{k}_0'')
\nonumber\\
 &&\hskip -0.6in 
        +\; (2\pi)^4\delta^4(k-k'-k'')
              F_{-s-s'}(k',k'')
              n_{\rm eq}(\hat{k}_0) \left(1-n_{\rm eq}(\hat{k}_0')\right)
               \left(1+ n^\phi_{\rm eq}(\hat{k}_0'')\right)
      \bigg\}
\nonumber\\
 &&\hskip -0.8in \times\;
     \Bigg\{\frac{\delta n^{s}(k) - \delta n^{-s}_{cp}(k)}
                 {n_{\rm eq}(\hat{k}_0)\big(1- n_{\rm eq}(\hat{k}_0)\big)}
       -    \frac{\delta n^{s'}(k') - \delta n^{-s'}_{cp}(k')}
                 {n_{\rm eq}(\hat{k}_0')\big(1- n_{\rm eq}(\hat{k}_0')\big)}
       +    \frac{\delta n^\phi(k'') - \delta n^\phi_{cp}(k'')}
                 {n^\phi_{\rm eq}(\hat{k}_0'')
                           \big(1+n^\phi_{\rm eq}(\hat{k}_0'')\big)}
     \Bigg\}
\,.
\nonumber
\eeqa

The on-shell $\delta$-functions in these two equations finally project the
differences of the deviation functions $\delta n - \delta n_{cp}$ onto
the CP-violating distributions $\delta f^\phi$ and $\delta f^a_s$. Of course,
the relaxation term of the kinetic equation for $\delta f^v_s$ is
\beq
 \int_0^\infty \frac{dk_0}{\pi}
  \Big[ {\cal K}^s_0(k) + {\cal K}^{-s}_0(-k) \Big]
\,,
\eeq
so that here we find $\delta f^v_s$ as well as on the right hand side. 
This shows that our kinetic equations~(\ref{scalar_pos_freq_deltaCP_part2}),
(\ref{boltzmann:deltaf-v_part2}) and~(\ref{boltzmann:deltaf-a_part2})
indeed form a closed set of equations, or in other words,
that in the relaxation terms of the kinetic equations for
the functions $\delta f^\phi$, $\delta f^v_s$ and $\delta f^a_s$
only these functions themselves do appear.
It might be surprising that the relaxation term for the scalar equation
contains the fermionic $\delta f^a_s$, but not $\delta f^v_s$. The reason
for this is the occurrence of the chiral projectors in the interaction 
Lagrangean.

The relaxation terms that we have discussed so far 
can be slightly simplified further by performing
the sums over those spins that do not appear in the deviation functions. 
But essentially the form for the rates~(\ref{relaxation-delta-scalar})
and~(\ref{relaxation-delta-ferm})
is as simple as it gets, since we have to integrate over
the unknown functions $\delta f$.
In the next section we make a fluid {\it Ansatz} for the CP-violating
distribution functions in order to facilitate numerical solutions
of the kinetic equations. The fluid {\it Ansatz} expresses the distribution
functions in terms of two unknown functions, 
the chemical potential and the plasma velocity, 
which, in the stationary limit and in the wall frame,
depend only on the spatial $z$-coordinate.
The momentum dependence is completely specified by the fluid {\it Ansatz}, 
so that the integrals can be performed and result in interaction rates, 
which multiply the chemical potentials and plasma
velocities of the respective particles in the relaxation term.
In appendix~\ref{app_rates} we show the results for the rates for the
fermionic quantities in the equation for $\delta f^a_s$.

 The rates we obtain from the one-loop self-energies do not always capture
the physically dominant processes responsible for thermalization, 
however. This is so because we calculate them by forcing the outgoing
particles on-shell, which, as a consequence of the energy-momentum
conservation, results in a kinematic suppression. Because of this suppression,
the rates vanish as any of the (tree level) masses approaches zero and, 
in general, they are suppressed when the masses are small,
that is when $|m|, M \ll T$. The processes described by the one loop 
approximation in figure~\ref{figure:scalar-reactions} constitute absorption and
emission processes only. The one-loop approximation completely misses
2-to-2 particle elastic scatterings, which are essential for particle 
transport.

When the self-energies are approximated at two-loop level, 
apart from radiative vertex corrections to the one-loop rates, one
also captures (tree level) 2-to-2 scatterings rates,
which involve off-shell (boson or fermion) exchange. These rates, even
though suppressed by two more powers of the coupling constant, 
do not in general vanish when (any of the) masses approach zero, and hence 
often dominate thermalization rates in a linear response approximation. 
For details of thermalization rate calculations
starting with an (on-shell) Boltzmann collision term, we refer to 
Refs.~\cite{JoyceProkopecTurok:1996,JoyceProkopecTurok:1996b,
MooreProkopec:1995,ClineJoyceKainulainen:2000+2001,
ArnoldMooreYaffe:2000+2003}.
A complete two-loop treatment of the collision term in the Schwinger-Keldysh
formalism is a much more complex task. Nevertheless, 
we can say something constructive on results of such an undertaking. 
Indeed, our findings in the first part of this section indicate that 
the relaxation rates calculated from two-loop self-energies
(with distribution functions projected on-shell)
can as well be obtained by writing down the naive Boltzmann collision term.
Then the scattering amplitude in the collision term
is calculated from the 2-to-2 scattering processes, 
with the important caveat that one must not average over spins
(or helicities), as it is usually done in
literature, but needs the expressions for the eigenstates of 
the spin operator~(\ref{Sz},\ref{Sz_part2}). (For an example of a helicity-flip rate
calculation, which does not involve averaging over helicities, 
we refer to Appendix B of Ref.~\cite{JoyceProkopecTurok:1996}.)
Recall however that in the super-relativistic limit the spin operator $S_z$
approaches the helicity operator~(\ref{helicity-operator})
(multiplied by $\hat{\vec k}\equiv\vec k/|\vec k|$).
The rates involving particles with a definite spin
are therefore in this limit identical to the rates 
for particles with definite helicities, defined as the projection of spin onto
the direction of particle motion.


\cleardoublepage
\section{Fluid equations}
\label{Fluid equations}

In Paper~I and in the previous sections we have shown that in the semiclassical
limit the equations of motion for the scalar and fermionic Wigner functions can
be reduced to on-shell conditions and Boltzmann equations.
In the first part of this section we closely examine the physical
interpretation of the quantities we have been dealing with.
Then we examine the highly relativistic limit of our Boltzmann
equations for the fermionic particles and compare with the corresponding
equations for particles with definite helicity.
In the third part we make a fluid {\it Ansatz}
for the CP-violating distribution
functions. The resulting fluid equations are numerically tractable,
and we solve them in a very simple model. Finally we reduce the
fluid equations to diffusion equations in order to make an estimate
of the importance of collisional sources in comparison to the
flow term sources.

\subsection{Currents}
\label{sec_fluid-currents}

We begin this section by establishing some links between the quantities
we dealt with in the previous sections and physical, measurable
quantities. The best way to do this is to study various currents.
The expectation value of the scalar current operator
\beq
    j_\phi(x)
  = \left< i \phi^\dagger(x) \stackrel{\leftrightarrow}{\!\partial} \phi(x)
    \right>
\label{current:scalar}
\eeq
can be written in terms of the Wigner function:
\beq
    j_\phi(x)
  = - (\del_{u}-\del_{v}) \Delta^<(u,v) \Big|_{u=v=x} 
  = 2 \int \frac{d^4k}{(2\pi)^4} k i\Delta^<(k,x)
\,.
\label{current:scalar:Delta}
\eeq
This current contains information about the density of scalar particles
(minus antiparticles) at some space-time point $x$.
With the on-shell {\it Ansatz}~(\ref{ce-diag-solution_part2}) for the 
scalar Wigner function, the components of the current can be written as
\beqa
  j^0_\phi(x) &=& \int \frac{d^3k}{(2\pi)^3}
                   \left[ f^\phi_+(\vec{k},x)
                         -f^\phi_-(\vec{k},x)
                         -1
                   \right]
               =  \int \frac{d^3k}{(2\pi)^3}
                   \left[ \delta f^\phi(\vec{k},x) - 1
                   \right]
\\
  \vec{j}_\phi(x) &=& \int \frac{d^3k}{(2\pi)^3}
                       \frac{\vec{k}}{\omega_\phi}
                        \left[ f^\phi_+(\vec{k},x)
                              -f^\phi_-(\vec{k},x)
                        \right]
\,.
\eeqa
This confirms that $f^\phi_\pm(\vec{k},x)$ indeed measures the density
of scalar particles and antiparticles with momentum $\vec{k}$. The
one in the first equation is the vacuum contribution and is therefore
ignored.

Similarly, the expectation value of the fermionic current operator
\beq
  j^\mu_\psi(x) = \left< \bar{\psi}(x) \gamma^\mu \psi(x) \right>
\,,
\label{current:fermionic}
\end{equation}
which can be rewritten as
\beq
    j^\mu_\psi(x)
  = -\Tr [\gamma^\mu i S^<(x,x)]
  = - \int \frac{d^4k}{(2\pi)^4} \Tr \gamma^\mu iS^<(k,x)
\,,
\label{current:fermionic:S}
\eeq
contains information about the density of fermions.
A careful consideration of the zero-component and the vector part of
this current, and the three-component of the spin-density
\beq
  {\cal S}_z(x) = \frac{\hbar}{2} 
                   \left<  \bar{\psi}(x) \gamma^3\gamma^5 \psi(x)   \right>
\eeq
expressed in terms of the on-shell distribution
functions~(\ref{onshell-density_part2}) and~(\ref{onshell-density-cp_part2})
tells that $f_{s\pm}(\vec{k},x)$ measures the density of particles
and antiparticles with momentum $\vec{k}$ and
spin $s$ (in the direction specified by~(\ref{spin-direction})).

When considering electroweak baryogenesis, the final step in the mechanism
that produces the baryon asymmetry is the sphaleron process: it turns the
CP-violating flows, whose kinetics we study in this paper, into a net
baryon density. To be more precise, the sphaleron acts only on the left
handed SU(2) doublets, which are eigenstates of the chirality projector $P_L$,
but leaves right handed particles unaffected. The densities of
left- and right-handed particles
\beq
  j^0_{L/R}(x) = \left<\bar{\psi}_{L/R} \gamma^0 \psi_{L/R}  \right>
\eeq
are linear combinations of the vector density and the axial density
$ j^5 = \left<  \bar{\psi}\gamma^0\gamma^5 \psi \right> $:
\beq
  j^0_{L/R} = \frac 12 \Big( j^0 \mp j^5 \Big)
\,.
\eeq
Finally, the vector density and the axial density are related to
the fermionic CP-violating densities~(\ref{distribution-fv_part2})
and~(\ref{distribution-fa_part2}) we defined earlier,
\beqa
     j^0(x)
 &=& \sum_s \int \frac{d^3k}{(2\pi)^3} \,
            \delta f^v_s(\vec{k}, x)
\\
     j^5(x)
 &=& \sum_s \int \frac{d^3k}{(2\pi)^3} \,
            \Big[  s\frac{k_z}{\tilde{\omega}_0}
                   \delta f^a_s(\vec{k}, x)
                 + s\frac{k_z}{\tilde{\omega}_sZ_s}
                   f^{\rm eq}_s(\vec{k}, x)
            \Big]
\,,
\label{j5-by-deltaf-s}
\eeqa
which explains the terminology. We omitted the vacuum contribution
to the vector density. It might seem strange that the equilibrium
distribution contributes to the axial density, but this is just a
consequence of defining $f^{\rm eq}$ as the projection of $n_{\rm eq}$
to the mass shell $\omega_s$. Had we defined
$f^{\rm eq} = n_{\rm eq}(\omega_0)$, then the axial density would be
expressed by $\delta f^a_s$ alone, but in turn there would be
additional source terms in the Boltzmann equation for this function.

Both CP-violating densities contribute to the left-handed density,
so both could lead to baryon production.
Recall however that in the equation~(\ref{boltzmann:deltaf-v_part2})
for $\delta f^v_s$ there is no source, neither from the flow term nor
from the collision term.
We therefore assume that this density, and consequently the vector
density, vanishes and then we have $j^0_L= -j^0_R = - j^5/2$.
%
%
The source in the equation for $\delta f^a_s$ is proportional to spin,
so that any contribution in $\delta f^a_s$ caused by this source is also
proportional to spin. 
Therefore the spin summation in the expression for $j^5$
does not cancel the contributions against each other, but adds them up,
wherefore the source in the equation for $\delta f^a_s$ eventually is
an effective source for electroweak baryogenesis.
The source in the scalar equation can contribute to baryogenesis
only indirectly. It first creates a CP-violating scalar density, which then
may be transferred into the fermionic CP-violating distribution function
{\it via} the collision term of the scalar kinetic equation.

Finally, a word on mixing particles. The sphaleron acts on the
particles in the electroweak interaction basis, while the densities
we are dealing with are those of particles in the mass basis. So in principle
we would have to rotate these densities into the interaction basis. This is
not necessary, however: the sphaleron is effective only in the symmetric phase,
where the Higgs vacuum expectation value vanishes, and in those regions of
the bubble wall where the expectation value is small. But for a vanishing
Higgs expectation value the mass basis and the
interaction basis coincide. So we can compute the contribution
to the axial density for each particle, using the densities in the mass
eigenbasis, and sum them up.

In the following we will not consider the equilibrium contribution to the
axial density any further, another consequence of the sphaleron's refusal to
work anywhere else than in the symmetric phase.
Electroweak baryogenesis is most effective when there are CP-violating
flows that are efficiently
transported away from the wall into the symmetric phase, where
they are converted into a baryon asymmetry.
The equilibrium
contribution to the axial density, however, sits right on top of the wall
and does not participate in diffusion.

\subsection{Spin or helicity?}

In the highly relativistic limit we can approximate $\tilde{\omega}_0$
by $|k_z|$. In this case the density of left-handed particles can be written
as
\beq
    j^0_L
  =  \frac 12 \sum_s \int \frac{d^3k}{(2\pi)^3}
       \left(1-\frac{sk_z}{|k_z|} \right)
       \left( \delta f_{s+}(\vec{k}) - \delta f_{-s-}(\vec{k}) \right)
\,,
\eeq
where we allowed for a possible contribution from the vector density.
But this is of course
precisely the density of particles with negative helicity minus the density
of antiparticles with positive helicity (for massless antiparticles
chirality is opposite to helicity):
\beq
  j^0_L \propto   \int_0^\infty dk_z
                \left( \delta f_{-+} - \delta f_{+-} \right)
             + \int_{-\infty}^0 dk_z
                \left( \delta f_{++} - \delta f_{--} \right)
\,.
\label{lefthanded_density_by_helicity}
\eeq
In fact, in every earlier work on EWB that used the WKB method, the
Boltzmann equations were written directly for the densities of particles
with a definite helicity (or chirality). The equation we derived for spin
states is
\beq
    \big(  \del_t
         + \vec v_{0i} \cdot \nabla_{\vec x} 
         + F_{0i} \partial_{k_z}
    \big)
          \delta f^a_{si} 
    + s{\cal S}_i
 =  \int_0^\infty \frac{dk_0}{\pi}
      \Big[ {\cal K}^s_{0i}(k) - {\cal K}^{cp-s}_{0i}(k) \Big]_{relax}
\,,
\label{repeat_boltzmann_f5}
\eeq
where ${\cal S}_i$ denotes all sources, both those from the flow
term~(\ref{boltzmann:deltaf-a_source}--\ref{boltzmann:deltaf-a_even})
and from the collision term~(\ref{C:psi:K}--\ref{ferm-coll-src}),
while on the right hand side only the relaxation part of the collision term
is left.
It is now not difficult to break this equation up into four equations
with $s=\pm1$ and $k_z>0$ and $k_z<0$, respectively, and then combine
those pieces with opposite $s$ and $k_z$ to the equation
\beq
    \big(  \del_t
         + \vec v_{0i} \cdot \nabla_{\vec x} 
         + F_{0i} \partial_{k_z}
    \big)
          \delta f_{hi} 
    + h \, \text{sign}[k_z] \, {\cal S}_i
 =  \int_0^\infty \frac{dk_0}{\pi}
      \Big[ {\cal K}^h_{0i}(k) - {\cal K}^{cp-h}_{0i}(k) \Big]_{relax}
\,,
\label{helicity_boltzmann_f5}
\eeq
where $\delta f_{hi}$ is the density of particles with helicity $h$ minus
the density of antiparticles with opposite helicity, precisely the quantity
needed for the left-handed density~(\ref{lefthanded_density_by_helicity}).
Formally, this equation for helicity states has almost identical appearance
as the one used in the heuristic WKB approach to baryogenesis
(the classical force in the flow of the CP-violating density is usually
omitted in these works), where the CP-violating source ${\cal S}_i$ 
is approximated by the semiclassical force. 
 What this analysis really shows is that the WKB approach, combined with a
good guess work, which involves a correct calculation of the dispersion 
relation (obtained by transforming from canonical to kinetic momentum),
and the right modus of insertion of the dispersion relation into
the kinetic equation, can lead to the correct CP-violating force. 
However, we emphasize that no work using the WKB approach has so far gotten
the full source completely right.
Another point is that the interpretation of $h$ as helicity really
makes sense only in the highly relativistic limit, or equivalently,
in the symmetric phase, in which the tree level masses vanish. 

 Furthermore, it is important to note that the explicit form
of the source found in those works differs from our result:
first, the contribution from the flow term to the
source there contains only the semiclassical force $\delta F_i$
from~(\ref{boltzmann:definitions}), but misses the other contributions
we found. In section~\ref{Solving the fluid equations} we make a numerical
estimate of the importance of these terms.
Second, the semiclassical WKB method is not in principle suitable
to calculate the source that originates from the collision term.

\subsection{Fluid equations}
\label{Fluid equations indeed}

We now have the Boltzmann equations for the CP-violating distribution
functions, and we know how to interpret them in terms of densities
and currents and what is their role in the creation of a baryon asymmetry.
A direct numerical solution of the Boltzmann
equations~(\ref{scalar_pos_freq_deltaCP_part2}) fails due to their complexity,
however, at least at present.
So we have to rely on approximations to the full Boltzmann equations.
A very convenient and with respect to the application to EWB quite popular
approximation is a fluid {\it Ansatz} for the distribution
functions~\cite{JoyceProkopecTurok:1996}:
\beq
  f_{si\pm}
  = \frac{1}{ {\rm e}^{\beta(\omega_{si} + v_w k_z
                        - \mu^s_{i\pm} + u^s_{i\pm} k_z )    } + 1}
\label{fluid_ansatz}
\eeq
for the fermions, and an analogous expression without spin indices
for the scalar distributions.
This form mimics the equilibrium distribution, but the chemical potential
$\mu$ and the plasma velocity $u$, which are only functions of the
spatial coordinates, allow for local fluctuations in the density and velocity
of the plasma. In the given form the velocity perturbation accounts only for
a net motion of particles in $z$-direction.
Because of the planar symmetry of the wall,
this should be sufficient. The term including the wall velocity
is due to the movement of the wall frame in the plasma rest frame.
%
With the {\it Ansatz}~(\ref{fluid_ansatz})
the momentum dependence is explicit, such that,
upon integrating the Boltzmann equations over momentum,
only spatial dependencies remain.
%

The CP-violating chemical potential and the velocity perturbation 
are caused by the interaction with the wall and hence are implicitly
of first order in gradients. When we expand
the fluid {\it Ansatz}~(\ref{fluid_ansatz}), keeping only terms linear
in $\mu$, $u$ and $v_w$, and compare it with our
decomposition~(\ref{distribution-fn_fsi_decompose}) of the fermionic
distribution function, leaving aside the CP-even correction, we can identify
\beq
  \delta f_{si\pm} = -\beta f_{0i}(1-f_{0i})
                       \left[1-\beta v_w k_z(1-2f_{0i})\right]
                        \left( -\mu_{si\pm} + k_z u_{si\pm} \right)
\,,
\label{deltaf-a-by-fluid}
\eeq
where $f_{0i} = 1/({\rm e}^{\beta \omega_{0i}} +1) $,
$\omega_{0i} = (\vec k^2 + |m_d|_i^2)^{1/2}$.
The form~(\ref{distribution-fa_part2}) of the CP-violating distribution $\delta f^a_s$
then suggests to define
\beqa
  \mu_{si} &\equiv& \mu_{si+} - \mu_{-si-}
\\
  u_{si}   &\equiv&  u_{si+}  - u_{-si-}
\,.
\eeqa
We insert the fluid {\it Ansatz} into the Boltzmann equation for the
fermionic CP-violating density $\delta f^a_s$ and keep all terms
up to second order in gradients and up to first order in $v_w$.
Then we take the zeroth and first moment with respect to $k_z$,
that is we multiply by $1$ and $k_z/\omega_0$, respectively, and
integrate over the momenta. Since the flow term source
is even in $k_z$, it contributes to the zeroth moment equation:
\beqa
&&
         I_{00,i} \beta\dot{\mu}_{si}
   + v_w J_{20,i} \dot{u}_{si}
   - v_w J_{21,i} \beta\mu_{si}'
   -     I_{21,i} u_{si}'
\nonumber\\
&&
   + v_w J_{01,i} \beta^3   \frac{{|m_d|_i^2}'}{2} \mu_{si}
   +     I_{01,i} \beta^2 \frac{{|m_d|_i^2}'}{2} u_{si}
   + \beta^3\int \frac{d^3k}{(2\pi)^3} s{\cal S}_i^{\rm flow}(\vec{k})
\nonumber\\
&=& 
    - \sum_{s'i'} \Big( \beta     \Gamma^{0\mu}_{si,s'i'} \mu_{s'i'}
                       + v_w \Gamma^{0u}_{si,s'i'}   u_{s'i'}
                  \Big) 
    - \sum_{i'}   \Big(  \beta    \Gamma^{0\mu}_{i,i'} \mu_{i'}
                       + v_w \Gamma^{0u}_{i,i'} u_{i'}
                  \Big)
\,.
\label{continuity-equation}
\eeqa
In the following we do not consider the 
contribution~(\ref{boltzmann:deltaf-a_even}) from the CP-even
distribution function $\delta f_{i}^{\rm even}$
to ${\cal S}_i^{\rm flow}$, which in principle can be
of the same order as~(\ref{boltzmann:deltaf-a_source}).
The source from the collision term is odd in $k_z$ and therefore
appears in the first moment equation:
\beqa
&&
     v_w J_{21,i} \beta\dot{\mu}_{si}
   +     I_{21,i} \dot{u}_{si}
   -     I_{22,i} \beta\mu_{si}'
   - v_w J_{42,i} u_{si}'
\nonumber\\
&&
   + v_w J_{22,i} \beta^2 {|m_d|_i^2}' u_{si}
   + \beta^3 \int \frac{d^3k}{(2\pi)^3}
                          \frac{k_z}{\omega_0}
                           s{\cal S}_i^{\rm coll}(\vec{k})
\nonumber\\
&=&
    - \sum_{s'i'} \Big( v_w \beta \Gamma^{1\mu}_{si,s'i'} \mu_{s'i'}
                       +     \Gamma^{1u}_{si,s'i'} u_{s'i'} 
                  \Big)
    - \sum_{i'}   \Big( v_w\beta \Gamma^{1\mu}_{i,i'} \mu_{i'}
                       +     \Gamma^{1u}_{i,i'} u_{i'} 
                  \Big)
\,.
\label{Euler-equation}
\eeqa
Here dots denote the derivative with respect to time, primes are $z$-derivatives,
and we introduced the dimensionless integrals
\beqa
  I_{ab,i} &=& \beta^{3+a-b}
             \int \frac{d^3k}{(2\pi)^3}
             f_{0i}(1-f_{0i})
             \frac{k^a_z}{\omega^b_{0i}}
\label{I_ab,i}
\\
  J_{ab,i} &=& \beta^{3+a-b}
             \int \frac{d^3k}{(2\pi)^3}
             f_{0i}(1-f_{0i})(1-2f_{0i})
             \frac{k^a_z}{\omega^b_{0i}}
\,.
\label{J_ab,i}
\eeqa
With the fluid {\it Ansatz} the momentum dependence of the $\delta f^a_s$
and $\delta f^\phi$ is
specified, so that the integrals in the relaxation
term~(\ref{relaxation-delta-ferm}) can be performed, leading
to the relaxation rates $\Gamma_{si,s'i'}$ and $\Gamma_{i,j}$,
which connect different fermionic species with each other and fermionic
particles with scalar particles, respectively.
Some explicit expressions for the rates calculated in our
one-loop approximation can be found in appendix~\ref{app_rates}.
Note however that with our conventions the above fluid equations
imply that the actual rates, by which the chemical potential
and the fluid velocity perturbations are attenuated, are given by
$\Gamma^{0\mu}_{si,s'i'}/I_{00,i}$
and $\Gamma^{1u}_{si,s'i'}/I_{21,i}$, respectively, which are about 
one order of magnitude larger than
$\Gamma^{0\mu}_{si,s'i'}$ and $\Gamma^{1u}_{si,s'i'}$.

The equations for the scalar particles can be obtained from the ones
for fermions in a simple way: one has to replace $|m|^2$ by $M^2$,
remove the source from the flow term, replace the collisional source
by the scalar one, use the appropriate relaxation rates for scalars,
and replace the coefficients $I$ and $J$ by their scalar counterparts,
whereby the Bose-Einstein distribution replaces the Fermi-Dirac function.

\vskip 0.1in

The fluid equations~(\ref{continuity-equation}--\ref{Euler-equation})
are well suited for a numerical treatment.
They consist of a system of first order differential equations,
which is a comparatively simple problem.
Note that only the time- and the $z$-derivative have survived. This is a
consequence of the fluid {\it Ansatz}, which in the above form effectively
allows for spatial variations only in the $z$-direction.
Since in the wall frame the mass depends only on the $z$-coordinate, the
problem is stationary in essence and we could drop the time derivative.
We argued in the beginning of section~\ref{The 3+1 dimensional (moving) frame}
that keeping the time derivative could allow a treatment of
non-equilibrium initial conditions. But in order not to violate
the spin, which is conserved by the symmetry of the problem, 
these initial conditions have to have quite a special form, given by
equation~(\ref{S<-tildeS<}), so that their physical relevance is at best
limited.
Keeping the time derivative is however a convenient tool to solve
the equations. At a first sight it seems to be simpler to have only
ordinary differential equations instead of partial ones. But to
solve these ordinary equations is actually quite tricky, since boundary
conditions at infinity have to be satisfied: at sufficiently
large distances apart from the wall we expect the system to be in chemical
equilibrium. Starting with some initial values at one side of the
wall usually does not lead to the correct chemical equilibrium on the other
side, so that the initial value has to be fine-tuned. A simpler procedure
is  to keep the time derivative and thus simulate relaxation 
of the system toward a stationary state, starting from some
well-chosen initial functions $\mu_0(z)$ and $u_0(z)$. 
Because of the interplay between the dissipative effects in
the relaxation term on the one hand and the sources on the other,
this procedure leads to stable stationary solutions relatively fast.

Once the solutions of the fluid equations are known, the axial
density can be calculated.
Inserting~(\ref{deltaf-a-by-fluid}) into~(\ref{j5-by-deltaf-s})
leads to the relation
\beq
  j^5 = - \sum_s s \int \frac{d^3k}{(2\pi)^3}
           \beta f_0(1-f_0) \frac{k_z^2}{\tilde{\omega}_0}
            \Big(  u_s
                 + v_w \beta(1-2f_0) \mu_s
            \Big)
\,.
\label{j5-by-fluid}                  
\eeq
It is interesting that the main contribution to the axial density
here comes from the plasma velocity, since the chemical potential
is suppressed by an extra factor $v_w$.

\subsection{Solving the fluid equations}
\label{Solving the fluid equations}

We now solve the fluid equations in a toy model consisting
of one fermionic particle species. Despite of the marked simplicity
of this model, the explicit solution of the equations will be
instructive, and in particular will help to answer three questions.
First, we would like to know how big is the collisional
source~(\ref{ferm-coll-src}), when compared to the flow term
source~(\ref{boltzmann:deltaf-a_source}). Since these sources
appear in different equations, a direct comparison is not possible.
But we can study the influence of the presence or absence of the
collisional source on the axial density. This should allow us to
judge if this contribution could be important in an actual
calculation of EWB in a realistic model.
Second, we found that previous work based on the WKB method missed
a part of the flow term source~(\ref{boltzmann:deltaf-a_source}).
Are the new terms relevant?
And finally, we have seen that in the highly relativistic limit
on the level of Boltzmann equations the use of spin states or
helicity states is equivalent. As soon as an approximation
to these equations is made, however, as for example the fluid
{\it Ansatz}, differences occur.
We can solve the fluid equations in both pictures, compute
the axial density and compare the results, in order to study by how much 
the equivalence between spin and helicity states is broken on
the level of fluid equations.

\vskip 0.1in

\begin{figure}[tbp]
\centerline{
\epsfig{figure=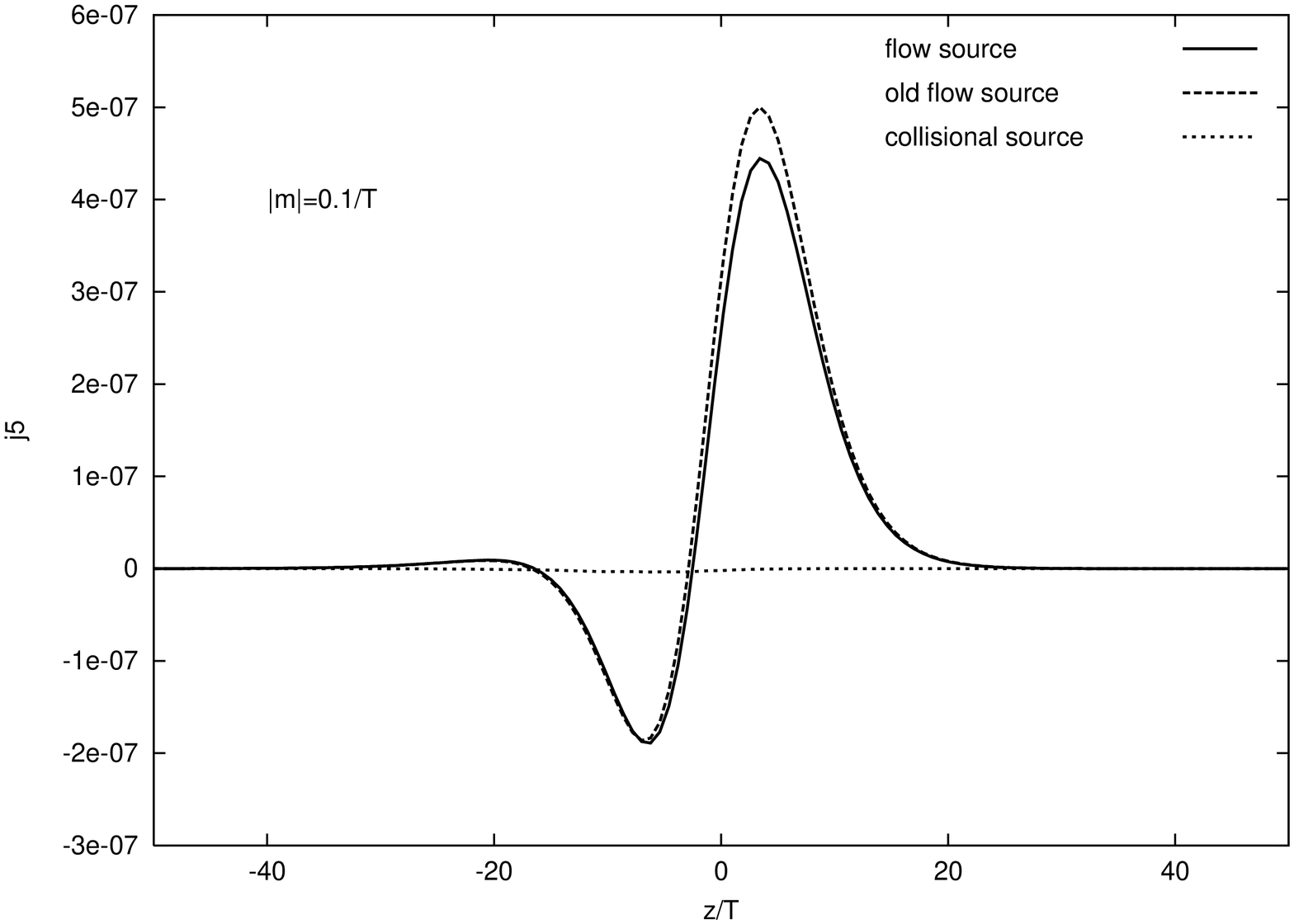, height=3.5in,width=2.6in,angle=-90}
\epsfig{figure=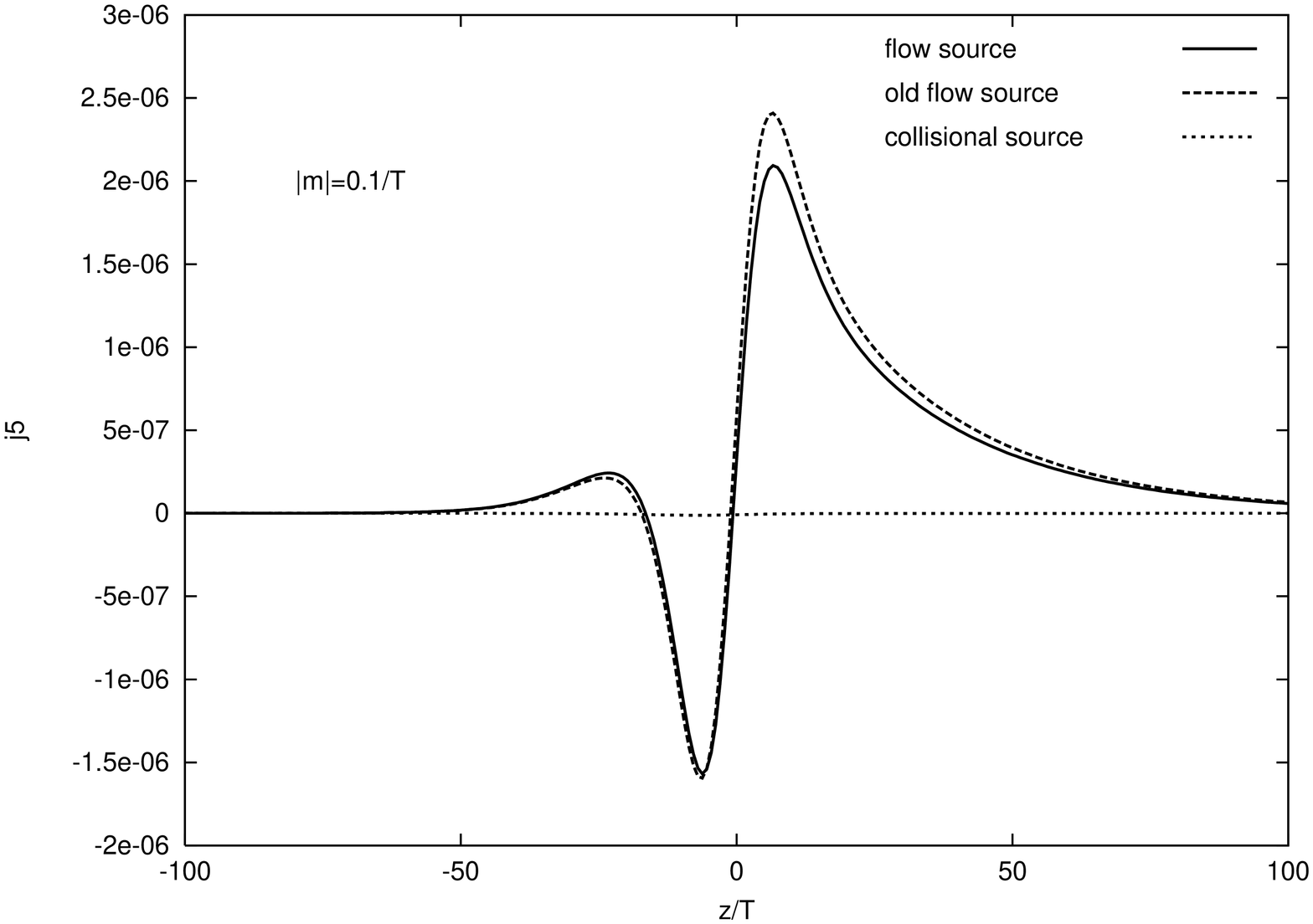,  height=3.5in,width=2.6in,angle=-90}
}
\centerline{
\epsfig{figure=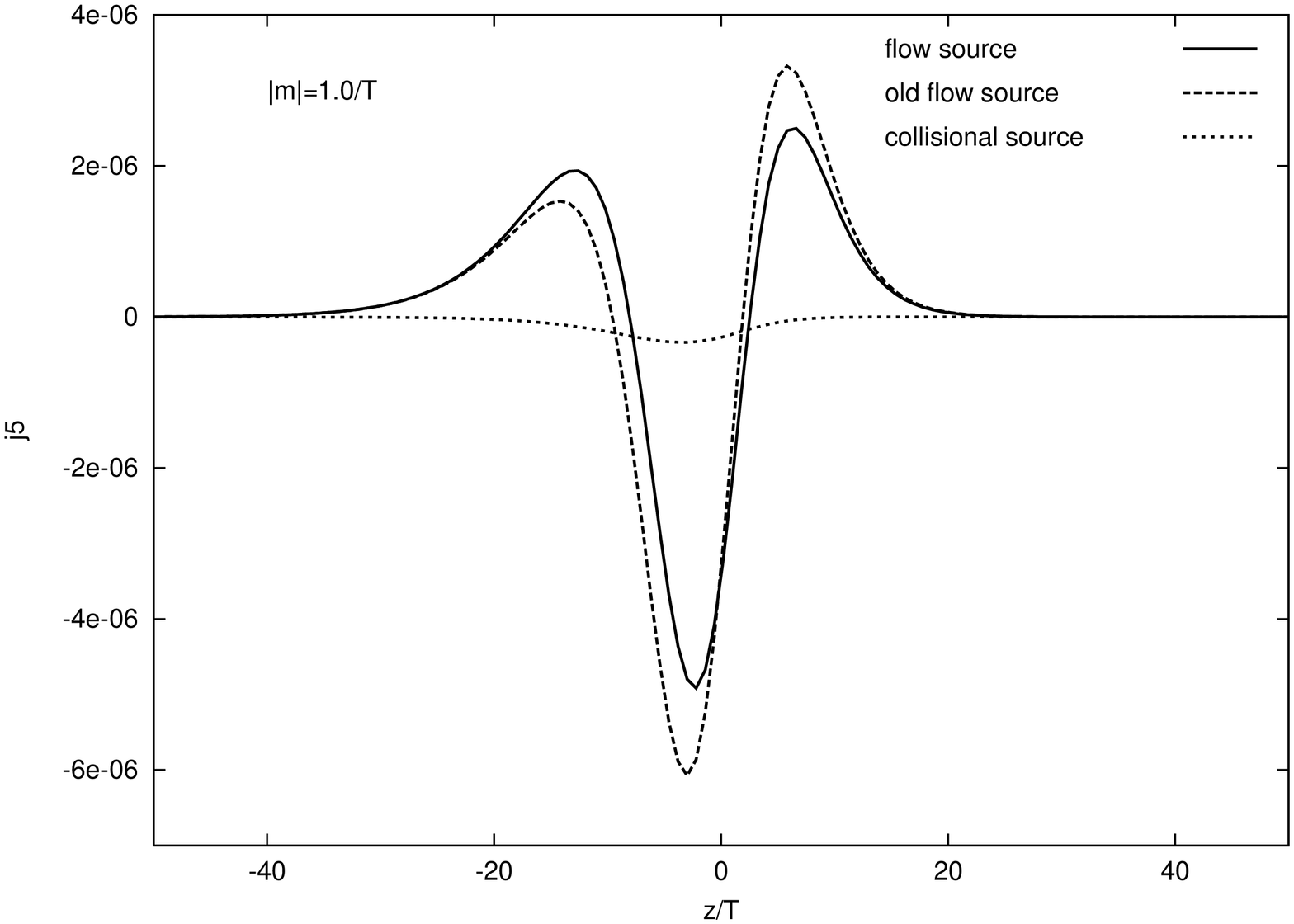, height=3.5in,width=2.6in,angle=-90}
\epsfig{figure=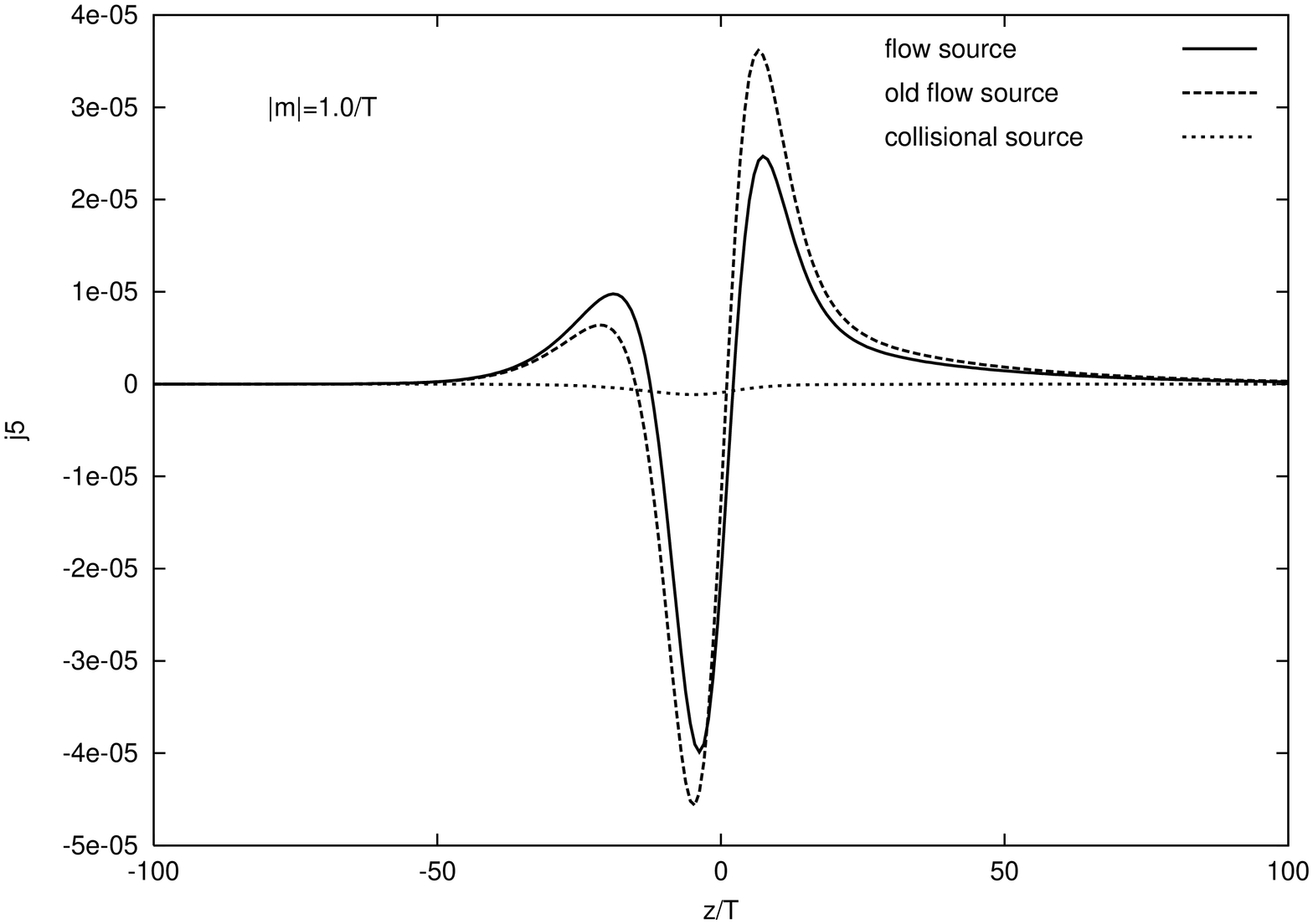,  height=3.5in,width=2.6in,angle=-90}
}
\centerline{
\epsfig{figure=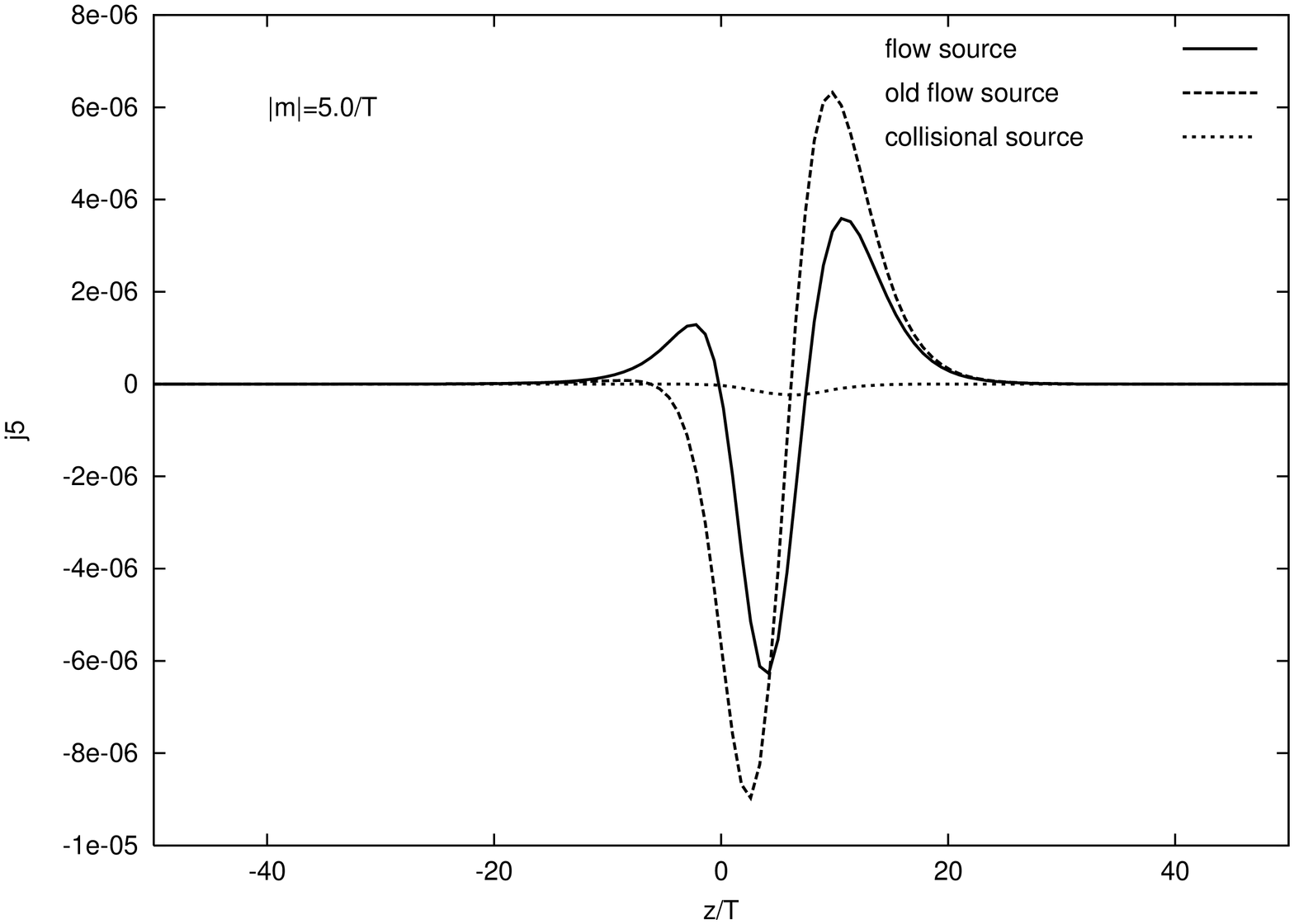, height=3.5in,width=2.6in,angle=-90}
\epsfig{figure=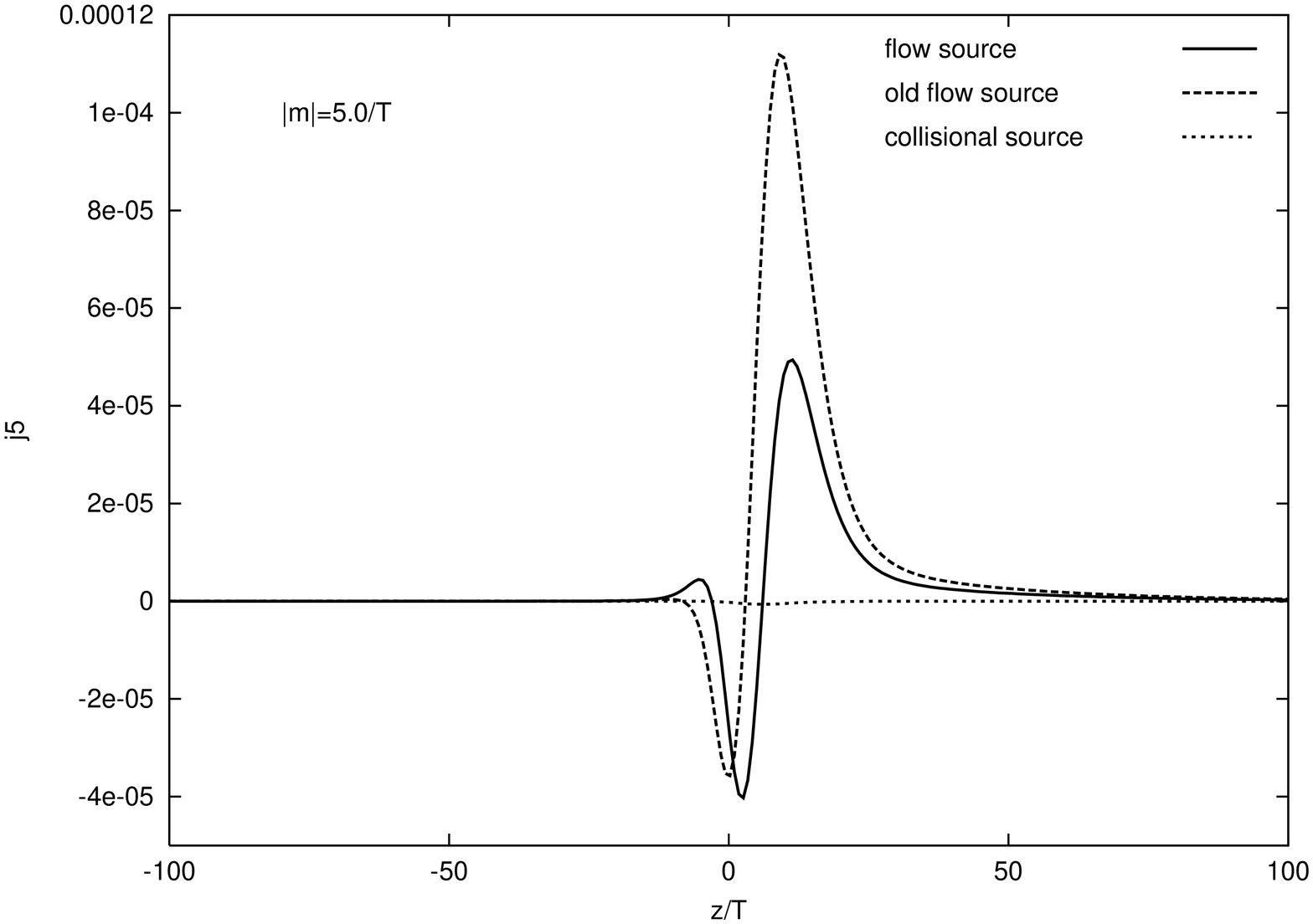,  height=3.5in,width=2.6in,angle=-90}
}
\caption{\small
The axial density generated by the flow term source,
the ``old'' flow term source, and the collisional source, respectively,
for different values of the fermionic mass parameter.
The pictures on the left hand side,
computed with the rates~(\ref{rates_big}), show no diffusion, while in the
pictures on the right, computed with~(\ref{rates_small}), the effects
of diffusion can be seen. 
For this choice of parameters 
the contribution of the collisional source is negligible.
The difference between the results for the two flow term sources is apparent.
The bubble wall with width $L_w=10/T$ sits at $z=0$, the broken phase
is at the left. Note the different scales of the $z$-axes.
}
\lbfig{fig:compare-sources}
\end{figure}

The bubble wall can be modeled by a hyperbolic tangent
\beq
 |m|    = m_0      \frac 12 \Big( 1 - \mbox{tanh} \frac{z}{L_w} \Big)
 \,,\quad
 \theta = \theta_0 \frac 12 \Big( 1 - \mbox{tanh} \frac{z}{L_w} \Big)
\eeq
for both the absolute value and the phase of the fermionic mass.
Here $L_w$ characterizes the wall thickness. 
In contrast to the fermionic mass parameter $m_0$, which influences the
form of the source and therefore also the form of the result, $\theta_0$
is simply a common prefactor to all source terms and therefore only
scales the result of the computation. We choose $\theta_0=0.1$.
The width of the bubble wall is set to $L_w=10/T$, a typical value for
the MSSM, and the wall velocity is chosen to be $v_w=0.1$.
We want to investigate the collisional source for the fermionic
equations, so the system must also contain scalar particles. But since we
are interested only in the dynamics of the fermionic particles, we
assume a thermal distribution for the scalars. Both fermions
and scalars get their masses from the interaction with the Higgs field,
and therefore the scalar mass is proportional to $|m|$.
To keep things simple, we set all relaxation rates
in~(\ref{continuity-equation}) and~(\ref{Euler-equation}) to zero,
except of $\Gamma^{0\mu}_{ss}$ and $\Gamma^{1u}_{ss}$. For these rates
we use two different sets of values,
\beq
  \Gamma^{0\mu}_{ss} =  T/25 \,,\quad
  \Gamma^{1u}_{ss}   =  T/20 \,,\quad
  s = \pm 1
\,,
\label{rates_big}
\eeq
which are so big that there will essentially be no diffusion of particles
in front of the wall, and
\beq
  \Gamma^{0\mu}_{ss} =  T/120 \,,\quad
  \Gamma^{1u}_{ss}   =  T/100 \,,\quad
  s = \pm 1
\,,
\label{rates_small}
\eeq
which are small enough to have a notable diffusion.
Note that both fluid equations contain the time derivative of
the chemical potential and of the plasma velocity. We first build
linear combinations of these two equations, such that we obtain
one equation which contains only the time derivative of $\mu$,
and another equation in which only the time derivative of $u$
appears. The temporal evolution of these equations can then
be simulated numerically in a straightforward manner, where we
choose $\mu$ and $u$ to be identical to zero as initial values.
We emphasize again that this evolution is a physical temporal
evolution only for the very special initial conditions~(\ref{S<-tildeS<}).
But in any case it is a convenient way of obtaining
the solutions of the ordinary differential equations in $z$,
which emerge by setting the time derivatives to zero, without
having to worry about boundary conditions.

\begin{figure}[btp]
\centerline{
\epsfig{figure=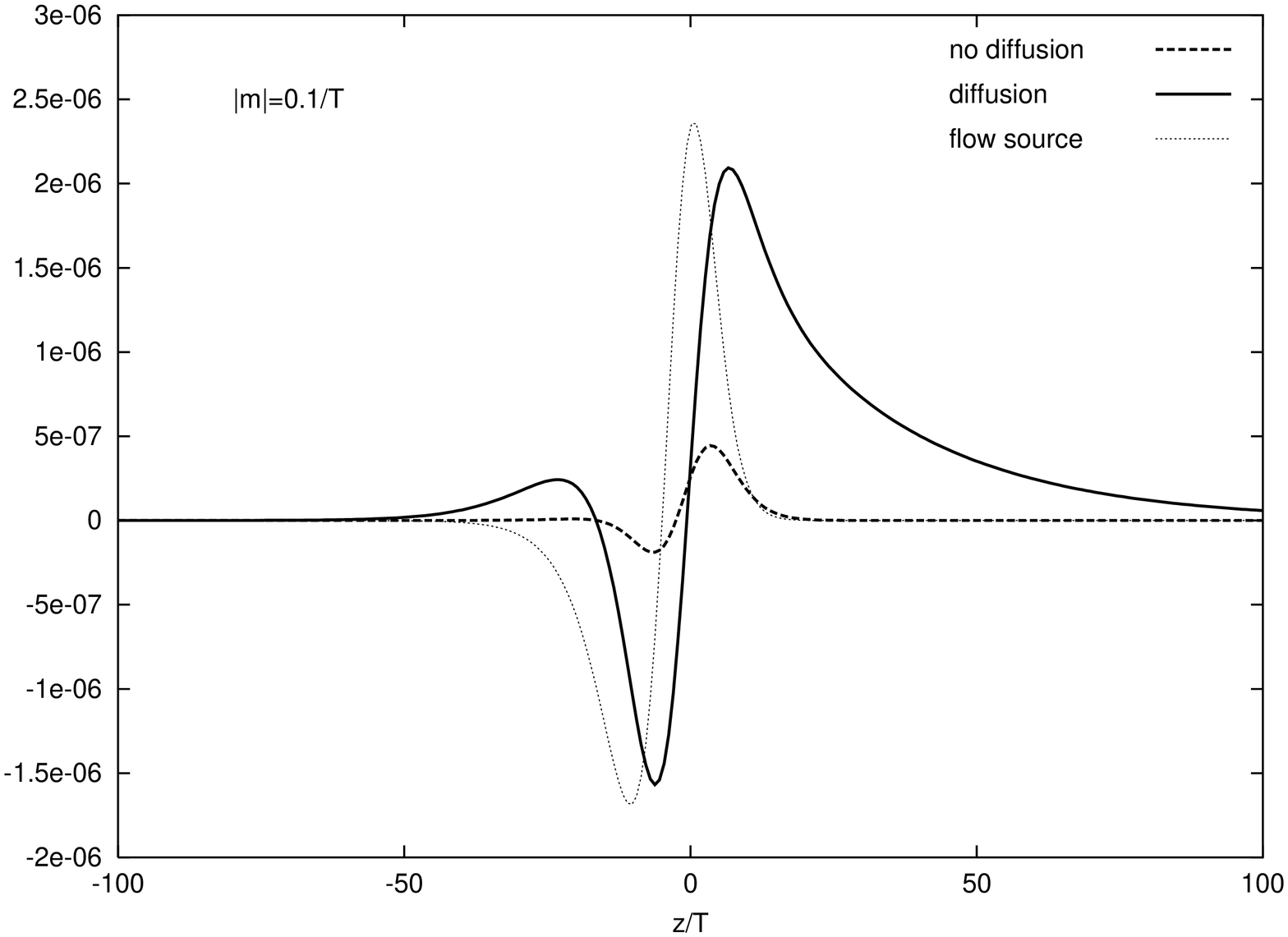, height=3.6in,width=2.8in,angle=-90}
\epsfig{figure=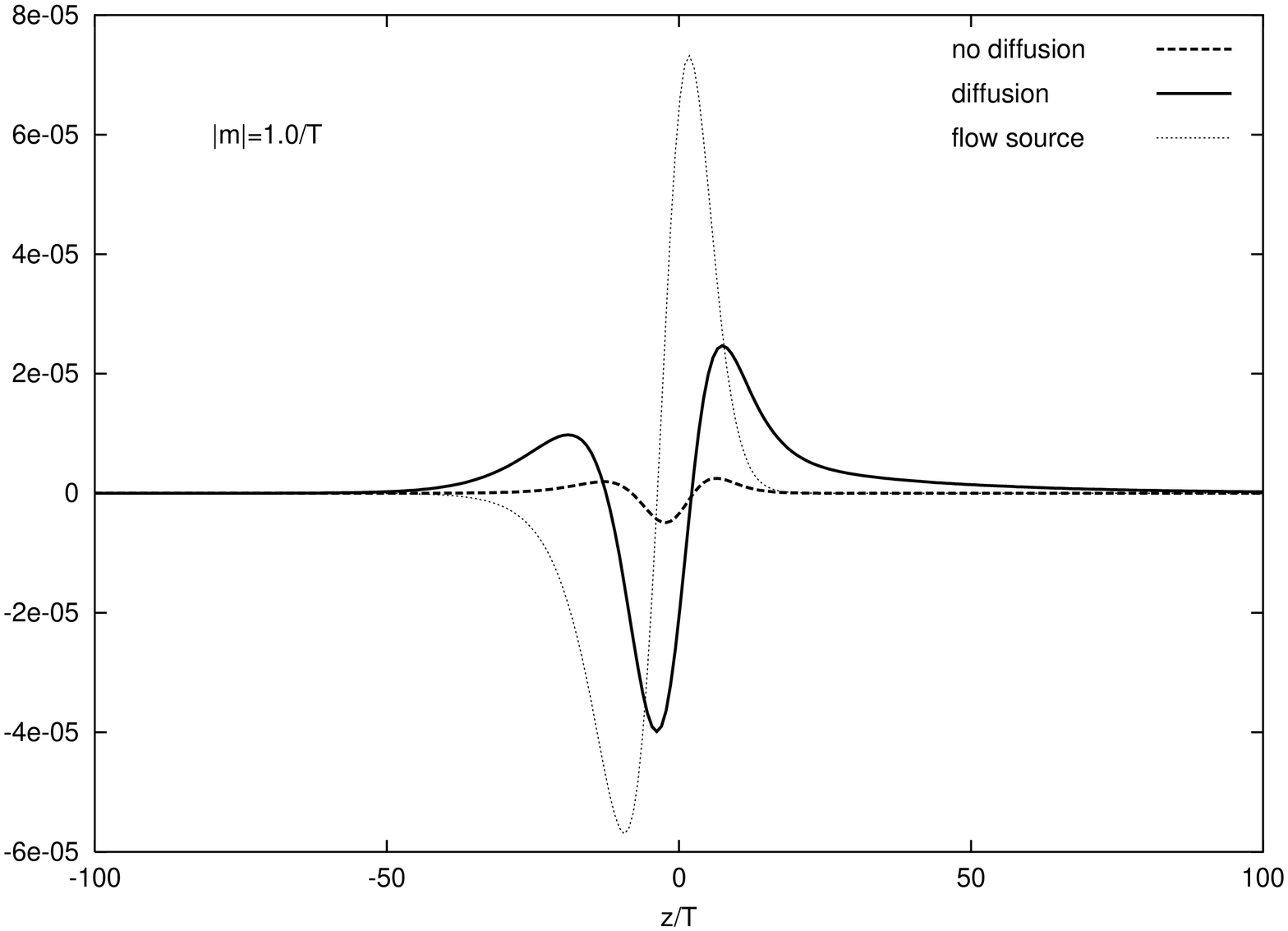,  height=3.6in,width=2.8in,angle=-90}
}
\caption{\small
The effects of diffusion can be nicely seen here, where we plotted the axial
density produced by the same source, but with different sets of relaxation
rates, (\ref{rates_big}) and~(\ref{rates_small}). For comparison we also show
the flow source. The axial density computed with the larger relaxation rates
is not transported into the symmetric phase, but sits right on top of the
source, while for the small rates a diffusion tail is clearly visible.
For small masses diffusion is more efficient.
}
\lbfig{fig:transport}
\end{figure}

To address the first two questions we mentioned above, we solve the
fluid equations for three different cases:
with the flow term source, with the ``old'' flow term source, that
means using only the first term of~(\ref{boltzmann:deltaf-a_source}),
as it was done in the WKB calculations, and with the collisional source
only.
The computation is done for three different values of the fermionic
mass parameter $m_0$, and for the collisional source we set $M=3.0|m|$
and use $|y|^2=0.1$ for the coupling constant.
In figure~\ref{fig:compare-sources} the resulting axial density is shown as a
function of the $z$-coordinate. On the left hand side are the results
obtained by using the larger relaxation rates~(\ref{rates_big}), while
for the pictures on the right hand side the smaller rates~(\ref{rates_small})
have been used.
For the parameters chosen, the collisional source has no notable
effect.
Repeating the calculation with different
wall velocities and scalar masses does not increase the collisional
contribution significantly.
Note that for smaller relaxation rates not only a diffusion tail develops,
but the amplitude of the axial density grows, too.

The axial density originating in the ``old'' flow term source
overestimates the one from the true flow term source,
with a difference increasing with the mass parameter.  
While for small mass parameters $m_0<T$ both sources
give about the same result, the difference becomes significant
for larger masses. We conclude that a computation of EWB with
the correct flow term source will most probably lead to different
results for the baryon density than the WKB calculations.

In order to illustrate the difference between the situations with
and without effective diffusion, we show in figure~\ref{fig:transport}
the axial densities calculated with the two sets of
rates~(\ref{rates_big}) and~(\ref{rates_small}) in direct comparison.

\vskip 0.1in

\begin{figure}[tbp]
\centerline{
\epsfig{figure=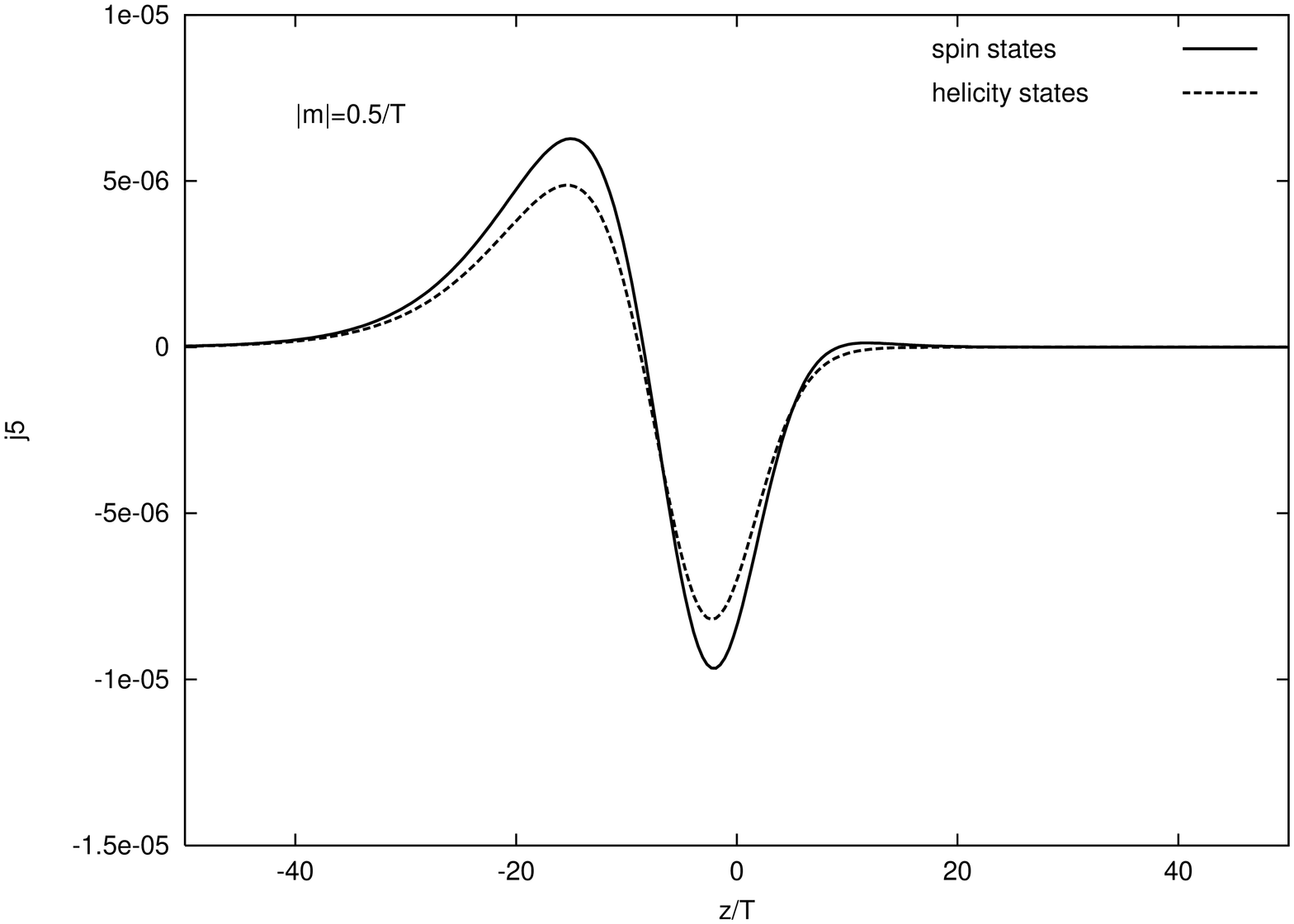, height=3.6in,width=3in,angle=-90}
\epsfig{figure=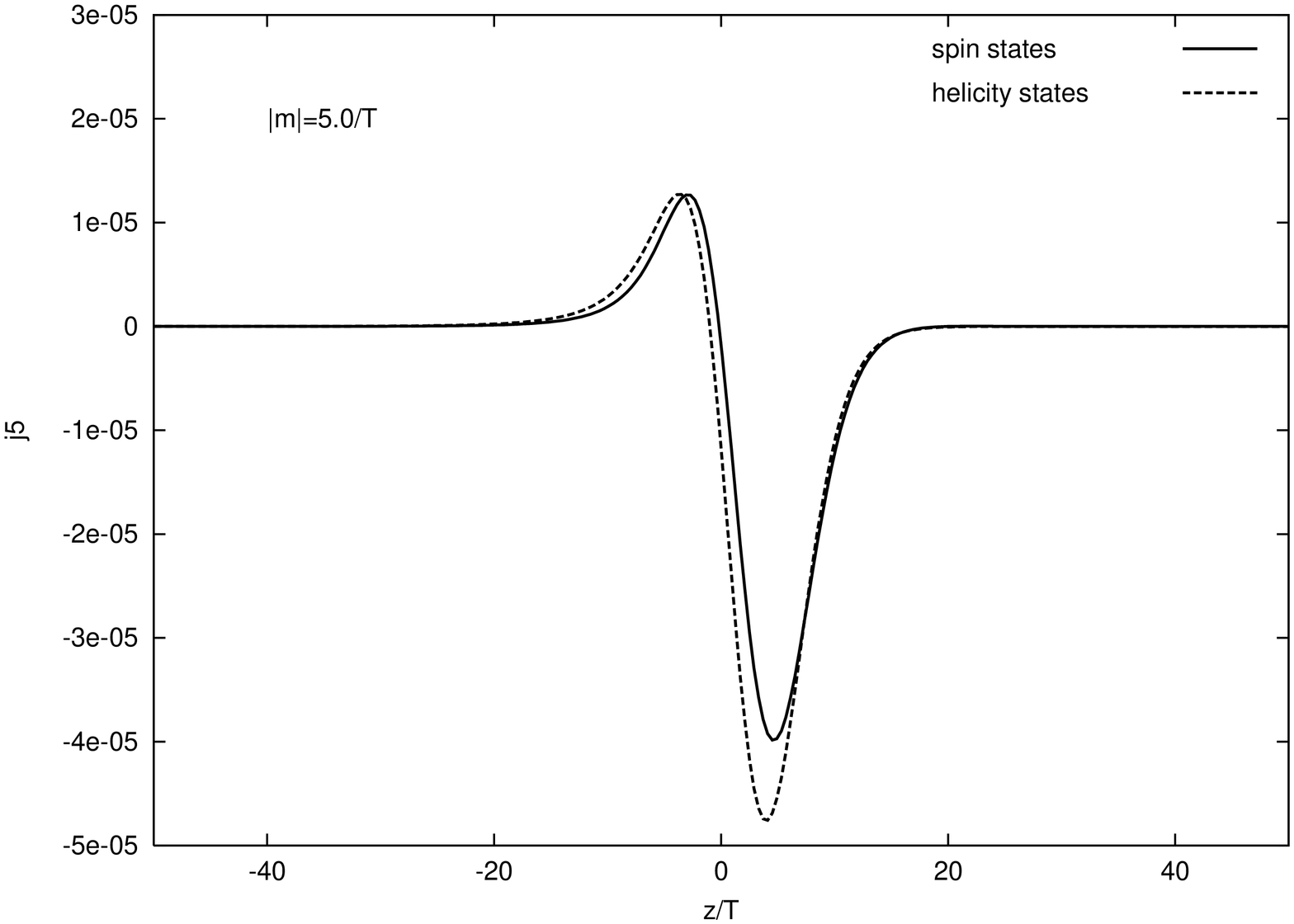, height=3.6in,width=3in,angle=-90}
}
\caption{\small
The axial density computed by solving the fluid equations
for spin states and helicity states, respectively.
The results are very similar, but for both values of the
fermion mass there is a clearly visible difference.
}
\lbfig{fig:spin-helicity}
\end{figure}
Working with spin states instead of helicity states leads to two
important changes on the level of the fluid {\it Ansatz}. Because in the
Boltzmann equation~(\ref{helicity_boltzmann_f5}) the source is multiplied
by $\mbox{sign}(k_z)$, the flow term source for helicity
states will appear in the first moment equation, and the
collisional source will go into the zeroth moment equation,
just opposite to the fluid equations for spin states.
The other difference concerns the axial density, which in terms of
chemical potential and plasma velocity now reads
\beq
  j^5 = \sum_h h \int \frac{d^3k}{(2\pi)^3}
           \beta f_0(1-f_0) 
            \Big(  \mu_h
                 + v_w \beta(1-2f_0) u_h
            \Big)
\,.
\label{j5-by-fluid-hel}                  
\eeq
Compared to~(\ref{j5-by-fluid}) for spin states, the roles of chemical
potential and plasma velocity have changed, in a way. The question
is now, whether the effects of both of these changes cancel each other.
So we solve the fluid equations also for helicity states and show the
result for the axial density in figure~\ref{fig:spin-helicity},
in direct comparison with the density obtained by working with spin states.
One can see that the results differ by about ten percent. For
small masses the fluid equations for spin states produce a larger
axial density, for high masses the equations for helicity states
produce the larger result. Since both cases are an approximation
to the full Boltzmann equations, this difference can to some extent
also be regarded as an estimate of the error one makes by going
from Boltzmann to fluid equations.

%
%
%
%
\subsection{Diffusion equation and sources} 
\label{Diffusion equation and sources}

 Here we make a rough analytic comparison between the flow and collision term
sources. In the stationary limit and neglecting the terms 
of order $v_w^2\ll 1$ (as usually, in the wall frame), 
Eqs.~(\ref{continuity-equation}--\ref{Euler-equation}) can be recast as 
\begin{eqnarray}
&&\!\!\!\!\!\!\!\!\!\!\!\!
  I_{21,i} u_{si}^\prime 
- I_{01,i}\beta^2\frac{{|m_d|_i^2}^\prime}{2} u_{si}
- v_w J_{01,i}\beta^3\frac{{|m_d|_i^2}^\prime}{2} \mu_{si}
- \beta^3\!\int\frac{d^3k}{(2\pi)^3} s {\cal S}^{\rm flow}_i 
\label{Diffusion:1a}
\\
&=& \!\sum_{s'i'}\Big[
                     \Gamma^{0\mu}_{si,s'i'}\beta\mu_{s'i'}
                   - v_w\Big(
                       \Gamma^{1u}_{si,s'i'}
                     \!-\!  \frac{J_{21,i}}{I_{22,i}}\Gamma^{0u}_{si,s'i'}
                   \Big) u_{s'i'}
                   \Big]
+ \!\sum_{i'}\Big[
                     \Gamma^{0\mu}_{i,i'}\beta\mu_{i'}
                   - v_w\Big(
                        \Gamma^{1u}_{i,i'}
                     \!-\!  \frac{J_{21,i}}{I_{22,i}}\Gamma^{0u}_{i,i'}
                   \Big) u_{i'}
                   \Big]
\nonumber\\
&&\!\!\!\!\!\!\!\!\!\!\!\!\!\!
  I_{22,i}\beta\mu_{si}^\prime 
+ v_w\Big(\frac{I_{01,i}J_{42,i}}{2I_{21,i}}\!-\!J_{22,i}\Big)
               \beta^2{|m_d|_i^2}^\prime u_{si}
-\beta^3\!\int\frac{d^3k}{(2\pi)^3}\frac{k_z}{\omega_{0i}}
                                    s{\cal S}^{\rm coll}_i
\label{Diffusion:1b}
\\
&=& \sum_{s'i'}\Big[v_w
                   \Big(
                        \Gamma^{1\mu}_{si,s'i'}
                     \!-\!  \frac{J_{42,i}}{I_{21,i}}\Gamma^{0\mu}_{si,s'i'}
                   \Big)\beta\mu_{s'i'}
                     + \Gamma^{1u}_{si,s'i'} u_{s'i'}
                   \Big]
 +\sum_{i'}\Big[v_w
                   \Big(
                        \Gamma^{1\mu}_{i,i'}
                     \!-\!  \frac{J_{42,i}}{I_{21,i}}\Gamma^{0\mu}_{i,i'}
                   \Big)\beta\mu_{i'}
                     + \Gamma^{1u}_{i,i'} u_{i'}
                   \Big]
\,,
\nonumber
\end{eqnarray}
where we kept only those sources which are not suppressed
by the wall velocity. (Recall that the sources ${\cal S}^{\rm flow}_i$ and 
${\cal S}^{\rm coll}_i$ are both linear in $v_w$.) From 
Eq.~(\ref{Diffusion:1b})
we can easily write an expression for the fluid velocity, 
\begin{eqnarray}
u_{si}\!&=&\! \frac{1}{\Gamma^{1u}_{si,si} 
              + v_w\Big(J_{22,i}\!-\!\frac{I_{01,i}J_{42,i}}{2I_{21,i}}\Big)
                             \beta^2{|m_d|_i^2}^\prime}
\Bigg[
   I_{22,i}\beta\mu_{si}^\prime 
   -\!\sum_{s'i'}v_w
                   \Big(
                        \Gamma^{1\mu}_{si,s'i'}
                     \!-\!  \frac{J_{42,i}}{I_{21,i}}\Gamma^{0\mu}_{si,s'i'}
                   \Big)\beta\mu_{s'i'}
\nonumber\\
&&\hskip 2in 
 + v_w\frac{J_{42,i}}{I_{21,i}}\beta^3\! 
        \int\frac{d^3k}{(2\pi)^3} s {\cal S}^{\rm flow}_i 
 -\beta^3\!\int\frac{d^3k}{(2\pi)^3}\frac{k_z}{\omega_{0i}}
                                    s{\cal S}^{\rm coll}_i 
\nonumber\\
&&\hskip 2in 
       - {\sum_{s'i'}}^\prime\Gamma^{1u}_{si,s'i'} u_{s'i'}
       -\sum_{i'}\Big[v_w
                   \Big(
                        \Gamma^{1\mu}_{i,i'}
                     \!-\!  \frac{J_{42,i}}{I_{21,i}}\Gamma^{0\mu}_{i,i'}
                   \Big)\beta\mu_{i'}
                     + \Gamma^{1u}_{i,i'} u_{i'}
                   \Big]
\Bigg]
\,,\qquad
\label{Diffusion:2}
\end{eqnarray}
and similarly for $\mu_{si}$. Upon inserting Eq.~(\ref{Diffusion:2}) 
and its derivative into the continuity equation~(\ref{Diffusion:1a}), 
we get a rather complicated 
second order diffusion equation for the chemical potential,
from which we list just a few relevant terms, 
\begin{eqnarray}
&&\!\!\!\!\!\!\!\!\!\!
  D_{si}\mu_{si}^{\prime\prime}
 - v_w D_{si} \sum_{s'i'}\Big(\frac{\Gamma^{1\mu}_{si,s'i'}}{I_{22,i}}
                           -  \frac{J_{42,i}}{I_{22,i}I_{21,i}}
                                    \Gamma^{0\mu}_{si,s'i'}
                         \Big)\beta\mu_{si}^{\prime}
\label{Diffusion:3}
\\
 &&\!\!\!\!\!\!\!\!\!
- \frac{D_{si}}{I_{22,i}} \beta^2
    \int\frac{d^3k}{(2\pi)^3} \frac{k_z}{\omega_{0i}}
                s{{\cal S}^{\rm coll}_i}^{\,\prime}
 - \frac{I_{01,i}}{I_{00,i}}\beta^4 \frac{{|m_d|^2}^\prime}{2}
    \int\frac{d^3k}{(2\pi)^3} \frac{k_z}{\omega_{0i}}s{\cal S}^{\rm coll}_i
-\frac{1}{I_{00,i}}\beta^2 \int\frac{d^3k}{(2\pi)^3} s{\cal S}^{\rm flow}_i
 + .. = 0
\,,
\quad
\nonumber
\end{eqnarray}
where we defined a diffusion coefficient 
\begin{equation}
 D_{si} \equiv \frac{I_{21,i}I_{22,i}}
                    {I_{00,i}
                    }
   \frac{1}{\Gamma^{1u}_{si,si}}
  + O(v_w)
\,.
\label{Diffusion:4}
\end{equation}
\begin{figure}[t]
  \unitlength=1in
  \begin{center}
    \psfrag{Lx}[r]{$|m|/T$}
    \psfrag{Ly}[r][][1][-90]{$\Gamma^{1u}_{s,s}/T$}
    \includegraphics[width=3.5in,angle=-90]{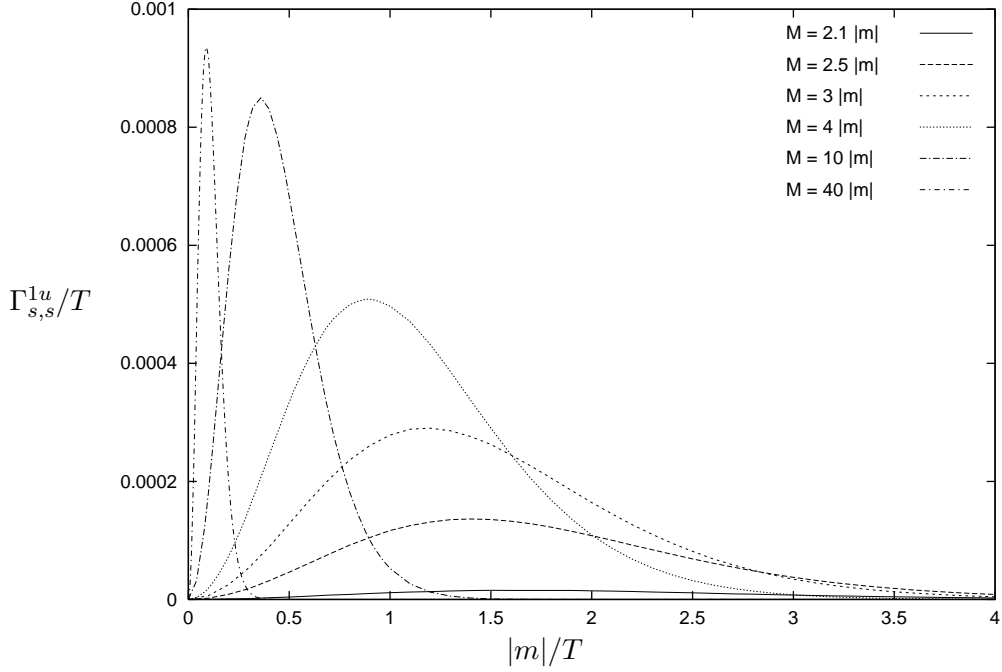}
  \end{center}
\lbfig{figure:GammaU1}
\caption{The rate $\Gamma^{1u}_{s,s}$ as a function of the fermion mass
$|m|$.
The scalar mass $M$ is a multiple of $|m|$. 
At the ratio $M/|m|$ increases, the rate peaks at an asymptotic value
$\Gamma^{1u}_{s,s}\rightarrow (0.9 - 1.0)\times 10^{-3}T$.
The spin-flip rate $\Gamma^{1u}_{s,-s}$ is negative and smaller
by a factor four, approximately. 
}
\end{figure}
We can neglect the contribution of the collisional source term
$\propto {|m_d|_i^2}^\prime$, unless the diffusion constant is very
small and $|m_d|_i \ll T$.
From Eq.~(\ref{Diffusion:4}) it is then clear that as the diffusion
parameter $D_{si}$ grows
(or equivalently, when $\Gamma^{1u}_{si,si}$ becomes small), 
the collisional source becomes more and more important.
To give  
a rough quantitative comparison between the sources, we first note that
the flow and the collision term
sources~(\ref{parametric-flow-source}), (\ref{1st_moment_of_coll_source}) 
can be approximated by 
\begin{eqnarray}
    \beta^4\! \int \frac{d^3k}{(2\pi)^3} {\cal S}_i^{\rm flow} 
   &\sim& 2\times 10^{-6} \frac{m_0^2}{T^2}\Big(
                                            -  {\cal I}_a 
                                            +  \frac{m_0^2}{T^2}{\cal I}_b 
                                           \Big)
\label{Sflow:estimate}
\\
\beta^4\!\int\frac{d^3k}{(2\pi)^3}
           \frac{k_z}{\omega_{0i}}{\cal S}_i^{\rm coll} 
            &\sim& 2\times 10^{-7} {\cal I}
\,,
\label{Scoll:estimate}
\end{eqnarray}
where we made use of $\partial_z \sim 1/L = T/10$,
$\theta \sim \theta_0 = 0.1$, $|m_d|_i \sim m_0$, and $v_w = 0.1$.
Now from figures~\ref{figure:flow-source} and~\ref{fig_coll}
we see that ${\cal I}_a$ and  ${\cal I}_b$ 
peak at about {\it one}, and are typically one order 
of magnitude larger than ${\cal I}$, and we conclude that 
the flow term source~(\ref{Sflow:estimate}) is typically 
about two orders of magnitude larger than the collision term 
source~(\ref{Scoll:estimate}). When viewed as 
sources in the diffusion equation~(\ref{Diffusion:3}), we first observe 
that both sources appear at second order in gradients, and are hence 
suppressed as $1/L_w^2$, so changing the wall thickness $L_w$ should not 
affect much the relative strength of the sources. 
Analogously, the relative strength of the sources is not changed by 
changing the wall velocity $v_w$, Yukawa coupling $|y|$,  
or CP-violating phase $\theta$.  
Second, making a comparison of the collisional and flow term sources
in Eq.~(\ref{Diffusion:3}) at a maximum
$|m_d|_i\sim T$ leads to the conclusion that the sources are comparable 
when the diffusion constant becomes $D_{si} \sim 300 T^{-1}$, such that 
$\Gamma^{1u}_{si,si} \sim 0.03 D_{si}^{-1} \sim 10^{-4}T$ is the rate 
below which the collisional source should dominate
\footnote{In making these estimates, we made use of the
integrals~(\ref{I_ab,i}--\ref{J_ab,i}), 
$I_{ab,i} = I_{ab}(|m_d|_i)$  and 
$J_{ab,i} = J_{ab}(|m_d|_i)$, 
which, when evaluated in the massless limit, yield:
$I_{22}(0) = \zeta(2)/(6\pi^2) = 1/36 = 0.02\dot{7}$, 
$I_{21}(0) = 3\zeta(3)/(4\pi^2)\simeq 0.091$, 
$I_{01}(0) = \ln(2)/(2\pi^2)\simeq 0.035$ and
$I_{00}(0) = \zeta(2)/(2\pi^2) = 1/12 = 0.08\dot{3}$.
}.
A self consistent quantitative comparison of the two sources  
can be inferred from the above analysis and 
Figure~\ref{figure:GammaU1}, where we show  
the rate $\Gamma^{1u}_{si,si}$ as a function of the masses $|m_d|_i$ and $M_i$.
In the parameter range where the sources peak, $|m_d|_i\simeq T$ and 
$M_i\simeq (3-4)T$, the rate $\Gamma^{1u}_{si,si}\simeq (3-5)\times 10^{-4}T$.
This then implies that, when compared with the flow term source,
in the region of parameter space where both sources peak,
the collisional source is a few times smaller.
This collisional source suppression can be attributed to a phase 
space suppression of ${\cal S}_i^{\rm coll}$, to which 
only particles with relatively small momenta contribute, which are not
so numerous. We expect that a similar conclusion can be reached 
when higher order loop processes are considered, both as regards
to the collisional source as well as diffusion constant contributions.


\cleardoublepage
\section{Conclusion}

With the present paper we complete our work on a controlled first principle
derivation of transport equations for chiral fermions Yukawa-coupled to
complex scalars by a discussion of the collision term, whereby we
approximate the self-energies by the one-loop expressions.

\vskip 0.05in

We calculate, from first principles, the CP-violating source appearing
in the collision term of the fermionic kinetic equation.
This represents a first calculation in which both semiclassical and
spontaneous sources are obtained within one formalism.
We also present a first computation of a CP-violating source from
the scalar collision term, which arises from nonlocal fermion loop
Wigner functions, and can be traced to the CP-violating split in
the dispersion relation for fermions. 
Next, we derive the fluid equations associated to the Boltzmann equations
found in Paper~I and solve them (by a novel method) for a simple case,
including only fermionic particles, which allows us to make a comparison
between different sources. 
It turns out that the collisional source is subdominant when compared with
the semiclassical source coming from the semiclassical force in the flow
term of the fermionic kinetic equation, unless the fermions diffuse very
efficiently. 

\vskip 0.05in

We discuss the relation between the helicity and the spin
quasiparticle states and show that in a highly relativistic limit
the Boltzmann equation for spin quasiparticles reduces to the one
for helicity quasiparticles. Fluid equations are, however, not the same. 
In particular, we show that in the fluid equations for spin
quasiparticles the semiclassical force sources the continuity equation,
while the collision term sources the Euler equation;
exactly the opposite is true for the fluid equations of helicity
quasiparticles.

\vskip 0.05in

The question of thermalization rates in the collision term of the
Kadanoff-Baym equations is also addressed.
We calculate the rates with the one-loop self-energies, for which 
only scalar absorption and emission processes are taken into account.
We point out at the shortcomings of these rates: 
as a consequence of the energy momentum conservation 
enforced in the on-shell limit, the rates are suppressed for small masses, and
vanish in the massless limit. By establishing a relation to more standard 
rates calculated with helicity states, we outline how 
one could approximate scattering rate calculations from the Kadanoff-Baym
equations with two-loop self-energies, which are needed for the
derivation of quantitatively correct fluid equations.

\vskip 0.05in

We conclude by making a few comments on further directions of the research
presented here and in Paper~I.
Solving the Kadanoff-Baym equations
with the self-energies truncated at two-loop order, 
needed for modeling baryogenesis at a first order electroweak phase
transition, would require a significantly more effort than what was 
needed in this work, and thus would be justified 
only when quantitative consistency checks of a particle physics
model would be needed.
Testing the accuracy of the fluid equations based on the Kadanoff-Baym
equations with one-loop self-energies is feasible, and should be
performed. Furthermore, it would be of interest to develop a basis independent
formalism for mixing fermions valid to order $\hbar$ in a derivative
expansion, and in particular to study the significance of the mixing
in the situations when mass eigenvalues are (nearly) degenerate. 
Moreover, an inclusion of the hermitean self-energies would be also 
desirable. Apart from providing thermal mass corrections, these self-energies
mix spin and may provide an additional source of CP violation
(which we expect to be a of a similar strength as the collisional source).
Finally, when fully developed, this formalism should be used
for quantitative baryogenesis calculations. Important examples comprise
baryogenesis mediated by charginos or neutralinos of supersymmetric
extensions of the Minimal Standard Model, baryogenesis mediated by 
(top) quarks or scalars in two Higgs doublet models, {\it etc}. 
The interest in these type of calculations is particularly well motivated
in the light of the next generation of accelerator experiments (LHC, NLC),
which will provide a new experimental input on the various parameters of
the electroweak scale physics. Moreover, novel experiments on electric
dipole moments will probe in more depth CP-violation in the electroweak
sector.

\vskip 0.1in

\section*{Acknowledgements}
We would like to acknowledge collaboration and engaged discussions
with Kimmo Kainulainen in earlier stages of this project.
We would also like to thank Thomas Konstandin for constructive discussions.
The work of SW was supported by the U.S. Department of Energy
under Grant No. DE-AC02-98CH10886.
SW also thanks the Alexander von Humboldt Foundation
for support by a Feodor Lynen Fellowship.


\begin{appendix}

\cleardoublepage

\newpage
\section{CP-violating sources in the scalar collision term}
\label{A:CP-violating sources in the scalar collision term}


In this Appendix we show how to perform analytic evaluations
of ({\it seven} out of {\it eight} integrals in) the integral~(\ref{calI})
\begin{eqnarray}
 {\cal I}^{<,>}_{\phi 0}(k) & = &
      \int \frac{d^4k'd^4k''}{(2\pi)^4} \delta^4(k+k'-k'')
\frac{k_0' k_0'' -  \vec k_\|'\cdot \vec k_\|''}
   {\tilde {k_0'}\tilde {k_0''}} 
      g_0^{s'>,<}(k') g_0^{s''<,>}(k'')  \frac{1}{(\tilde {k_0'})^2}
\,.
\label{A:calI}
\end{eqnarray}
Recall that this integral is important because it yields
a CP-violating source in the scalar collision term 
${\cal C}_{\phi 0}$~(\ref{C-phi0-e}).

We argued in section~\ref{Scalar collision term} that, 
when evaluating~(\ref{A:calI}), it suffices to work with 
the thermal equilibrium particle densities~(\ref{g0<>1}), which we rewrite as 
\begin{eqnarray}
{g^{s<}_0}_{\rm eq} 
 &=& \;\;\pi\frac{\tilde \omega_0}{\omega_0}
     \left[\delta(k_0-\omega_0)f_0 + \delta(k_0+\omega_0)(1-\bar f_0)\right]
\nonumber\\
{g^{s>}_0}_{\rm eq}  
 &=& - \pi\frac{\tilde \omega_0}{\omega_0}
    \left[\delta(k_0-\omega_0) (1-f_0) + \delta(k_0+\omega_0)\bar f_0\right]
\label{g<>0f}
\end{eqnarray}
where $\omega_0=\sqrt{\vec k^2 +|m|^2}$  and we used the following
equilibrium definitions for the particle and antiparticle 
distribution functions
\begin{eqnarray}
f_0 &=& n_{\rm eq}(\omega_0) = \frac{1}{e^{\beta\omega_0}+1}
\qquad\qquad\;\,\,
  ({\tt plasma\;\, frame\;\,\rm particle\;\, distribution\;\,function})
\nonumber\\
\bar f_0 &=& 1-n_{\rm eq}(-\omega_0) = \frac{1}{e^{\beta\omega_0}+1}
\qquad  
({\tt plasma\;\, frame\;\,\rm antiparticle\;\, distribution\;\,function})
\,.
\label{fbarf0}
\end{eqnarray}
Note that in thermal equilibrium in plasma frame and to leading order 
in gradients we have, $\bar f_0 = f_0$.

We can now easily evaluate the $k_0'$ and $k_0''$ integrals in
${\cal I}^{>,<}_{\phi 0}$. The result is 
\begin{eqnarray}
{\cal I}_{\phi 0}(k)  &\equiv & {\cal I}^>_{\phi 0}(k) =
    -  \frac{1}{4} \int \frac{d^3k'd^3k''}{(2\pi)^2} 
  \delta^3(\vec k+\vec k'-\vec k'')  \frac{1}{({\tilde \omega_0}')^2}
\nonumber\\
&\times&\Bigg\{  \frac{\omega_0'\omega_0'' - \vec k'_\|\cdot \vec k''_\|}
       {\omega_0'\omega_0''} 
\Big[\delta(k_0+\omega_0'-\omega_0'')f_0' (1-f_0'')
+\delta(k_0-\omega_0'+\omega_0'')(1-\bar f_0')\bar f_0''\Big]
\nonumber\\
&&\,\, + \,\, 
  \frac{\omega_0'\omega_0'' + \vec k'_\|\cdot \vec k''_\|}
       {\omega_0'\omega_0''} 
\Big[\delta(k_0+\omega_0'+\omega_0'')f_0'\bar f_0''
+\delta(k_0-\omega_0'-\omega_0'')(1-\bar f_0')(1-f_0'')
\Big]\Bigg\}
,\quad
\label{calI-b}
\end{eqnarray}
where $\omega_0' = \sqrt{ \vec {k'}^2 + |m|^2}$, 
 $\omega_0'' = \sqrt{ \vec {k''}^2 + |m|^2}$, 
$\tilde \omega_0' = \sqrt{ {k'_z}^2 + |m|^2}$, 
and $f_0' = n_0(\omega_0')$, $f_0'' = n_0(\omega_0'')$, 
$\bar f_0' = 1 - n_0(-\omega_0')$, $\bar f_0'' = 1 - n_0(-\omega_0'')$.
The former two terms in~(\ref{calI-b}) are contributions from absorption and 
emission of a scalar particle. The latter two arise from 
annihilation of a scalar particle with a fermion-antifermion pair
and the fermion-antifermion pair creation from a scalar particle, 
respectively. These four processes are illustrated in 
figure~\ref{figure:scalar-reactions}.

%
%

Upon performing the $\vec k''$-integral, Eq.~(\ref{calI-b}) can be reduced 
to the following integral
\begin{eqnarray}
{\cal I}_{\phi 0}(k)
 &=& -  \frac{1}{16\pi^2}\int d^3k' \frac{1}{({\tilde \omega_0}')^2}
\nonumber\\
&\times& \Bigg\{
  \left(1 - \frac{{k'_\|}^2 + \vec k_\|\cdot\vec k'_\|}{\omega_0'\omega_0''}
    \right)
\Big[\delta(k_0+\omega_0'-\omega_0'')f_0' (1-f_0'')
+\delta(k_0-\omega_0'+\omega_0'')(1-\bar f_0')\bar f_0''\Big]
\nonumber\\
&+&  \left(1 + \frac{{k'_\|}^2 + \vec k_\|\cdot\vec k'_\|}{\omega_0'\omega_0''}
     \right)
\Big[\delta(k_0+\omega_0'+\omega_0'')f_0'\bar f_0''
+\delta(k_0-\omega_0'-\omega_0'')(1-\bar f_0')(1-f_0'')
\Big]\Bigg\},\quad
\label{calI-c}
\end{eqnarray}
where now $\omega_0' = \sqrt{k'^2  + |m|^2}$,  
 $\tilde \omega_0' = \sqrt{k'_z + |m|^2}$,
 $\omega_0'' = \sqrt{ (\vec k+\vec k')^2  + |m|^2}$,  
$\vec k'_\|\cdot \vec k''_\| = {k'_\|}^2 + \vec k_\|\cdot\vec k'_\|$.
It is convenient to break the integral 
\begin{equation}
\int d^3k' = \int_{-\infty}^{\infty} dk_z' \int_0^\infty k'_\| dk'_\| 
              \int_0^{2\pi} d\phi'
\label{int-k'}
\end{equation}
where $\vec k_\|\cdot\vec k'_\| = k_\| k'_\| \cos \phi'$.
We can perform the angular $\phi$-integral in~(\ref{calI-c}) 
with the help of the $\delta$-functions. 
Note first that
\begin{eqnarray}
\delta(\pm k_0+\omega_0'-\omega_0'')
                 &=& \theta(\mp k_0) 
                     \theta\big(k^2{k'}^2- (\beta_0-k_zk_z')^2\big) 
\nonumber\\
&\times&          \frac{(\omega_0'-\omega_\phi)\theta(\omega_0'-\omega_\phi)}
                       {k_\|k'_\|\sqrt{1-{x'}_0^2}}
                   \Big[\delta(\phi'-\phi'_0) + \delta(\phi'+\phi'_0)\Big]
\nonumber\\
             x'_0 &\equiv& \cos\phi'_0 = 
                  \frac{(M^2/2)-\omega_\phi\omega_0'-k_zk'_z}{k_\|k'_\|}
\nonumber\\
     \omega_0''  &=& \omega_0'-\omega_\phi
\,,
\label{delta-omega}
\end{eqnarray}
such that in the first $\delta$-function only the antiparticle pole,
$\theta(-k_0)$, contributes, while in the second $\delta$-function only 
the particle pole, $\theta(k_0)$, contributes.
Similarly, we have 
\begin{eqnarray}
\delta(\pm k_0+\omega_0'+\omega_0'')
                 &=& \theta(\mp k_0) 
                     \theta\big(k^2{k'}^2- (\beta_0-k_zk_z')^2\big) 
\nonumber\\
&\times&          \frac{(\omega_\phi-\omega_0')\theta(\omega_\phi-\omega_0')}
                       {k_\|k'_\|\sqrt{1-{x'}_0^2}}
                   \Big[\delta(\phi'-\phi'_0) + \delta(\phi'+\phi'_0)\Big]
\nonumber\\
     \omega_0''  &=& \omega_\phi-\omega_0'
\,,
\label{delta-omega2}
\end{eqnarray}
The product of $\theta$-functions in~(\ref{delta-omega}-\ref{delta-omega2})
stems from the requirement that $|x'_0|\leq 1$, and can be conveniently 
expressed as 
\begin{eqnarray}
M &\geq& 2|m|
\nonumber\\
k'_{z_1} &\leq& k_z' \leq k'_{z_2}
,\qquad
k^\prime_{1,2} = \frac{k_z}{k^2}\beta_0
        \mp \frac{k_\|}{k^2}\big(k^2{k^\prime}^2 - \beta_0^2 \big)^\frac 12
\nonumber\\
k{k^\prime} &\geq& |\beta_0|
\,,\qquad
  \beta_0 \equiv \frac{M^2}{2} -\omega_\phi\omega_0^\prime
\,.
\label{a-pm}
\end{eqnarray}

The result of integration over the $\delta$-functions is then
\begin{eqnarray}
{\cal I}_{\phi 0}(k) &=&
    -  \frac{1}{8\pi^2 k}
           \theta(M-2|m|)
           \int_{0}^{\infty} \frac{k'dk'}{\omega_0'}
                       \theta(k^2{k^\prime}^2-\beta_0^2)
\int_{k'_{z_1}}^{k'_{z_2}} \frac{dk^\prime_z}{\tilde \omega_0^{\prime^2}}
    \frac{-(M^2/2)+k_zk_z'+\tilde\omega_0^{\prime 2}}
       {\big[(k'_{z_2} - k^\prime_z)(k^\prime_z - k'_{z_1})\big]^\frac 12}
\nonumber\\
&\times&\bigg[ \theta(\omega_0^\prime-\omega_\phi)
         \left[\theta(-k_0) f_0' (1-f_0'') 
           +   \theta(k_0) (1-\bar f_0') \bar f_0'' \right]
               _{\omega_0''=\omega_0^\prime-\omega_\phi}
\nonumber\\
&-&   \theta(\omega_\phi-\omega_0^\prime)
      \left[\theta(-k_0)f_0'\bar f_0'' 
          + \theta(k_0)(1-\bar f_0')(1-f_0'')\right]
               _{\omega_0''=\omega_\phi-\omega_0^\prime}
\bigg]
\,.
\label{calI-e}
\end{eqnarray}
To evaluate the $k_z'$-integrals in~(\ref{calI-e}) the following
integrals are useful.
They can for example be performed by deforming 
the contour of complex integration -- the closed curve that contains
the poles $k_z' = k'_{z_{2,1}}$ --  into a contour around the complex poles 
$k_z' = \pm i |m|$, and evaluating the sum of the associated residues:
\begin{eqnarray}
 {\cal I}_0 &=&  \int_{k'_{z_1}}^{k'_{z_2}} \frac{dk_z'}{ {k_z'}^2 + |m|^2 } 
                  \frac{1}{\sqrt{(k_z'-k'_{z_1})(k'_{z_2}-k_z')}}
    = \frac{\pi}{|m|}
      \Re\frac{1}{\sqrt{(|m|^2 - k'_{z_2}k'_{z_1}) +i|m|(k'_{z_1} + k'_{z_2})}}
\nonumber\\
 {\cal I}_1 &=&  \int_{k'_{z_1}}^{k'_{z_2}} \frac{k_z'dk_z'}{ {k_z'}^2 + |m|^2 } 
                  \frac{1}{\sqrt{(k_z'-k'_{z_1})(k'_{z_2}-k_z')}}
  = - \Im\frac{\pi}{\sqrt{(|m|^2 - k'_{z_2}k'_{z_1}) +i|m|(k'_{z_1} + k'_{z_2})}}
\nonumber\\
 {\cal I}_2 &=&  \int_{k'_{z_1}}^{k'_{z_2}} 
     \frac{dk_z'}{\sqrt{(k_z'-k'_{z_1})(k'_{z_2}-k_z')}}
     \qquad\qquad\;\;\,\, =  \pi
\nonumber\\
 {\cal I}_3 &=&  \int_{k'_{z_1}}^{k'_{z_2}} 
     \frac{k_z'dk_z'}{\sqrt{(k_z'-k'_{z_1})(k'_{z_2}-k_z')}}
     \qquad\qquad\;\;\,\, =   \frac{\pi k_z}{k^2}\beta_0
\,,
\label{integrals-f-kz'}
\end{eqnarray}
where 
\beq
   k'_{z_2}k'_{z_1} = \frac{{k'}^2k_z^2 + (\beta_0^2 - k^2{k'}^2)}{k^2}
\,,\qquad
   k'_{z_1} + k'_{z_2}  = \frac{2k_z\beta_0}{k^2}
\,.
\label{a+a-}
\eeq
The integrals ${\cal I}_0$ and ${\cal I}_1$ are generally expressed in 
terms of complex functions of the form 
\begin{equation}
 \frac{1}{\sqrt{(|m|^2 - k'_{z_1}k'_{z_2}) +i|m|(k'_{z_1}+k'_{z_1})}} 
    = \frac{k}{k'}\frac{1}{\sqrt{(k_z - c_-)(c_+ - k_z)}}
\label{a+a-2}
\end{equation}
where $c_-$ and $c_+$ are the (complex) roots of the equation 
$(|m|^2 - k'_{z_1}k'_{z_2}) +i|m|(k'_{z_1}+k'_{z_1})=0$, that is 
\begin{equation}
 c_\pm = \frac{i|m|\beta_0}{{k'}^2} 
  \pm \frac{\omega_0'}{{k'}^2} \sqrt{k^2{k'}^2 - \beta_0^2}
\label{c+c-}
\end{equation}
We can thus reduce~(\ref{calI-e}) to the form
\begin{eqnarray}
&& {\cal I}_{\phi 0}(k) =
     \frac{1}{8\pi}
           \theta(M-2|m|)
           \int_{0}^{\infty} \frac{dk'}{\omega_0'}
                       \theta(k^2{k^\prime}^2-\beta_0^2)
\nonumber\\
&\times&
\bigg[
  \frac{M^2}{2|m|}\Re \frac{1}{\sqrt{(k_z - c_-)(c_+ - k_z)}}
 + \Im \frac{k_z}{\sqrt{(k_z - c_-)(c_+ - k_z)}} - \frac{k'}{k}
\bigg]
\nonumber\\
&\times&\bigg[ \theta(\omega_0^\prime-\omega_\phi)
         \left[\theta(-k_0) f_0' (1-f_0'') 
           +   \theta(k_0) (1-\bar f_0') \bar f_0'' \right]
               _{\omega_0''=\omega_0^\prime-\omega_\phi}
\nonumber\\
&& -\;   \theta(\omega_\phi-\omega_0^\prime)
      \left[\theta(-k_0)f_0'\bar f_0'' 
          + \theta(k_0)(1-\bar f_0')(1-f_0'')\right]
               _{\omega_0''= \omega_\phi - \omega_0^\prime}
\bigg]
.
\label{calI-f}
\end{eqnarray}
The final $k'$-integral cannot be performed analytically. 
In section~\ref{The scalar collision term with no mixing} we use
this integral expression to compute the CP-violating collisional source in 
the scalar kinetic equation.

\newpage
\section{Consistency of the fermionic kinetic equations}
\label{Consistency of the fermionic kinetic equations}

In section~\ref{Kinetics of fermions: tree-level analysis}
we have presented a derivation of the fermionic kinetic equations. 
As a consequence of spin conservation, the Clifford algebra of 16 functions
is reduced to four (a priori different) functions $g_a^s$ 
(for each spin $s=\pm 1$). Furthermore, the constraint
equations~(\ref{ce0d1}--\ref{ce3d1}) relate different
$g_a^s$, resulting in one independent function, 
for which the standard choice is the vector particle density on phase space,
$g^s_0$. Here we prove that all kinetic equations~(\ref{ked0}-\ref{ked3})
are equivalent both at the level of flow, as well as collision, terms,
implying consistency of our kinetic equations. 
First we show the equivalence of the flow terms and then we consider 
the collision terms.

We shall not give a proof for all three equations, but instead
restrict ourselves to the kinetic equation~(\ref{ked3}) for $g^s_3$:
\begin{equation}
     \frac{1}{\tk} (k_0\del_t + \vec{k}_\|\cdot\vec{\del}) g^s_3
   +  s \del_z g^s_0
   + 2(m_a - \frac 18 m''_a \delk^2 ) g^s_1
   - 2(m_h - \frac 18 m''_h \delk^2 ) g^s_2
 = {\cal K}^s_{3}  
\,.
\end{equation}
We insert the expressions (\ref{ced:diag1})-(\ref{ced:diag2}) for 
$g^s_1$ and $g^s_2$ (accurate to second order in gradients,
since the kinetic equation for $g^s_3$ contains zeroth order terms), 
to get
\begin{eqnarray}
&&    \frac{1}{\tk} \left( sk_z + \frac{1}{2\tk}|m|^2\theta'\delk \right)
    \frac{1}{2\tk}(k_0\del_t + \vec{k}_\|\cdot\vec{\del}) g^s_0
\label{kin_g3_by_g0}\\
&&\hphantom{X}
 + \frac 12 s\del_zg^s_0
 + \frac{s}{2\tk^2} \bigg( -{|m|^2}'
                           -\frac 12 {|m|^2}' k_z \delk
                           - |m|^2\del_z                 \bigg) g^s_0
\nonumber\\
&& \hphantom{X}
    +\frac{1}{4\tk^3} \bigg( -\frac 12 {|m|^2}'|m|^2\theta' \delk^2
                             + (|m|^2\theta')' + 2|m|^2\theta'\del_z
                + \left( (|m|^2\theta')' + |m|^2\theta'\del_z \right)
                            \left( 1 + k_z\delk \right)   \bigg) g^s_0
\nonumber\\
&&\hphantom{X}
 = {\cal K}^s_{3} - \frac{m_a}{\tk}{\cal C}^s_{1}  
                 + \frac{m_h}{\tk}{\cal C}^s_{2}
\,,
\nonumber
\end{eqnarray}
where ${\cal K}^s_{3}$, $ {\cal C}^s_{1}$ and ${\cal C}^s_{2}$ are defined 
in Eqs.~(\ref{ked:collision-terms:0-3}) and~(\ref{ced:collision-terms:0-3}),
respectively.
Now we take the derivative of the constraint equation (\ref{ced:diag0b})
with respect to $z$ (the collisional contribution to this term can then be
neglected), and we find
\begin{equation}
  - {|m|^2}'g^s_0 
  + \frac{s}{\tk}(|m|^2\theta')' g^s_0
  + \left( k^2 - |m|^2 +\frac{s}{\tk}|m|^2\theta' \right) g^s_0 = 0
\,.
\end{equation}
Next, we use this relation to replace the term
$-{|m|^2}'g^s_0$ in the second line of~(\ref{kin_g3_by_g0}). 
After some algebra this leads to
\begin{eqnarray}
  \frac{1}{\tk}\!\!
  \left(\! sk_z +\frac{1}{2\tk}|m|^2\theta'\delk \right)\!
  \left[ \frac{1}{\tk} k\cdot \del  
        - \frac{1}{2\tk} {|m|^2}' \delk 
        + \frac{s}{2\tk^2}(|m|^2\theta')' \delk\! \right]\! g^s_0 
= {\cal K}^s_{3} - \frac{m_a}{\tk}{\cal C}^s_{1}  + \frac{m_h}{\tk}{\cal C}^s_{2}
\,.\quad\;
 \label{AppC:g3s:consistency}
\end{eqnarray}
We recognize this equation as some operator {\it times} the left hand side of
the familiar kinetic equation for $g^s_0$, thereby proving that the
flow term of the kinetic equation for $g_3^s$ is consistent with the
one for $g_0^s$, which thus carries no new physical information.
The analysis of Eqs.~(\ref{ked1}) and~(\ref{ked2}) for $g_1^s$
and $g_2^s$ is somewhat more complicated, but the final conclusion
is the same.

\bigskip

 From Eqs.~(\ref{ked0}) and~(\ref{AppC:g3s:consistency}) 
it follows that one of the conditions required for the consistency 
of the collisional sources is
\begin{equation}
{\cal K}^s_{3} - \frac{m_a}{\tk}{\cal C}^s_{1}  + \frac{m_h}{\tk}{\cal C}^s_{2}
            = \frac{sk_z}{\tilde k_0}{\cal K}^s_{0}
\,.
\label{AppC:consistency:3}
\end{equation}
In oder to show that this relation is indeed satisfied, 
we make use of the zeroth order collisional source~(\ref{calCpsi0sum})
and the first order collisional source~(\ref{coll_ferm_first}), 
whose precise form is
\begin{eqnarray}
  {\cal C}_{\psi 1} &=& \frac{|y|^2}{4} \beta\gamma_w v_w
   \int \frac{d^4k'd^4k''}{(2\pi)^4}\,\delta^4(k-k'+k'') 
     \sum_{ss'} P_{s'}(k') P_s(k)
\label{C:psi1:2}
\\
&\times&\Bigg(g_0^{s<}(k)g_0^{s'>}(k')i\Delta^{<}(k'')
                  \bigg\{
                        s\gamma^3\gamma^5\bigg[iss'm_h'\frac{1}{\tilde k_0}
                                             -sm_a'\frac{1}{\tilde k_0}
                                                   \frac{k_z'}{\tilde k_0'}
                                          \bigg]
                     -   is\gamma^3\bigg[
                                         iss'm_a'\frac{1}{\tilde k_0}
                                       +  sm_h'\frac{1}{\tilde k_0}
                                               \frac{k_z'}{\tilde k_0'}
                                    \bigg]
                  \bigg\}
\nonumber\\
&&+\;\partial_z\Big(g_0^{s<}(k)g_0^{s'>}(k')i\Delta^{<}(k'')\Big)
        \bigg\{\mathbbm{1}\bigg[iss' - i\frac{k_z}{\tilde k_0}
                                         \frac{k_z'}{\tilde k_0'}
                          \bigg]
  + s\gamma^3\gamma^5\bigg[iss'm_h\frac{1}{\tilde k_0}
                             - sm_a\frac{1}{\tilde k_0}
                                    \frac{k_z'}{\tilde k_0'}
                      \bigg]
\nonumber\\
    &&\hskip 5.3cm -\;   is\gamma^3\bigg[
                                       iss'm_a\frac{1}{\tilde k_0}
                                    +  sm_h\frac{1}{\tilde k_0}
                                           \frac{k_z'}{\tilde k_0'}
                                  \bigg]
                         -\gamma^5\bigg[
                                        is'\frac{k_z}{\tilde k_0}
                                     -  is\frac{k_z'}{\tilde k_0'}
                                   \bigg]
         \bigg\}
     \Bigg)
\,.
\nonumber
\end{eqnarray}
%
%
%
We find
\begin{eqnarray}
  {\cal K}^s_{0} &=& {\cal K}^s_{k}\bigg[|m|^2\theta^\prime
                                   \bigg(
                                    - s'\frac{k_z}{\tilde k_0}
                                          \frac{1}{{\tilde k_0}^{\,\prime\,2}}
                                    + s\frac{1}{{\tilde k_0}^{2}}
                                       \frac{k'_z}{\tilde k_0'}
                                   \bigg)
                                  \bigg]
\nonumber\\
  {\cal K}^s_{3} &=& {\cal K}^s_{k}\bigg[
                                    - ss'|m|^2\theta'
                                    \bigg(\frac{1}{{\tilde k_0}^{2}}
                                       + \frac{1}{{\tilde k_0}^{\,\prime\,2}}
                                    \bigg)
                                  \bigg]
\nonumber\\
  {\cal C}^s_{1} &=&  {\cal K}^s_{k}\bigg[
                                     - ss'|m|^2\theta'
                                         \frac{m_a}{\tilde k_0}
                                         \frac{1}{{\tilde k_0}^{\,\prime\,2}}
                                     + m_h'\frac{k_z}{{\tilde k_0}^2}
                                         \frac{k_z'}{\tilde k_0^\prime}
                                     + ss'm_h'\frac{1}{\tilde k_0}
                                  \bigg]
                  +  {\cal K}^{s'}_{k}\bigg[
                                               ss'\frac{m_h}{\tilde k_0}
                                          \bigg]
\nonumber\\
  {\cal C}^s_{2} &=&  {\cal K}^s_{k}\bigg[
                                      ss'|m|^2\theta'
                                         \frac{m_h}{\tilde k_0}
                                         \frac{1}{{\tilde k_0}^{\,\prime\,2}}
                                     + m_a'\frac{k_z}{{\tilde k_0}^2}
                                         \frac{k_z'}{\tilde k_0^\prime}
                                     + ss'm_a'\frac{1}{\tilde k_0}
                                  \bigg]
                  +  {\cal K}^{s'}_{k}\bigg[
                                               ss'\frac{m_a}{\tilde k_0}
                                          \bigg]
\,,
\label{AppC:collisional:sources}
\end{eqnarray}
where ${\cal K}^s_{k}[]$ and ${\cal K}^{s'}_{k}[]$ denote the functionals 
\begin{eqnarray}
{\cal K}^s_{k}[\cdot] \!\! &\equiv&\!\! 
                     \frac{|y|^2}{4} \beta\gamma_w v_w
   \!\!\int\!\! \frac{d^4k'd^4k''}{(2\pi)^4}\delta^4(k\!-\!k'\!+\!k'') 
     \sum_{s'} \bigg(1+ss'\frac{k_0k_0'-\vec k_\|\cdot \vec k'_\|}
                               {\tilde k_0\tilde k_0'}
               \bigg)
         g_{0\rm eq}^{s<}(k)g_{0\rm eq}^{s'>}(k')i\Delta_{\rm eq}^{<}(k'')
                  [\cdot]
\nonumber\\
{\cal K}^{s'}_{k}[\cdot]\!\! &\equiv&\!\! 
              \frac{|y|^2}{4} \beta\gamma_w v_w
   \!\!\int\!\! \frac{d^4k'd^4k''}{(2\pi)^4}\delta^4(k\!-\!k'\!+\!k'') 
     \sum_{s'} \bigg(1+ss'\frac{k_0k_0'-\vec k_\|\cdot \vec k'_\|}
                               {\tilde k_0\tilde k_0'}
               \bigg)
                 [\del_z g_{0\rm eq}^{s<}(k)g_{0\rm eq}^{s'>}(k')i\Delta_{\rm eq}^{<}(k'')]
                  [\cdot]
.
\nonumber\\
\label{AppC:functional:K}
\end{eqnarray}
From this it immediately follows that the consistency 
condition~(\ref{AppC:consistency:3}) is satisfied. 
Note that the last two terms in each of the sources
${\cal C}^s_{1}$ and ${\cal C}^s_{2}$ originate from the first order 
collisional source~(\ref{C:psi1:2}).

\medskip

 By considering the kinetic equations for $g_1^s$ and  $g_2^s$,
one can show that similar consistency relations hold, which completes
the consistency proof for the fermionic kinetic equations.

\newpage
\section{CP-violating sources in the fermionic collision term}
\label{Collisional source in the fermionic kinetic equation}

In this appendix we show the details of the calculation of the fermionic
collisional source term, which we omitted in the main text. We show here
the calculation in the mixing case, the non-mixing case is simply obtained
by setting $|m|^2_i=|m|^2_j=|m|^2$. We begin with equation
(\ref{coll_ferm_src_0_b_mixing}):
\beqa
  {\cal K}^s_{0,i}  
    &=& \frac 14 \beta v_w
    \left[ \delta(k_0+\omega_{0,i}) - \delta(k_0-\omega_{0,i}) \right]
    \sum_j   \frac{|y_{ij}|^2+|y_{ji}|^2}{2}
\label{coll_src_app}\\
&&
      \int\frac{d^3k'}{(2\pi)^3} \frac{\pi^2}{2}
      \frac{1}{\omega''_\phi}
     \delta(\omega_{0i}+\omega'_{0j}-\omega''_\phi)
       f_{0i} f'_{0j} (1+f^\phi_0)   \nonumber\\
&&
         \left( s\big(|m|^2(\theta'+\Delta_z)\big)_i
                  \frac{k'_z}{\omega_{0i}\tw_{0i}\omega'_{0j}}
               +s\big(|m|^2(\theta'+\Delta_z)\big)_j
                  \frac{k_z}{ \omega_{0i}\tw_{0i}\omega'_{0j}}
                           \frac{\omega_{0i}\omega'_{0j}
                                 + \vec{k}_\|\cdot\vec{k}'_\|}{\tw_{0j}^{'2}}
         \right)   \,.
\nonumber
\eeqa
The $\vec{k}'$ integration is performed in cylindrical coordinates,
$d^3k' = dk'_z k'_\| dk'_\| d\phi$,
where $\phi$ is the angle between $\vec{k}_\|$ and $\vec{k}'_\|$, and
$k_\|$ and $k'_\|$ are the absolute values of the parallel momenta, respectively.
 We substitute
\beq
  x = \vec{k}\cdot\vec{k}' = k_\| k'_\| \cos\phi
\eeq
and instead of integrating over the absolute value of the parallel momentum
we integrate over $k'=|\vec{k}'|$:
\beq
   \int d^3k'
= \int_{-\infty}^\infty dk'_z \frac 12 \int_0^\infty{dk'}^2 \theta(k'^2-k_z^2) 
   \int_{-\infty}^\infty \frac{2\,dx}{\sqrt{(k_\|k'_\|)^2-x^2}} \, \theta(k_\|k'_\|-|x|) \,.
\eeq
Since $\omega''_\phi=\big( (\vec{k}-\vec{k}')^2 + M^2 \big)^{-\frac 12}
                     =\big( k^2+{k'}^2-2x-2k_zk'_z+M^2 \big)^{-\frac 12}$
is a function of $x$, we can rewrite the remaining $\delta$-function
\beq
    \delta(\omega_0+\omega'_0-\omega''_\phi)
  = \omega''_\phi \,  \delta\left(x - (\beta_0 - k_zk'_z)\right) \,,
\eeq
where we used the abbreviation
\beq
  \beta_0 = \frac{M^2 - |m_i|^2 - |m_j|^2}{2} - \omega_{0i} \omega_{0j}' \,.
\eeq
With the $\delta$-function the $x$-integration can now be performed trivially.
With a not so trivial calculation one can show that for arbitrary $\beta_0$ the
relation
\beq
  \theta \big( k_\|k'_\| - (\beta_0 - k_zk'_z) \big)
  \theta \big( {k'}^2-k_z'^2 \big)
   =
  \theta \big( k^2{k'}^2 - \beta_0^2 \big)
  \theta \big( k'_{z_2}-k'_z \big)
  \theta \big( k'_z-k'_{z_1} \big)
\eeq
holds, where $k'_{z_{1,2}}$ are the roots of $(k_\|k_\|')^2-(\beta_0-k_zk'_z)^2=0$,
see~(\ref{a-pm}).
The function $\theta \big( k^2{k'}^2 - \beta_0^2 \big)$ contains a mass-threshold. It
can be found by trying to determine those values of $k'$ for which the
$\theta$-function is non-vanishing:
\beq
  \beta_0^2 < k^2{k'}^2 \Leftrightarrow -kk' < \beta_0 < kk' \,.
\eeq
The latter two conditions can only be satisfied if the quadratic equations
\beqa
&&  {k'}^2|m_i|^2 \mp k'k(M^2-|m_i|^2-|m_j|^2)^2  \\
&&\hphantom{XXXXXX}
       + k^2|m_j|^2 + |m_i|^2|m_j|^2 - \frac 14 (M^2 - |m_i|^2 - |m_j|^2)^2 
   =   0  \,,
\nonumber
\eeqa
have real solutions, which is only the case if
\beq
  M^2 > |m_i|^2 + |m_j|^2 \,.
\eeq
Now we can write the source as
\beqa
  {\cal K}^s_{0,i}  
    &=& \frac{1}{32\pi} \beta v_w
    \sum_j   \frac{|y_{ij}|^2+|y_{ji}|^2}{2}
    \left[ \delta(k_0+\omega_{0i}) - \delta(k_0-\omega_{0i}) \right]
      \frac{1}{\omega_{0i}\tw_{0i}}
\label{coll_src_app_b}\\
&&
      \int k'dk' \theta(k^2{k'}^2-\beta_0^2)
      \frac{1}{\omega'_{0j}}
       f_{0i} f'_{0j} (1+f^\phi_0)
      \int_{k'_{z_1}}^{k'_{z_2}} \frac{1}{k\sqrt{(k'_{z_2}-k'_z)(k'_z-k'_{z_1})}}
   \nonumber\\
&&
         \bigg[
               s\big(|m|^2(\theta'+\Delta_z)\big)_j
                 k_z \frac{M^2-|m_i|^2-|m_j|^2}{2(k_z'^2+|m_j|^2)} \nonumber\\
&&\hphantom{XXXXXX}
               +k_z'\Big( s\big(|m|^2(\theta'+\Delta_z)\big)_i
                         -s\big(|m|^2(\theta'+\Delta_z)\big)_j
                            \frac{k_z^2}{k_z'^2+|m_j|^2}        \Big)
         \bigg]   \,.
\nonumber
\eeqa
For the fluid equations the integral over all momenta is needed. With the observation
that the integrand is odd under $k_z\rightarrow-k_z$ and $k'_z\rightarrow-k'_z$, we
can immediately write
\beq
  \int_\pm \frac{d^4k}{(2\pi)^4} {\cal K}^s_{0,i} = 0 \,,
\eeq
where the index $\pm$ denotes the integral over positive and negative frequencies,
respectively. The integral with an additional factor $k_z/\omega_{0i}$ can
be evaluated further by making use of the integrals~(\ref{integrals-f-kz'}):
\beqa
&& \int_\pm \frac{d^4k}{(2\pi)^4} \frac{k_z}{\omega_{0i}}  {\cal K}^s_{0,i}
\label{coll_src_app_c}\\
&=&  \pm \frac{1}{64\pi^3} \beta v_w
    \sum_j   \frac{|y_{ij}|^2+|y_{ji}|^2}{2}
      \int kdk \int k'dk' \theta(k^2{k'}^2-\beta_0^2)
       f_{0i} f'_{0j} (1+f^\phi_0)
 \nonumber\\
&&\hphantom{X}
      \int_{-k}^k dk_z  \frac{k_z^2}{\omega_{0i}^2\omega'_{0j}\tw_{0i}}
         \bigg[ s\big(|m|^2(\theta'+\Delta_z)\big)_i \frac{\beta_0}{k^3}      \nonumber\\
&&
               +s\big(|m|^2(\theta'+\Delta_z)\big)_j \Big(k_z 
                \Im\left(k^2(|m_j|^2+{k'}^2)-k_z^2{k'}^2-\beta_0^2+2i|m_j|\beta_0 k_z \right)^{-\frac 12}    \nonumber\\
&&\hphantom{XX}
              + \frac{M^2-|m_i|^2-|m_j|^2}{2|m_j|}
                \Re\left(k^2(|m_j|^2+{k'}^2)-k_z^2{k'}^2-\beta_0^2+2i|m_j|\beta_0 k_z  \right)^{-\frac 12} \Big)
         \bigg]   \,.
\nonumber
\eeqa


\newpage
\section{Collisional Rates}
\label{app_rates}

With the fluid {\it Ansatz}~(\ref{fluid_ansatz}) the momentum dependence of
$\delta g$ is completely specified.

The evaluation of the integrals runs along the same lines as for the
collisional source in the first part of this appendix. We just give
the results. The contribution to the zeroth moment equation is
\beq
  2\beta^3 \int_{k_0=0}^\infty \frac{d^4k}{(2\pi)^4}
   \Big[ {\cal K}_0^s(k) +  {\cal K}_0^s(-k) \Big]_{relax}
 = - \sum_{s'}
     \Big(
             \beta \Gamma^{0\mu}_{ss'} \mu_{s'} 
           + v_w   \Gamma^{0u}_{ss'}   u_{s'}
     \Big)
 \,,
\eeq
where the rates are given by
\beqa
    \Gamma^{0\mu}_{ss'}
  &=&
    \frac{1}{4\pi^3}\beta^3\frac{y^2}{4}
     \int_0^\infty k\,dk\,k'\,dk'\,
      \theta(k^2{k'}^2-\beta_0^2)
      \frac{\omega_0\omega_0'+\beta_0}{\omega_0\omega_0'}
      f_0 f_0' (1+{f^\phi_0}'')
      \left(\delta_{ss'} - \frac 12 \right)
\nonumber\\
 && - \frac{1}{16\pi^4}\beta^3\frac{y^2}{4}   
       \int_0^\infty dk \, k' dk'
        \theta(k^2{k'}^2-\beta_0^2)
        f_0 f_0' (1+{f^\phi_0}'')
\nonumber\\
&&\hphantom{XXX}
      \int_{-\infty}^\infty dk_z \, dk_z'
       \frac{ \theta(k^2-k_z^2)\theta(k_{z_2}'-k_z')\theta(k_z'-k_{z_1}')}
            { \sqrt{(k_{z_2}'-k_z')(k_z'-k_{z_1}')}}
\nonumber\\
&&\hphantom{XXX}
 \times \left(  \frac{\tilde{\omega}_0\tilde{\omega}_0'}
                     {\omega_0\omega_0'}
              + \frac{\omega_0\omega_0'+\beta_0 - k_zk_z'}
                     {\omega_0\omega_0'\tilde{\omega}_0\tilde{\omega}_0'}k_zk_z'
        \right) ss'
\label{Gamma0mu}
\eeqa
and
\beqa
    \Gamma^{0u}_{ss'}
  &=&
    \frac{1}{4\pi^3}\beta^3\frac{y^2}{4}
     \int_0^\infty k\,dk\,k'\,dk'\,
      \theta(k^2{k'}^2-\beta_0^2)
      \frac{\omega_0\omega_0'+\beta_0}{\omega_0\omega_0'}
      f_0 f_0' (1+{f^\phi_0}'')
\nonumber\\
&&\hphantom{XXXX}
      \frac 13 \beta^2
      \left( (1-f_0+{f^\phi_0}'')k^2 - (1-f'_0+{f^\phi_0}'')\beta_0 \right)
      \left(\delta_{ss'} - \frac 12 \right)
\nonumber\\
 && + \frac{1}{16\pi^4}\beta^3\frac{y^2}{4}   
       \int_0^\infty dk \, k' dk'
        \theta(k^2{k'}^2-\beta_0^2)
        f_0 f_0' (1+{f^\phi_0}'')
\nonumber\\
&&\hphantom{XXX}
      \int_{-\infty}^\infty dk_z \, dk_z'
       \frac{ \theta(k^2-k_z^2)\theta(k_{z_2}'-k_z')\theta(k_z'-k_{z_1}')}
            { \sqrt{(k_{z_2}'-k_z')(k_z'-k_{z_1}')}}
\nonumber\\
&&\hphantom{XXX}
 \times \left(  \frac{\tilde{\omega}_0\tilde{\omega}_0'}
                     {\omega_0\omega_0'}
              + \frac{\omega_0\omega_0'+\beta_0 - k_zk_z'}
                     {\omega_0\omega_0'\tilde{\omega}_0\tilde{\omega}_0'}k_zk_z'
        \right)
\nonumber\\
&&\hphantom{XXXX}
        \beta^2
        \left( (1-f_0+{f^\phi_0}'')k_zk'_z - (1-f'_0+{f^\phi_0}''){k_z'}^2 \right) ss'
\,.
\label{Gamma0u}
\eeqa
\newpage
In the first moment equation the collision term is 
\beq
  -2\beta^3 \int_{k_0=0}^\infty \frac{d^4k}{(2\pi)^4} \frac{k_z}{\omega_0}
   \Big[ {\cal K}_0^s(k) +  {\cal K}_0^s(-k) \Big]_{relax}
 = - \sum_{s'}
     \Big(
             \Gamma^{1u}_{ss'} u_{s'} 
           + v_w \beta  \Gamma^{1\mu}_{ss'} \mu_{s'}
     \Big)
\,,
\eeq
where, besides the source term, there are the rates
\beqa
    \Gamma^{1\mu}_{ss'}
  &=&
    \frac{|y|^2}{48\pi^3}\beta^4
     \int_0^\infty k\,dk\,k'\,dk'\,
      \theta(k^2{k'}^2-\beta_0^2)
      \frac{\omega_0\omega_0'+\beta_0}{\omega_0\omega_0'}
      f_0 f_0' (1+{f^\phi_0}'')
\nonumber\\
&&\hphantom{XXXXXXX}
      \times\, \frac{1}{\omega_0}
      \left( (1-f_0+{f^\phi_0}'')k^2 - (1-f'_0+{f^\phi_0}'')\beta_0 \right)
      \left(\delta_{ss'} - \frac 12 \right)
\nonumber\\
 && - \frac{|y|^2}{64\pi^4}\beta^4
       \int_0^\infty dk \, k' dk'
        \theta(k^2{k'}^2-\beta_0^2)
        f_0 f_0' (1+{f^\phi_0}'')
\nonumber\\
&&\hphantom{XXXX}
      \int_{-\infty}^\infty dk_z \, dk_z'
       \frac{ \theta(k^2-k_z^2)\theta(k_{z_2}'-k_z')\theta(k_z'-k_{z_1}')}
            { \sqrt{(k_{z_2}'-k_z')(k_z'-k_{z_1}')}}
\nonumber\\
&&\hphantom{XXXXX}
 \times \left(  \frac{\tilde{\omega}_0\tilde{\omega}_0'}
                     {\omega_0\omega_0'}
              + \frac{\omega_0\omega_0'+\beta_0 - k_zk_z'}
                     {\omega_0\omega_0'\tilde{\omega}_0\tilde{\omega}_0'}k_zk_z'
        \right)
\nonumber\\
&&\hphantom{XXXXXX}
        \frac{k_z}{\omega_0}
        \left( (1-f_0+{f^\phi_0}'')k_z - (1-f'_0+{f^\phi_0}'')k'_z \right) ss'
\label{Gamma1mu}
\eeqa
and
\beqa
    \Gamma^{1u}_{ss'}
  &=&
    \frac{|y|^2}{48\pi^3}\beta^4
     \int_0^\infty k\,dk\,k'\,dk'\,
      \theta(k^2{k'}^2-\beta_0^2)
      \frac{\omega_0\omega_0'+\beta_0}{\omega_0\omega_0'}
      f_0 f_0' (1+{f^\phi_0}'')
      \frac{1}{\omega_0}
      \left(k^2\delta_{ss'} + \frac 12 \beta_0 \right)
\nonumber\\
&& +\; \frac{|y|^2}{64\pi^4}\beta^4
       \int_0^\infty dk \, k' dk'
        \theta(k^2{k'}^2-\beta_0^2)
        f_0 f_0' (1+{f^\phi_0}'')
\nonumber\\
&&\hphantom{XXXXXX}
      \int_{-\infty}^\infty dk_z \, dk_z'
       \frac{ \theta(k^2-k_z^2)\theta(k_{z_2}'-k_z')\theta(k_z'-k_{z_1}')}
            { \sqrt{(k_{z_2}'-k_z')(k_z'-k_{z_1}')}}
\nonumber\\
&&\hphantom{XXXXXX}
 \times \left( \frac{\tilde{\omega}_0\tilde{\omega}_0'}
                    {\omega_0\omega_0'}
             + \frac{\omega_0\omega_0'+\beta_0 - k_zk_z'}
                    {\omega_0\omega_0'\tilde{\omega}_0\tilde{\omega}_0'}k_zk_z'
        \right)
        \frac{1}{\omega_0} k_zk'_z ss'
\,.
\label{Gamma1u}
\eeqa

\end{appendix}


\cleardoublepage

\cleardoublepage

\end{document}